\documentclass[12pt]{article}
\usepackage{epsf}
\usepackage{amsmath,amssymb}
\usepackage{cancel}
\usepackage{cite}
\usepackage{url}

\input colordvi

\usepackage[dvips]{graphicx}

\usepackage{comment}

\begin{document}

\newcommand{\mathhyphen}{\mathchar"712D} 
\newcommand{\tr}{{\rm Tr}}
\newcommand{\eps}[1]{\varepsilon_{#1}}
\newcommand{\epsD}{\bar{\varepsilon}_{\rm D}}
\newcommand{\msbar}{\overline{\rm MS}}
\renewcommand{\det}{{\rm Det}}

\newcommand{\fnc}{{\cal G}}
\newcommand{\cc}{{C}}

\renewcommand{\theequation}{\thesection.\arabic{equation}}

\renewcommand{\thefootnote}{\fnsymbol{footnote}}
\setcounter{footnote}{0}

\begin{titlepage}

\begin{center}

\hfill UT-18-04\\
\hfill March, 2018\\

\vskip .75in

{\Large\bf
  Decay Rate of Electroweak Vacuum\\[1mm]
  in the Standard Model and Beyond\\
}

\vskip .5in

{\large
  So Chigusa$^{\rm (a)}$, 
  Takeo Moroi$^{\rm (a)}$ and
  Yutaro Shoji$^{\rm (b)}$
}

\vskip 0.25in

$^{\rm (a)}${\em 
  Department of Physics, University of Tokyo, Tokyo 113-0033, Japan}

\vskip 0.1in
$^{\rm (b)}${\em 
Institute for Cosmic Ray Research, The University of Tokyo, 
Kashiwa 277-8582, Japan}

\end{center}

\vskip .5in

\begin{abstract}

  We perform a precise calculation of the decay rate of the electroweak
  vacuum in the standard model as well as in models beyond the standard
  model.  We use a recently-developed technique to calculate the decay
  rate of a false vacuum, which provides a gauge invariant calculation
  of the decay rate at the one-loop level.  We give a prescription to
  take into account the zero modes in association with translational,
  dilatational, and gauge symmetries. We calculate the decay rate per
  unit volume, $\gamma$, by using an analytic formula.  The decay rate
  of the electroweak vacuum in the standard model is estimated to be
  $\log_{10}\gamma\times{\rm Gyr~Gpc^3} =
  -582^{+40~+184~+144~+2}_{-45~-329~-218~-1}$, where the 1st, 2nd, 3rd,
  and 4th errors are due to the uncertainties of the Higgs mass, the top
  quark mass, the strong coupling constant and the choice of the
  renormalization scale, respectively.  The analytic formula of the
  decay rate, as well as its fitting formula given in this paper, is
  also applicable to models that exhibit a classical scale invariance at
  a high energy scale.  As an example, we consider extra fermions that
  couple to the standard model Higgs boson, and discuss their effects on
  the decay rate of the electroweak vacuum.

\end{abstract}

\end{titlepage}

\setcounter{page}{1}
\renewcommand{\thefootnote}{\#\arabic{footnote}}
\setcounter{footnote}{0}

\section{Introduction}
\label{sec_intro}
\setcounter{equation}{0}

In the standard model (SM) of particle physics, it has been known that
the Higgs quartic coupling may become negative at a high scale through
quantum corrections, so that the Higgs potential develops a deeper
vacuum.  The detailed shape of the Higgs potential depends on the Higgs
and the top masses; with the recently observed Higgs mass of $\sim 125\
{\rm GeV}$, it has been known that the electroweak (EW) vacuum is not
absolutely stable if the SM is valid up to $\sim 10^{10}\ {\rm GeV}$ or
higher.\footnote
{For the absolute stability of the EW vacuum in the SM, see
  \cite{Cabibbo:1979ay, Hung:1979dn, Lindner:1988ww, Ford:1992mv,
    Casas:1994qy, Casas:1996aq, Einhorn:2007rv, Ellis:2009tp,
    Degrassi:2012ry, Alekhin:2012py, Bezrukov:2012sa,
    Andreassen:2014gha, DiLuzio:2014bua, Bednyakov:2015sca}.}
In such a case, the EW vacuum can decay into the deeper vacuum through
tunneling in quantum field theory.  The lifetime of the EW vacuum has
been one of the important topics in particle physics and cosmology.

The decay rate of the EW vacuum has been discussed for a long time.
The calculation of the decay rate at the one-loop level first appeared
in \cite{Isidori:2001bm} and was also discussed in other literature
\cite{Arnold:1991cv, DiLuzio:2015iua, Espinosa:1995se,
  ArkaniHamed:2008ym, EliasMiro:2011aa, Plascencia:2015pga,
  Espinosa:2016nld, Lalak:2016zlv, Andreassen:2017rzq,
  Chigusa:2017dux}.  However, there are subtleties in the treatment of
zero modes related to the gauge symmetry breaking, which make it
difficult to perform a precise and reliable calculation of the decay
rate.  The lifetime of a vacuum can be evaluated through a rate of
bubble nucleation in unit volume and unit time as formulated in
\cite{Coleman:1977py,Callan:1977pt}.  The rate is expressed in the
form of
\begin{align}
  \gamma = \mathcal Ae^{-\mathcal B},
\end{align}
where $\mathcal B$ is the action of a so-called bounce solution, and
prefactor $\mathcal A$ is quantum corrections having mass dimension 4.
Bounce solution is an $O(4)$ symmetric solution of the Euclidean
equations of motion, connecting the two vacua. Although the dominant
suppression of the decay rate comes from $\mathcal B$, the prefactor
$\mathcal A$ is also important. This is because of large quantum
corrections from the top quarks and the gauge bosons. Thus, it is
essential to calculate both $\mathcal A$ and $\mathcal B$ to determine
the decay rate precisely. In the SM, there are infinite bounce solutions
owing to (i) the classical scale invariance at a high energy scale, (ii)
the global symmetries corresponding to $SU(2)_L\times U(1)_Y/U(1)_{\rm
EM}$, as well as (iii) the translational invariance.  For the
calculation of the prefactor $\mathcal A$, a proper procedure to take
account of the effects of the zero modes related to (i) and (ii) were
not well understood until recently.  In addition, the previous
calculations of $\mathcal A$ were not performed in a gauge-invariant
way, which made the gauge invariance of the result unclear.

Recently, a prescription for the treatments of the gauge zero modes
was developed \cite{Endo:2017gal,Endo:2017tsz}, based on which a
complete calculation of the decay rate of the EW vacuum became
possible.  The calculation has been performed by the present authors
in a recent publication \cite{Chigusa:2017dux} and also by
\cite{Andreassen:2017rzq}.  The purpose of this paper is to give more
complete and detailed discussion about the calculation of the decay
rate.  In \cite{Chigusa:2017dux}, we have numerically evaluated the
functional determinants of fluctuation operators, which are necessary
for the calculation of the decay rate.  Here, we perform the
calculation analytically; a part of the analytic results was first
given in \cite{Andreassen:2017rzq}.  The effects of the zero modes and
the modes with the lowest angular momentum are carefully taken into
account, on which previous works had some confusion.  We give fitting
formulae of the functional determinants based on analytic results,
which are useful for the numerical calculation of the decay rate.  We
also provide a C++ package to study the \verb|EL|ectroweak
\verb|VA|cuum \verb|S|tability, \verb|ELVAS|, which is available at
\cite{ELVAS}.

In this paper, we discuss the calculation of the decay rate of the EW
vacuum in detail.  We first present a detailed formulation of the
calculation of the decay rate at the one-loop level.  We derive a
complete set of analytic formulae that can be used for any models that
exhibit classical scale invariance at a high energy scale like the SM.
Then, as one of the important applications, we calculate the decay
rate of the EW vacuum in the SM. We find that the lifetime of the EW
vacuum is much longer than the age of the universe. There, we see that
one-loop corrections from the top quark and the gauge bosons are very
large although there is an accidental cancellation. It shows the
importance of $\mathcal A$ for the evaluation of a decay rate.  We
also evaluate the decay rates of the EW vacuum for models with extra
fermions that couple to the Higgs field.  In such models, the EW
vacuum tends to be destabilized compared with that of the SM since the
quartic coupling of the Higgs field is strongly driven to a negative
value.  (For discussion about the stability of the EW vacuum in models
with extra particles, see \cite{Casas:1999cd, Gogoladze:2008ak,
  He:2012ub, Rodejohann:2012px, Chakrabortty:2012np, Chao:2012mx,
  Masina:2012tz, Khan:2012zw, Dev:2013ff, Kobakhidze:2013pya,
  Datta:2013mta, Chakrabortty:2013zja, Xiao:2014kba, Hamada:2014xka,
  Khan:2014kba, Bambhaniya:2014hla, Khan:2015ipa, Salvio:2015cja,
  Lindner:2015qva, Rose:2015fua, Espinosa:2016nld, Haba:2016zbu,
  Bambhaniya:2016rbb, Khan:2016sxm, Garg:2017iva}.)  We consider three
models that contain, in addition to the SM particles, (i) vector-like
fermions having the same SM charges as the left-handed quark and the
right-handed down quark, (ii) vector-like fermions with the same SM
charges as left-handed lepton and right-handed electron, and (iii) a
right-handed neutrino. We give constraints on their couplings and
masses, requiring that the lifetime of the EW vacuum be long enough.

This paper is organized as follows. In Section \ref{sec_formulation},
we summarize the formulation for the decay rate at the one-loop level,
where we provide an analytic formula for each field that couples to
the Higgs boson. The detail of the calculation is given in Appendices
\ref{apx_determinantJ} $-$ \ref{apx_divpart}.  In Section
\ref{sec_SM}, we evaluate the vacuum decay rate in the SM.  Readers
who are interested only in the results can skip over the former
section to this section. In Section \ref{sec_extra}, we analyze decay
rates in models with extra fermions. Finally, we conclude in Section
\ref{sec_conclusion}.

\section{Formulation}
\label{sec_formulation}
\setcounter{equation}{0}

We first discuss how we calculate the decay rate of the EW vacuum.  In
the SM, the EW vacuum becomes unstable due to the renormalization
group (RG) running of the quartic coupling constant of the Higgs
boson, which makes the quartic coupling constant negative at a high
scale.  In the SM, the instability occurs when the Higgs amplitude
becomes much larger than the EW scale.  Since the typical field value
for the bounce configuration is around that scale, we can neglect the
quadratic term in the Higgs potential.

In this section, we use a toy model with $U(1)$ gauge symmetry to
derive relevant formulae.  The calculation of the decay rate of the EW
vacuum is almost parallel to that in the case with $U(1)$ gauge
symmetry; the application to the SM case will be explained in the next
section.

\subsection{Setup}

Let us first summarize the setup of our analysis.  We study the decay
rate of a false vacuum whose instability is due to an RG running of
the quartic coupling constant of a scalar field, $\Phi$.  We assume that
$\Phi$ is charged under the $U(1)$ gauge symmetry (with charge $+1$);
the kinetic term includes
\begin{align}
  {\cal L}_{\rm kin} \ni
  \left[
    \left(
      \partial_\mu - i g A_\mu 
    \right) \Phi
  \right]^\dagger
  \left(
    \partial_\mu - i g A_\mu 
  \right) \Phi,
\end{align}
where $A_\mu$ is the gauge field and $g$ is the gauge coupling
constant, while we consider the following scalar potential:
\begin{align}
  V(\Phi) = \lambda(\Phi^\dagger\Phi)^2.
  \label{V(Phi)}
\end{align}
The quartic coupling, $\lambda$, depends on the renormalization scale,
$\mu$, and is assumed to become negative at a high scale due to the RG
effect.  As we have mentioned before, we neglect the quadratic term
assuming that $\lambda$ becomes negative at a much higher scale.  In
this setup, the scalar potential has scale invariance at the classical
level.  In the application to the case of the SM, $\Phi$ corresponds
to the Higgs doublet and $\lambda$ corresponds to the Higgs quartic
coupling constant.

Hereafter, we perform a detailed study of the effects of the fields
coupled to $\Phi$ on the decay rate of the false vacuum.  We consider
a Lagrangian that contains the following interaction terms:
\begin{align}
  \mathcal L_{\rm int} \ni
  \kappa\sigma^2|\Phi|^2
  + ( y \Phi \bar{\psi}_L \psi_R + {\rm h.c.} )
  + V(\Phi),
\end{align}
where $\sigma$ is a real scalar field, and $\psi_L$ and $\psi_R$ are
chiral fermions (with relevant $U(1)$ charges).\footnote
{We assume that there exist other chiral fermions that cancel out
  the gauge anomaly.}
We take $y$ real and $\kappa>0$.  We neglect dimensionful parameters
that are assumed to be much smaller than the typical scale of the
bounce.  In addition, gauge fixing is necessary to take into account
the effects of gauge boson loops.  Following \cite{Kusenko:1996bv,
Endo:2017tsz}, we take the gauge fixing function of the following
form:
\begin{align}
  \mathcal F = \partial_\mu A_\mu.
\end{align}
Then, the gauge fixing term and the Lagrangian of the ghosts (denoted
as $\bar{c}$ and $c$) are given by
\begin{align}
  \mathcal L_{\rm GF} & = \frac{1}{2\xi}\mathcal F^2, \\
  \mathcal L_{\rm FP} & = - \bar{c} \partial_\mu \partial_\mu c,
  \label{L_FP}
\end{align}
where $\xi$ is the gauge fixing parameter.\footnote
{In the non-Abelian case, one of $\partial_\mu$ in eq.\ \eqref{L_FP}
  is replaced by the covariant derivative. The interaction of the
  ghosts with the gauge field does not affect the following
  discussion.}

The bounce solution is an $O(4)$ symmetric object \cite{Coleman:1977th,
Blum:2016ipp}, and transforms under a global $U(1)$ symmetry.  Thus,
choosing the center of the bounce at $r=0$ (with $r\equiv\sqrt{x_\mu
x_\mu}$), we can write the bounce solution as
\begin{align}
  \Phi|_{\rm bounce} = \frac{1}{\sqrt{2}} e^{i\theta}
  \bar\phi(r),
  \label{bounce}
\end{align}
with our choice of the gauge fixing function, without loss of
generality.  Here, $\theta$ is a real parameter.  The function,
$\bar{\phi}$, obeys
\begin{align}
  \partial_r^2\bar\phi(r)
  +\frac{3}{r}\partial_r\bar\phi(r)
  -\lambda\bar\phi^3(r) = 0,
  \label{eq_Bounce}
\end{align}
with boundary conditions $\partial_r\bar\phi(0) = 0$ and
$\bar\phi(\infty) = 0$.  For a negative $\lambda$, we have a series of
Fubini-Lipatov instanton solutions
\cite{Fubini:1976jm,Lipatov:1976ny}:
\begin{align}
  \bar\phi(r) = \bar\phi_C
  \left(1+\frac{|\lambda|}{8}\bar\phi_C^2r^2\right)^{-1},
  \label{BounceSolution}
\end{align}
which is parameterized by $\bar\phi_C$ ({\it i.e.}, the field value at
the center of the bounce).  We also define $R$, which gives the size
of the bounce, as
\begin{align}
  R \equiv \sqrt{\frac{8}{|\lambda|}} \phi_C^{-1}.
\end{align}
The action of the bounce is given by
\begin{align}
 \mathcal B = \frac{8\pi^2}{3|\lambda|}.
\end{align}
Notice that the tree level action is independent of $\bar\phi_C$ owing to
the classical scale invariance.

Once the bounce solution is obtained, we may integrate over the
fluctuation around it.  We expand $\Phi$ as
\begin{align}
  \Phi = \frac{1}{\sqrt{2}} 
  e^{i\theta}
  \left( 
    \bar\phi + h + i \varphi
  \right),
\end{align}
where $h$ and $\varphi$ are the physical Higgs mode and the Nambu-Goldstone
(NG) mode, respectively.  At the one-loop level, the prefactor can be
decomposed as
\begin{align}
  \mathcal A e^{-\mathcal B} = \frac{\mathcal A^{(h)}}{\mathcal V_{\rm 4D}}
  \mathcal A^{(\sigma)}\mathcal A^{(\psi)}\mathcal A^{(A_\mu,\varphi)}
  \mathcal A^{(c,\bar{c})}e^{-\mathcal B},
  \label{eq_AeB}
\end{align}
where $\mathcal V_{\rm 4D}$ is the volume of spacetime, and $\mathcal
A^{(X)}$ is the contribution from particle $X$.  Each of the factors
has a form of
\begin{align}
 \mathcal A^{(X)} = \left(
  \frac{\det\mathcal M^{(X)}}{\det \widehat{\mathcal M}^{(X)}}
 \right)^{w^{(X)}},
\end{align}
where $\mathcal M^{(X)}$ and $\widehat{\mathcal M}^{(X)}$ are the
fluctuation operators around the bounce solution and around the false
vacuum, respectively.  Here, $w^{(X)} = 1$ for Dirac fermions and
Faddeev-Popov ghosts, and $w^{(X)} = -1/2$ for the other bosonic
fields.  The fluctuation operator is defined as second derivatives of
the action:
\begin{align}
 \mathcal M^{(X)} \delta^{(4)}(x-y) = \left\langle
 \frac{\delta^2S(X)}{\delta X(x)\delta X(y)}
 \right\rangle,
\end{align}
where the brackets indicate the evaluation around the bounce solution.
In addition, $\widehat{\mathcal M}^{(X)}$ can be obtained from
$\mathcal M^{(X)}$ with replacing $\bar\phi$ by the vacuum expectation
value at the false vacuum, {\it i.e.}, $\bar\phi\to0$.

Since the Faddeev-Popov ghosts do not couple to the Higgs boson with
the present choice of the gauge fixing function, ${\mathcal
  M}^{(c,\bar{c})}= \widehat{\mathcal M}^{(c,\bar{c})}$.  Thus, we
have
\begin{align}
  \mathcal A^{(c,\bar{c})}=1.
\end{align}

\subsection{Functional determinant}

For the evaluation of the functional determinants, we first decompose
fluctuations into partial waves, making use of the $O(4)$ symmetry of
the bounce \cite{Isidori:2001bm}. The basis is constructed from
$Y_{J,m_A,m_B}(\Omega)$, the hyperspherical function on $S^3$ with
$\Omega$ being a coordinate on $S^3$. The decomposition of each
fluctuation is given in Appendix \ref{apx_determinantJ}.  Here, $J=0,1/2,1,\cdots$ is
a non-negative half-integer that labels the total angular momentum in
four dimension, and $m_A$ and $m_B$ are the azimuthal quantum numbers
for the $A$-spin and the $B$-spin of $so(4)\simeq su(2)_A\times
su(2)_B$, respectively. The four-dimensional Laplacian operator acts
on the hyperspherical function as
\begin{align}
 -\partial^2 Y_{J,m_A,m_B}(\Omega) = \frac{L^2}{r^2}Y_{J,m_A,m_B}(\Omega),
\end{align}
with
\begin{align}
 L \equiv \sqrt{4J(J+1)}.
\end{align}
In addition, 
$Y_{J,m_A,m_B}(\Omega)$ is normalized as
\begin{align}
  \int d\Omega |Y_{J,m_A,m_B}|^2 = 1.
  \label{Y-normalization}
\end{align}

With the above expansion, the functional determinants can be expressed
as
\begin{align}
  \frac{\det\mathcal M^{(X)}}{\det \widehat{\mathcal M}^{(X)}}
  = \prod_{J=0}^{\infty}
  \left(
    \frac{\det \mathcal M^{(X)}_J}{\det \widehat{\mathcal M}^{(X)}_J}
  \right)^{n^{(X)}_J},
\end{align}
where $\mathcal M^{(X)}_J$ is the fluctuation operator for each
partial wave labeled by $J$.  In addition, $n^{(\psi)}_J =
(2J+1)(2J+2)$ for fermions, and $n^{(X)}_J = (2J+1)^2$ for the others.
The explicit forms of the fluctuation operators are shown in Appendix
\ref{apx_determinantJ}. Using a theorem \cite{Gelfand:1959nq,
  Dashen:1974ci, Kirsten:2003py, Kirsten:2004qv, Endo:2017tsz}, the
ratio of the functional determinants can be calculated as
\begin{align}
 \frac{\det \mathcal M^{(X)}_J}{\det \widehat{\mathcal M}^{(X)}_J} =
 \left(
  \lim_{r\to\infty}\frac{\det\Psi}{\det\hat\Psi}
 \right)
 \left(
  \lim_{r\to0}\frac{\det\Psi}{\det\hat\Psi}
 \right)^{-1},
 \label{eq_theorem}
\end{align}
where $\Psi = (\Psi_1,\Psi_2,\cdots)$ and $\hat\Psi =
(\hat\Psi_1,\hat\Psi_2,\cdots)$ are sets of independent solutions of
\begin{align}
  \mathcal M^{(X)}_J\Psi_i = &\, 0,
  \label{eqforPsi}\\
  \widehat{\mathcal M}^{(X)}_J\hat\Psi_i = &\, 0,
  \label{eqforPsihat}
\end{align}
and are regular at $r = 0$. Notice that, when $\mathcal M^{(X)}_J$ is
an $n\times n$ object, there are $n$ independent solutions that are
regular at $r = 0$. Since $\Psi_i$ and $\hat\Psi_i$ obey the same
linear differential equation at $r\to\infty$, the ratio of the two
determinant converges for each $J$.

With a Fubini-Lipatov instanton, we can calculate the ratio
analytically, as first pointed out in \cite{Andreassen:2017rzq}.  For
the convenience of readers, we give the details of the calculation in
Appendix \ref{apx_determinantJ}.  The ratios are given by
\begin{align}
  \frac{\det \mathcal M^{(h)}_J}{\det \widehat{\mathcal M}^{(h)}_J}
  & = \frac{2J(2J-1)}{(2J+3)(2J+2)},
  \label{detM^h}
  \\
  \frac{\det \mathcal M^{(\sigma)}_J}{\det \widehat{\mathcal M}^{(\sigma)}_J}
  & = \frac{\Gamma(2J+1)\Gamma(2J+2)}{\Gamma(2J+1-z_\kappa)\Gamma(2J+2+z_\kappa)},
  \\
  \frac{\det \mathcal M^{(\psi)}_J}{\det \widehat{\mathcal
      M}^{(\psi)}_J}
  & =
  \left[
    \frac{[\Gamma(2J+2)]^2}{\Gamma\left(2J+2-z_y\right)\Gamma\left(2J+2+z_y\right)}
  \right]^2,
  \\
  \frac{\det \mathcal M^{(A_\mu,\varphi)}_J}{\det \widehat{\mathcal M}^{(A_\mu,\varphi)}_J}
  & = \frac{J}{J+1}
  \left[
    \frac{\Gamma(2J+1)\Gamma(2J+2)}{\Gamma(2J+1-z_g)\Gamma(2J+2+z_g)}
  \right]^3.
  \label{detM^gauge}
\end{align}
Here, 
\begin{align}
  z_\kappa = &\, -\frac{1}{2}
  \left(
    1-\sqrt{1-8\frac{\kappa}{|\lambda|}}
  \right),\\
  ~z_g = &\, -\frac{1}{2}
  \left(
    1-\sqrt{1-8\frac{g^2}{|\lambda|}}
  \right),\\
  ~z_y = &\,
  i\frac{y}{\sqrt{\lambda}}.
\end{align}
where $g$ is the gauge coupling constant, and $\Gamma(z)$ is the gamma
function.

\subsection{Zero modes}
\label{subsec:zeromodes}

In the calculation of the decay rate of the EW vacuum with the present
setup, there show up zero modes in association with dilatation,
translation, and global transformation of the bounce solution.
Consequently, $\mathcal M_0^{(h)},\mathcal M_{1/2}^{(h)}$ and
$\mathcal M_0^{(A_\mu,\varphi)}$ have zero eigenvalues.  Their
determinants vanish as shown in eqs.\ \eqref{detM^h} and
\eqref{detM^gauge} (see also eq.~\eqref{eq_gaugezero}); a naive
inclusion of those results gives a divergent behavior of the decay
rate, which requires a careful treatment of the zero modes.  In the
present case, we can consider the effect of each partial wave (labeled
by $J$) separately.  Thus, in this subsection, we consider the case
where the fluctuation operator $\mathcal M_J^{(X)}$ for a certain
value of $J$ has a zero eigenvalue and discuss how to take account of
its effect.  Because only bosons have zero modes in the calculation of
vacuum decay rates, $\mathcal M_J^{(X)}$ is considered to be a
fluctuation operator of a bosonic field.

We first decompose the fluctuation of the bosonic field (denoted as $X$)
into eigenstates of the angular momentum:
\begin{align}
  X \ni \sum_i c_i \fnc_i (r) Y_{J,m_A,m_B} (\Omega),
\end{align}
where $c_i$ is the expansion coefficient while $\fnc_i$ obeys
\begin{align}
  \mathcal M_J^{(X)} \fnc_i = \omega_i \fnc_i,
\end{align}
with $\omega_i$ being the eigenvalue.  We normalize them as
\begin{align}
  \langle \fnc_i| \fnc_j \rangle = \frac{\delta_{ij}}{\mathcal N_i^2},
\end{align}
where the inner product is defined by
\begin{align}
  \langle \fnc_i| \fnc_j\rangle = \int dr r^3 \fnc_i^\dagger(r) \fnc_j(r).
\end{align}
We leave $\mathcal N_i$'s unspecified since, as we see below, the final
result is independent of them.  We denote the zero mode as $\fnc_0$ (and
hence $\omega_0=0$).

When there is a zero mode, the saddle point approximation for the path
integral breaks down and we need to go back to the original path
integral. The path integral is defined as the integration over the
coefficient $c_i$, and the functional determinant of $\mathcal
M_J^{(X)}$ originates from the following integral:
\begin{align*}
  \int \prod_i\frac{dc_i}{\sqrt{2\pi}\mathcal N_i}
  e^{-\frac{1}{2} c_j c_k \langle \fnc_j|\mathcal M_J^{(X)}|\fnc_k \rangle},
\end{align*}
where the summations over the indices $j$ and $k$ are implicit.
Applying the saddle point approximation for the modes other than the
zero mode, we get
\begin{align}
  \int \prod_i\frac{dc_i}{\sqrt{2\pi}\mathcal N_i}
  e^{-\frac{1}{2} c_j c_k \langle \fnc_j|\mathcal M_J^{(X)}|\fnc_k \rangle}
  = \int \frac{dc_0}{\sqrt{2\pi}\mathcal N_0}
  \prod_{i\neq0}\frac{1}{\sqrt{\omega_i}}.
\end{align}
Notice that $\mathcal M^{(X)}_J$, $\fnc_i$ and $\omega_i$ may depend on
$c_0$.

We are interested in the case where there exists a symmetry (at least at
the classical level) and the Lagrangian is invariant under a
transformation, which we parameterize by $z$.  The zero mode is in
association with such a symmetry.  Then, the transformation of the bounce
solution with $z\to z+\delta z$ can be seen as a shift of $X$ as
\begin{align}
  X \to X + \delta z \tilde{\fnc}_0 Y_{J,m_A,m_B}
  + \mathcal O(\delta z^2),
\end{align}
where the function $\tilde{\fnc}_0$ is proportional to $\fnc_0$.  The
integration over $c_0$ can thus be regarded as the integration over
the collective coordinate $z$:
\begin{align}
  \frac{dc_0}{\sqrt{2\pi}\mathcal N_0}
  \rightarrow
  \sqrt{\frac{\langle\tilde{\fnc}_0|\tilde{\fnc}_0\rangle}{2\pi}} dz,
  \label{eq_dzIntegral}
\end{align}
and hence
\begin{align}
  \int \prod_i\frac{dc_i}{\sqrt{2\pi}\mathcal N_i}
  e^{-\frac{1}{2} c_j c_k \langle \fnc_j|\mathcal M_J^{(X)}|\fnc_k \rangle}
  \to
  \int dz \sqrt{\frac{\langle\tilde{\fnc}_0|\tilde{\fnc}_0\rangle}{2\pi}}
  \prod_{i\neq0}\frac{1}{\sqrt{\omega_i}}.
  \label{FuncDetForZeromode}
\end{align}

Next, we discuss how we evaluate the integrand of the right-hand side
of eq.\ \eqref{FuncDetForZeromode}.  To omit the zero eigenvalue from
the functional determinant, we introduce a regulator to the
fluctuation operator:
\begin{align}
  \mathcal M_J^{(X)} \to \mathcal M_J^{(X)} + \nu \rho(r),
\end{align}
where $\nu$ is a small positive number, and $\rho(r)$ is an arbitrary
function that satisfies
\begin{align}
  \langle \tilde{\fnc}_0|\rho|\tilde{\fnc}_0\rangle = 2\pi.
  \label{rho-normalization}
\end{align}
Then, we have
\begin{align}
  \int \prod_{i}\frac{dc_i}{\sqrt{2\pi}\mathcal N_i}
  e^{-\frac{1}{2}c_jc_k \langle \fnc_j|({\mathcal M}_J^{(X)}+\nu \rho) |\fnc_k \rangle}
  = \sqrt{\frac{\langle\tilde{\fnc}_0|\tilde{\fnc}_0\rangle}{2\pi}}
  \frac{1}{\sqrt{\nu+\mathcal O(\nu^2)}}
  \prod_{i\neq 0}\frac{1}{\sqrt{\omega_i+\mathcal O(\nu)}},
\end{align}
which gives
\begin{align}
  \sqrt{\frac{\langle\tilde{\fnc}_0|\tilde{\fnc}_0\rangle}{2\pi}}
  \prod_{i\neq0}\frac{1}{\sqrt{\omega_i}}
  = \lim_{\nu\to0}\sqrt{\nu}\int \prod_{i}\frac{dc_i}{\sqrt{2\pi}\mathcal N_i}
  e^{-\frac{1}{2}c_jc_k\langle \fnc_j|({\mathcal M}_J^{(X)}+\nu \rho)|\fnc_k\rangle}.
\end{align}
The integration in the above expression is nothing but $\det
({\mathcal M}^{(X)}+\nu \rho)$, and can be evaluated with the use of
eq.~\eqref{eq_theorem}.  If $\mathcal M_J^{(X)}$ is a $1\times1$
object, for example, we obtain
\begin{equation}
  \frac{\det
  \left[
    \mathcal M^{(X)}_J+\nu\rho
  \right]
  }{\det\widehat{\mathcal M}^{(X)}_J}
  =
  \nu\lim_{r\to\infty}\frac{\check\Psi(r)}{\hat\Psi(r)}+\mathcal O(\nu^2),
\end{equation}
where
\begin{align}
  \mathcal M_J^{(X)}\check\Psi(r) = - \rho(r)\Psi(r),
  \label{eq_detPrimeFunc}
\end{align}
while the functions $\Psi$ and $\hat{\Psi}$ obey eqs.\
\eqref{eqforPsi} and \eqref{eqforPsihat}, respectively.  Then, we
interpret the functional determinant of the fluctuation operator with
the zero eigenvalue as
\begin{align}
  \left(
    \frac{\det\mathcal M^{(X)}_J}{\det\widehat{\mathcal M}^{(X)}_J}
  \right)^{-1/2}
  \to
  \int dz
  \left[
    \lim_{r\to\infty}\frac{\check\Psi(r)}{\hat\Psi(r)}
  \right]^{-1/2}.
  \label{eq_detPrime}
\end{align}

The above argument can be applied to the zero modes of our interest.
In Appendix \ref{apx_zeromode}, we obtain the following replacements
to take care of the dilatational, translational, and gauge zero modes:
\begin{align}
  \left|
    \frac{\det \mathcal M^{(h)}_0}{\det \widehat{\mathcal M}^{(h)}_0}
  \right|^{-1/2}
  & \to\int d \ln R
  \sqrt{\frac{16\pi}{|\lambda|}},
  \label{eq_dilatationalZeromode}
  \\
  \left(
    \frac{\det \mathcal M^{(h)}_{1/2}}{\det \widehat{\mathcal M}^{(h)}_{1/2}}
  \right)^{-2}
  & \to \left(
    \frac{32\pi}{|\lambda|}\right)^2\frac{
  {\mathcal V}_{\rm 4D}}{R^4},
  \\
  \left(
    \frac{\det \mathcal M^{(A_\mu,\varphi)}_0}
    {\det \widehat{\mathcal M}^{(A_\mu,\varphi)}_0}
  \right)^{-1/2}
  & \to \int d\theta \sqrt{\frac{16\pi}{|\lambda|}}.
\end{align}

\subsection{Renormalization}

After taking the product over $J$, we have a UV divergence and thus we
need to renormalize the result.  In this paper, we use the
$\msbar$-scheme, which is based on the dimensional regularization.  In
this subsection, we explain how the divergences can be subtracted
using counter terms in the $\msbar$-scheme.

Since the dimensional regularization cannot be directly used in this
evaluation, we first regularize the result by using the angular
momentum expansion as
\begin{align}
  \left[
    \ln\mathcal A^{(X)}
  \right]_{\eps{X}}
  \equiv w^{(X)}\sum_{J=0}^{\infty}
  \frac{n^{(X)}_J}{(1+\eps{X})^{2J}}\ln
  \left(
    \frac{\det \mathcal M^{(X)}_J}{\det \widehat{\mathcal M}^{(X)}_J}
  \right).
\end{align}
We call this regularization as ``angular momentum regularization.''
Here, $\eps{X}$ is a positive number, which will be taken to be zero at
the end of the calculation. Since the divergence is at most a power of
$J$, the regularized sum converges.  In Appendix \ref{apx_infiniteSum},
we calculate the sum analytically, and obtain
\begin{align}
  \left[
    \ln\mathcal A^{(\sigma)}
  \right]_{\eps{\sigma}}
  = &\, 
  - \frac{\kappa}{|\lambda|}
  \left(
    \frac{1}{\eps{\sigma}^2}+\frac{2}{\eps{\sigma}}
    +\frac{\kappa}{3|\lambda|}\ln\eps{\sigma}
  \right)
  -\frac{1}{2}\mathcal S_\sigma(z_\kappa)
  +\mathcal O(\eps{\sigma}),
  \\
  \left[
    \ln\mathcal A^{(\psi)}
  \right]_{\eps{\psi}}
  = &\, \frac{y^2}{|\lambda|}
  \left(
    \frac{2}{\eps{\psi}^2}+\frac{5}{\eps{\psi}}
    +\frac{1}{3}\frac{y^2+2|\lambda|}{|\lambda|}\ln\eps{\psi}
  \right)+\mathcal S_\psi(z_y)+\mathcal O(\eps{\psi}),
\end{align}
where the functions, $\mathcal S_\sigma$ and $\mathcal S_\psi$, are given in
eqs.\ \eqref{fnS_B} and \eqref{fnS_F}, respectively.  In addition,
we define
\begin{align}
  \mathcal A^{(h)}
  & = {\mathcal V}_{\rm 4D} \int d \ln R
  \frac{1}{R^4}
  \mathcal A'^{(h)},
  \\
  \mathcal A^{(A_\mu,\varphi)}
  & = \int d\theta \mathcal A'^{(A_\mu,\varphi)}.
\end{align}
Then, the primed quantities are given by
\begin{align}
  \left[
    \ln\mathcal A'^{(h)}
  \right]_{\eps{h}}
  = &\,
  -\frac{1}{2}
  \left[
    \left(
      \ln\frac{|\lambda|}{16\pi}
    \right)+4
    \left(
      \ln\frac{|\lambda|}{32\pi}
    \right)+\sum_{J=1}^{\infty}\frac{(2J+1)^2}{(1+\eps{h})^{2J}}
    \ln\frac{2J(2J-1)}{(2J+3)(2J+2)}
  \right]
  \nonumber \\ = &\,
  3 \left(
    \frac{1}{\eps{h}^2}+\frac{2}{\eps{h}}-\ln\eps{h}
  \right)
  - \frac{3}{4} -3 \gamma_E - 6\ln A_G 
  + \frac{5}{2}\ln\frac{\pi}{3} - \frac{5}{2} \ln\frac{|\lambda|}{8}
  +\mathcal O(\eps{h}),
  \\
  \left[
    \ln\mathcal A'^{(A_\mu,\varphi)}
  \right]_{\eps{A}}
  = &\,
    -\frac{1}{2}
    \Bigg[
    \left(
      \ln\frac{|\lambda|}{16\pi}
    \right)
    \nonumber \\ &\,
    +\sum_{J=1/2}^{\infty}\frac{(2J+1)^2}{(1+\eps{A})^{2J}}
    \ln\frac{J}{J+1}
    \left(
      \frac{\Gamma(2J+1)\Gamma(2J+2)}{\Gamma(2J+1-z_g)\Gamma(2J+2+z_g)}
    \right)^3 \Bigg]
  \nonumber \\ = &\,
  - \left(
    \frac{3g^2}{|\lambda|}-1
  \right)
  \left(
    \frac{1}{\eps{A}^2}+\frac{2}{\eps{A}}
  \right)
  - \left(
    \frac{1}{3}+\frac{g^4}{|\lambda|^2}
  \right) \ln\eps{A}
  \nonumber \\ &\,
  + \frac{3}{4} - \frac{1}{3}\gamma_E - 2\ln A_G
  -\frac{1}{2}
  \ln\frac{|\lambda|}{8\pi}
  \nonumber \\ &\,
  -\frac{3}{2} \mathcal S_\sigma(z_g)
  -\frac{3}{2} \ln\Gamma(1-z_g)\Gamma(2+z_g)
  +\mathcal O(\eps{A}),
\end{align}
where $A_G\simeq 1.282$ is the Glaisher number.  

Next, we relate the above results with those based on the dimensional
regularization in $D$ dimension, which contain the regularization
parameter, $\epsD$, defined as
\begin{align}
  \frac{1}{\epsD} = \frac{2}{4-D}+\ln4\pi-\gamma_E,
\end{align}
with $\gamma_E$ being the Euler's constant.  We convert the results
based on two different regularizations by calculating the following
quantity:
\begin{align}
 \left[\ln\mathcal A^{(X)}\right]_{\rm div}
  & = w^{(X)}\tr
 \left[
  \left(
   \widehat{\mathcal M}^{(X)}
  \right)^{-1}\delta\mathcal M^{(X)}
 \right]-\frac{w^{(X)}}{2}\tr
 \left[
  \left(
   \widehat{\mathcal M}^{(X)}
  \right)^{-1}\delta\mathcal M^{(X)}
  \left(
   \widehat{\mathcal M}^{(X)}
  \right)^{-1}\delta\mathcal M^{(X)}
 \right]\nonumber                         \\
  & = w^{(X)}\sum_{J=0}^{\infty}n^{(X)}_J
 \left[
  \ln\frac{\det
   \left[
    \widehat{\mathcal M}^{(X)}_J+\delta\mathcal M_J^{(X)}
   \right]}{\det \widehat{\mathcal M}^{(X)}_J}
 \right]_{\mathcal O(\delta\mathcal M_J^2)},
 \label{eq_deltaMExpansion}
\end{align}
where
\begin{align}
 \delta\mathcal M^{(X)} = \mathcal M^{(X)}-\widehat{\mathcal M}^{(X)},
\end{align}
and $\delta \mathcal M^{(X)}_J$ is defined similarly. Here,
$[\cdots]_{\mathcal O(\delta)}$ indicates the expansion up to
$\delta$. The most important point is that
$\left[\ln\frac{\det\mathcal M^{(X)}}{\det\widehat{\mathcal
      M}^{(X)}}\right]_{\rm div}$ has the same divergence as
$\left[\ln\frac{\det\mathcal M^{(X)}}{\det\widehat{\mathcal
      M}^{(X)}}\right]$ does when $\delta\mathcal M^{(X)}$ does not
have a derivative operator.

The first line of eq.~\eqref{eq_deltaMExpansion} can be calculated by
directly evaluating the traces in a momentum space with using the
dimensional regularization; the result is denoted as $\left[\ln
  \mathcal A^{(X)}\right]_{{\rm div}, \epsD}$.  On the other hand, the
second line of eq.~\eqref{eq_deltaMExpansion} can be evaluated as
\begin{align}
  \left[
    \ln\frac{\det
      \left[
        \widehat{\mathcal M}^{(X)}_J+\delta\mathcal M_J^{(X)}
      \right]}{\det \widehat{\mathcal M}^{(X)}_J}
  \right]_{\mathcal O(\delta\mathcal M_J^2)}
  = &\,
  \tr
  \left[
    \hat\Psi^{-1}\Psi^{(1)}
  \right]+\tr
  \left[
    \hat\Psi^{-1}\Psi^{(2)}
  \right]
  \nonumber \\ &\,
  -\frac{1}{2}\tr
  \left[
    \hat\Psi^{-1}\Psi^{(1)}\hat\Psi^{-1}\Psi^{(1)}
  \right],
  \label{eq_expansionEachJ}
\end{align}
where
\begin{align}
 \widehat{\mathcal M}^{(X)}_J\Psi^{(p)} = -\delta\mathcal M_J^{(X)}\Psi^{(p-1)},
\end{align}
for $p=1$ and $2$, and $\Psi^{(0)} = \hat\Psi$. Then, we calculate
\begin{align}
  \left[
    \ln\mathcal A^{(X)}
  \right]_{{\rm div}, \eps{X}}
  \equiv w^{(X)}\sum_J\frac{n^{(X)}_J}{(1+\eps{X})^{2J}}
  \left[
    \ln\frac{\det\left[\widehat{\mathcal M}^{(X)}_J+\delta\mathcal M_J^{(X)}
      \right]}{\det \widehat{\mathcal M}^{(X)}_J}
  \right]_{\mathcal O(\delta\mathcal M_J^2)}.
  \label{eq_regDiv}
\end{align}

We relate the results based on two regularizations by replacing
\begin{align}
  \left[
    \ln\mathcal A^{(X)}
  \right]_{{\rm div}, \eps{X}}
  \rightarrow
  \left[
    \ln\mathcal A^{(X)}
  \right]_{{\rm div}, \epsD}.
\end{align}
The expressions of $[\ln\mathcal A^{(X)}]_{{\rm div}, \eps{X}}$ and
$[\ln\mathcal A^{(X)}]_{{\rm div}, \epsD}$ for each field are given in
Appendix \ref{apx_divpart}, where we further simplify the relation so
that the left-hand side only includes terms that are divergent at the
limit of $\eps{X}\to 0$.  We summarize the
results below:
\begin{itemize}
\item Scalar field:
  \begin{align}
    \left(
      \frac{1}{\eps{\sigma}^2}+\frac{2}{\eps{\sigma}}
      +\frac{\kappa}{3|\lambda|}\ln\eps{\sigma}
    \right)
    \rightarrow
    -1-\frac{\kappa}{3|\lambda|}
    \left(
      \frac{1}{2\epsD}+1+\gamma_E+\ln\frac{\mu R}{2}
    \right),
    \label{AMR2DMR_sigma}
  \end{align}
\item Higgs field:
  \begin{align}
    \left(
      \frac{1}{\eps{h}^2}+\frac{2}{\eps{h}}-\ln\eps{h}
    \right)
    \rightarrow
    \frac{1}{2\epsD}+\gamma_E+\ln\frac{\mu R}{2},
    \label{AMR2DMR_higgs}
  \end{align}
\item Fermion:
  \begin{align}
    & \left(
      \frac{2}{\eps{\psi}^2}+\frac{5}{\eps{\psi}}
      +\frac{1}{3}\frac{y^2+2|\lambda|}{|\lambda|}\ln\eps{\psi}
    \right)
    \nonumber \\
    & \hspace{3ex}\to -\frac{y^2}{3|\lambda|}
    \left(
      \frac{1}{2\epsD}+1+\gamma_E+\ln\frac{\mu R}{2}
    \right)-\frac{2}{3}
    \left(
      \frac{1}{2\epsD}+\frac{25}{4}+\gamma_E+\ln\frac{\mu R}{2}
    \right), 
    \label{AMR2DMR_fermion} 
  \end{align}
\item Gauge and NG fields:
  \begin{align}
    & \left(
      \frac{3g^2}{|\lambda|}-1
    \right)
    \left(
      \frac{1}{\eps{A}^2}+\frac{2}{\eps{A}}
    \right)+
    \left(
      \frac{1}{3}+\frac{g^4}{|\lambda|^2}
    \right)\ln\eps{A}
    \nonumber \\
    & \hspace{3ex}\to
    -
    \left(
      \frac{1}{3}+\frac{2g^2}{|\lambda|}+\frac{g^4}{|\lambda|^2}
    \right)
    \left(
      \frac{1}{2\epsD}+1+\gamma_E+\ln\frac{\mu R}{2}
    \right)
    +1+\frac{g^4}{|\lambda|^2}
    \left(
      \frac{31}{3}-\pi^2
    \right)
    \label{AMR2DMR_gaugeNG}
  \end{align}
\end{itemize}
Subtracting $1/\epsD$, we obtain the renormalized prefactor in the
$\msbar$-scheme.

\subsection{Dilatational zero mode}

In the calculation of the decay rate of the EW vacuum, we have an
integral over $R$ in association with the classical scale invariance,
as we saw in eq.~\eqref{eq_dilatationalZeromode}.  So far, we have
performed a one-loop calculation of the decay rate, based on which the
decay rate is found to behave as
\begin{align}
  \gamma^{\rm (one\mathhyphen loop)}
  \propto\int d\ln R
  \frac{1}{R^4}
  \exp \left[
    - \frac{8\pi^2}{3|\lambda (\mu)|}
    - \frac{8\pi^2 \beta_\lambda^{(1)}}{3|\lambda (\mu)|^2} \ln \mu R
  \right],
  \label{gamma(1-loop)}
\end{align}
where $\beta_\lambda^{(1)}$ is the one-loop $\beta$-function of
$\lambda$.  (Here, we only show the $R$- and $\mu$-dependencies of the
one-loop corrections.)  Thus, using the purely one-loop result, the
integral does not converge.

We expect, however, that the integration can converge once higher
order effects are properly included.  To see the detail of the path
integral over the dilatational zero mode, let us denote the decay rate
as
\begin{align}
  \gamma^{\rm (full)} = \int d\ln R
  \frac{e^{-\mathcal B_{\rm eff}}}{R^4},
\end{align}
where $\mathcal{B}_{\rm eff}$ fully takes account of all the effects of higher
order loops.

In order to discuss how $\mathcal{B}_{\rm eff}$ should behave, it is
instructive to rescale the coordinate variable as
\begin{align}
  \tilde x_\mu \equiv \sqrt{\frac{|\lambda|}{8}}\bar\phi_C x_\mu
  = \frac{x_\mu}{R},
\end{align}
as well as the fields as
\begin{align}
  \tilde \Phi \equiv &\, \bar{\phi}_C^{-1} \Phi,
  \\
  \tilde \sigma \equiv &\, \bar{\phi}_C^{-1} \sigma,
  \\
  \tilde A_\mu \equiv &\, \bar{\phi}_C^{-1} A_\mu,
  \\
  \tilde{c} \equiv &\, \bar{\phi}_C^{-1} c,
  \\
  \tilde{\bar{c}} \equiv &\, \bar{\phi}_C^{-1} \bar{c},
  \\
  \tilde \psi \equiv &\,
  \left(\frac{|\lambda|}{8}\right)^{-1/4} \bar\phi_C^{-3/2}
  \psi.
\end{align}
Using the rescaled fields, all the explicit scales disappear from the
action as a result of scale invariance:
\begin{align}
 \frac{1}{\hbar}
 \int d^4 x {\cal L} = 
 \frac{1}{\tilde{\hbar}}
 \int d^4 \tilde{x} 
 \tilde{\cal L}  \left(
   \frac{\kappa}{|\lambda|}, 
   \frac{y}{\sqrt{|\lambda|}},
   \frac{g}{\sqrt{|\lambda|}}
 \right),
\end{align}
where ${\cal L}$ is the total Lagrangian,
\begin{align}
 \tilde{\hbar} = \frac{|\lambda|}{8}\hbar,
\end{align}
and $\tilde{\cal L}$ includes canonically normalized rescaled fields
and depends only on the combinations $\frac{\kappa}{|\lambda|}$,
$\frac{y}{\sqrt{|\lambda|}}$ and $\frac{g}{\sqrt{|\lambda|}}$.  In
addition, the rescaled bounce solution is given by
\begin{align}
 \frac{1}{1+\tilde x^2}.
 \label{eq_rescaledBounce}
\end{align}

Based on $\tilde{\cal L}$ and $\tilde{\hbar}$, we expect:
\begin{itemize}
\item Only positive powers of $\frac{\kappa}{|\lambda|}$,
  $\frac{y}{\sqrt{|\lambda|}}$ and $\frac{g}{\sqrt{|\lambda|}}$ appear
  in the decay rate since there is no singularity when any of these
  goes to zero. In paticular, they cannot be in a logarithmic function.
\item When we renormalize divergences using dimensional
  regularization, we introduce a renormalization scale $\tilde\mu$.
  It is always in a logarithmic function and is related to the
  original renormalization scale as
  \begin{align}
    \tilde\mu = \mu R.
  \end{align}
\item In subtracting zero modes associated with transformations of
  eq.~\eqref{eq_rescaledBounce}, the result should be again a
  polynomial of $\frac{\kappa}{|\lambda|}$,
  $\frac{y}{\sqrt{|\lambda|}}$ and
  $\frac{g}{\sqrt{|\lambda|}}$. Notice that, for each zero mode, we
  have $\sqrt{1/\tilde{\hbar}}$ since eq.~\eqref{eq_dzIntegral}
  implies
  \begin{align}
    \frac{dc_0}{\sqrt{2\pi\tilde{\hbar}}\mathcal N_0}
    = dz
    \sqrt{\frac{\langle\tilde \fnc_0|\tilde \fnc_0\rangle}{2\pi\tilde{\hbar}}}.
  \end{align}
\item Quantum corrections have $\tilde{\hbar}^{\ell -1}$ at the
  $\ell$-th loop since the loop expansion is equivalent to the
  $\tilde{\hbar}$ expansion.
\end{itemize}

Based on the above arguments, $\mathcal B_{\rm eff}$ is expected to be
expressed as
\begin{align}
 \mathcal B_{\rm eff}
  = 
  \frac{8\pi^2}{3|\lambda (\mu)|}
  + \frac{n_{\rm zero}}{2} \ln\frac{|\lambda (\mu)|}{8}
  + \sum_{\ell =1}^{\infty}
  \left(
    \frac{|\lambda (\mu)|}{8}
  \right)^{\ell-1}\mathcal P_\ell
  \left(
    \frac{\kappa (\mu)}{|\lambda (\mu)|},
    \frac{y (\mu)}{\sqrt{|\lambda (\mu)|}},
    \frac{g (\mu)}{\sqrt{|\lambda (\mu)|}},
    \ln \mu R
  \right),
  \label{B_eff}
\end{align}
where $\mathcal P_\ell$ is the contribution at the $\ell$-loop level,
and $n_{\rm zero}$ is the number of zero modes.

If the effects of higher order loops are fully taken into account,
$\mathcal{B}_{\rm eff}$ should be independent of $\mu$ because the decay
rate is a physical quantity; in such a case, we may choose any value of
the renormalization scale $\mu$.  In the perturbative calculation, the
$\mu$-dependence is expected to cancel out order-by-order; as shown in
eq.\ \eqref{gamma(1-loop)}, we can explicitly see the cancellation of
the $\mu$-dependence at the one-loop level \cite{Endo:2015ixx}.  In our
calculation so far, however, we only have the one-loop result, in which
$\mu$ dependence remains.  As indicated in eq.\ \eqref{B_eff}, the $\mu$
dependence shows up in the form of $\ln^p \mu R$ with $p=1$, $2$,
$\cdots$.  If $|\ln\mu R|\gg 1$, the logarithmic terms from higher order
loops may become comparable to the tree-level bounce action and the
perturbative calculation breaks down.  In order to make our one-loop
result reliable, we should take $\mu\sim O(1/R)$, {\it i.e.}, we use
$R$-dependent renormalization scale $\mu$.\footnote
{This is equivalent to summing over large logarithmic terms appearing
  in higher loop corrections if we work with a fixed $\mu$. Since we
  have calculated the decay rate at the one-loop level, it is
  preferable to use, at least, the two-loop $\beta$-function to
  include the next-to-leading logarithmic corrections.}
With such a choice of $\mu$ (as well as with the use of coupling
constants evaluated at the renormalization scale $\mu$), the integration
over the size of the bounce is dominated only by the region where
$|\lambda (1/R)|$ becomes largest.  In the case of the SM, the
integration over the size of the bounce converges with this prescription
as we show in the following section.

\subsection{Final result}

Here, we summarize the results obtained in the previous subsections
and Appendices.  The decay rate with a resummation of important
logarithmic terms is given by
\begin{align}
  \gamma = \int d\ln R \frac{1}{R^{4}}
  \left[
    \mathcal A'^{(h)}\mathcal A^{(\sigma)}\mathcal A^{(\psi)}
    \mathcal A^{(A_\mu,\varphi)}e^{-\mathcal B}
  \right]_{\msbar,~\mu\sim 1/R},
\end{align}
where
\begin{align}
  \left[
    \ln\mathcal A'^{(h)}
  \right]_{\msbar}
  & = -\frac{3}{4}-6\ln A_G+\frac{5}{2}\ln\frac{\pi}{3}
      -\frac{5}{2}\ln\frac{|\lambda|}{8}+3\ln\frac{\mu R}{2},
  \label{A'^(h)}
  \\
  \left[
    \ln\mathcal A^{(\sigma)}
  \right]_{\msbar}
  & =
    -\frac{1}{2}\mathcal S_\sigma(z_\kappa)+\frac{\kappa}{|\lambda|}+\frac{\kappa^2}{3|\lambda|^2}
    \left(
      1+\gamma_E+\ln\frac{\mu R}{2}
    \right), \\
  \left[
    \ln\mathcal A^{(\psi)}
  \right]_{\msbar}
  & = -\frac{y^4}{3|\lambda|^2}
  \left(
    1+\gamma_E+\ln\frac{\mu R}{2}
  \right)-\frac{2y^2}{3|\lambda|}
  \left(
    \frac{25}{4}+\gamma_E+\ln\frac{\mu R}{2}
  \right)+\mathcal S_\psi(z_y),   \\
  \left[
    \ln\mathcal A^{(A_\mu,\varphi)}
  \right]_{\msbar}
  & = \ln\mathcal V_G+
  \left[
    \ln\mathcal A'^{(A_\mu,\varphi)}
  \right]_{\msbar},\label{A(A,NG)}
\end{align}
with
\begin{align}
  \left[
    \ln\mathcal A'^{(A_\mu,\varphi)}
  \right]_{\msbar}
  = &\,
  \left(
    \frac{1}{3}+\frac{2g^2}{|\lambda|}+\frac{g^4}{|\lambda|^2}
  \right)
  \left(
    1+\gamma_E+\ln\frac{\mu R}{2}
  \right)
  \nonumber
  \\ & 
  -\frac{g^4}{|\lambda|^2}
  \left(
      \frac{31}{3}-\pi^2
  \right)
  -\frac{1}{4}-\frac{1}{3}\gamma_E-2\ln
  A_G-\frac{1}{2}\ln\frac{|\lambda|}{8}+\frac{1}{2}\ln\pi
  \nonumber
  \\ & 
  -\frac{3}{2}\mathcal S_\sigma(z_g)-\frac{3}{2}\ln\Gamma(1-z_g)\Gamma(2+z_g).
  \label{A'(A,NG)}
\end{align}
Here, $\mathcal V_G$ is the volume of the group space generated by the
broken generators.  The definitions of $\mathcal S_\sigma(z)$,
$\mathcal S_\psi(z)$ can be found in Appendix \ref{apx_infiniteSum}.
Here, we note that the analytic results for the scalar, Higgs, and
fermion contributions were first given in \cite{Andreassen:2017rzq}
with different expression.  We emphasize that the final result does
not depend on the gauge parameter, $\xi$, and hence our result is
gauge invariant. The above result is also applicable to the case where
the $U(1)$ symmetry is not gauged as explained in Appendix
\ref{apx_global}.

We have also derived fitting formulae of the functions necessary for
the calculation of the decay rate; the result is given in Appendix
\ref{apx_recipe}.  The fitting formulae are particularly useful for the
numerical calculation of the decay rate.  In addition, a C++ package to
study the \verb|EL|ectroweak \verb|VA|cuum \verb|S|tability,
\verb|ELVAS|, is available at \cite{ELVAS}, which is also applicable
to various models with (approximate) classical scale invariance.

\section{Decay Rate of the EW Vacuum in the SM}
\label{sec_SM}
\setcounter{equation}{0}

\subsection{Decay rate}

Now, we are in a position to discuss the decay rate of the EW vacuum
in the SM.  As we have discussed, the decay of the EW vacuum is
induced by the bounce configuration whose energy scale is much higher
than the EW scale.  Thus, we approximate the Higgs potential
as\footnote
{We assume that the Higgs potential given in eq.\ \eqref{V(H)} is
  applicable at a high scale.  In particular, we assume that the
  effect of Planck suppressed operators, which may arise from the
  effect of quantum gravity, is negligible.  For the discussion about
  the effect of Planck suppressed operators, see \cite{Burgess:2001tj,
    Branchina:2013jra, Lalak:2014qua, Branchina:2014rva,
    Branchina:2014usa, Branchina:2015nda, Branchina:2016bws,
    Salvio:2016mvj, Bentivegna:2017qry}.}
\begin{align}
  V(H) = \lambda(H^\dagger H)^2,
  \label{V(H)}
\end{align}
where $H$ is the Higgs doublet in the SM and $\lambda$ is the Higgs
quartic coupling constant.  Notice that $\lambda$ depends on the
renormalization scale $\mu$; in the SM, $\lambda$ becomes negative
when $\mu$ is above $\sim 10^{10}$ GeV with the best-fit values of the
SM parameters.  In addition, the relevant part of the Yukawa couplings
are given by
\begin{align}
  \mathcal L_{\rm Yukawa} \ni
  y_t H \bar{q}_L t_R^{c} + {\rm h.c.},
\end{align}
where $q_L$ is the left-handed 3rd generation quark
doublet, $t_R^c$ is the right-handed anti-top quark, and $y_t$
is the top Yukawa coupling constant.

Assuming that $\lambda<0$, the bounce solution for the SM is given
by
\begin{align}
  H|_{\rm bounce} = \frac{1}{\sqrt{2}} e^{i\theta^a \sigma^a}
  \begin{pmatrix}
    0 \\
    \bar\phi(r)
  \end{pmatrix},
\end{align}
where $\sigma^a$ is the Pauli matrices and function $\bar\phi$ is given
by eq.\ \eqref{BounceSolution}.  In particular, remember that $\bar\phi$
contains a free parameter, which we choose $R$, because of the classical
scale invariance.

The results given in the previous section can be easily applied to the
case of the SM.  Taking account of the effects of the (physical) Higgs
boson, top quark, and weak bosons (as well as NG bosons),\footnote
{We checked that the effect of the bottom quark is numerically
  unimportant.}
the decay rate of the EW vacuum in the SM can be written in the
following form:
\begin{align}
  \gamma = \int d \ln R
  \frac{1}{R^4}
  \left[
    \mathcal A'^{(h)} \mathcal A^{(t)}
    \mathcal A^{(W,Z,\varphi)}e^{-\mathcal B}
  \right]_{\msbar}.
\end{align}
As we have mentioned, the relevant renormalization scale of the
integrand is $\mu\sim O(1/R)$; in the following numerical analysis, we
take $\mu=1/R$ unless otherwise stated.  If $\lambda(\mu)$ is positive,
there is no bounce solution; the integrand is taken to be zero in such a
case.  In addition, since we neglect the mass term in the Higgs
potential, $1/R$ should be much larger than the EW scale.  This
condition is automatically satisfied in the present analysis because
$\lambda$ becomes negative at the scale much higher than the electroweak scale.

The Higgs contribution $A'^{(h)}$ is given in eq.\ \eqref{A'^(h)},
while the top-quark contributions is given by
\begin{align}
  \left[
    \ln \mathcal A^{(t)}
  \right]_{\msbar}
  & = \left.3
    \left[
      \ln \mathcal A^{(\psi)}
    \right]_{\msbar}
  \right|_{y\to y_t},
\end{align}
where the factor of $3$ is the color factor.

As for the gauge contributions, we have $SU(2)_L\times U(1)_Y/U(1)_{\rm
EM}$ broken symmetries, instead of $U(1)$ in our previous example. Thus,
we have a different volume of the group space, $\mathcal V_G$.  To
calculate $\mathcal V_G$, we first expand $H$ around the bounce solution
with $\theta^a=0$ as
\begin{align}
  H = \frac{1}{\sqrt{2}}
  \begin{pmatrix}
    i (\varphi^1 -i \varphi^2)
    \\
    \bar\phi - i \varphi^3
  \end{pmatrix}.
\end{align}
Here, $\varphi^1$ and $\varphi^2$ are NG bosons absorbed by charged
$W$-bosons while $\varphi^3$ is that by $Z$-boson.  With the change of
$\theta^a$, the NG modes are transformed as
\begin{align}
  \varphi^a \to 
  \varphi^a + \theta^a \tilde{\fnc}_G Y_{0,0,0} + \mathcal O(\theta^2),
\end{align}
where
\begin{align}
  \tilde{\fnc}_G = \sqrt{2\pi^2}\bar\phi.
\end{align}
The integration over the zero modes in association with
the gauge transformation of the bounce solution can be replaced by the 
integration over the parameter $\theta^a$:
\begin{align}
  \int d^3 \theta = 2\pi^2 \equiv {\cal V}_{SU(2)},\label{220155_21Feb18}
\end{align}
with the above definition of $\theta^a$.
Then, following the argument given in Appendix \ref{apx_zeromode}, the
gauge contribution is evaluated as
\begin{align}
  \left[
    \ln \mathcal A^{(W,Z,\varphi)}
  \right]_{\msbar}
  = &\, 
  \ln\mathcal V_{SU(2)}
  + 2
  \left. 
    \left[
      \ln \mathcal A'^{(A_\mu,\varphi)}
    \right]_{\msbar}
  \right|_{g\to g_W}
  + 
  \left. 
    \left[
      \ln \mathcal A'^{(A_\mu,\varphi)}
    \right]_{\msbar}
  \right|_{g\to g_Z},
\end{align}
where $\mathcal A'^{(A_\mu,\varphi)}$ is given in eq.\
\eqref{A'(A,NG)}, and
\begin{align}
  g_W = \frac{g_2}{2},~
  g_Z = \frac{\sqrt{g_Y^2+g_2^2}}{2},
\end{align}
with $g_2$ and $g_Y$ being the gauge coupling constants of $SU(2)_L$
and $U(1)_Y$, respectively.

\subsection{Numerical results}

Now, let us evaluate the decay rate of the EW vacuum in the SM.  The
decay rate of the EW vacuum is very sensitive to the coupling
constants in the SM.  In our numerical analysis, we use
\cite{Patrignani:2016xqp}:
\begin{align}
 m_h & = 125.09\pm0.24,   \\
 m_t & = 173.1\pm0.6,     \\
 \alpha_s (m_Z)  & = 0.1181\pm0.0011,
\end{align}
where $m_h$ and $m_t$ are the Higgs mass and the top
mass, respectively, while $\alpha_s$ is the strong coupling constant.
Following \cite{Buttazzo:2013uya}, the gauge couplings, the top Yukawa
coupling, and the Higgs quartic coupling are determined at $\mu =
m_t$; the calculations are done in the on-shell scheme at
NNLO precision. In addition, we use three-loop QCD and one-loop QED
$\beta$-functions \cite{Gorishnii:1990zu, Tarasov:1980au,
  Gorishnii:1983zi} together with values in \cite{Patrignani:2016xqp}
in order to determine the bottom and the tau Yukawa couplings at $\mu
= m_t$.

\begin{figure}[t]
 \begin{center}
   \hspace{1.6ex}\includegraphics[width=0.595\linewidth]{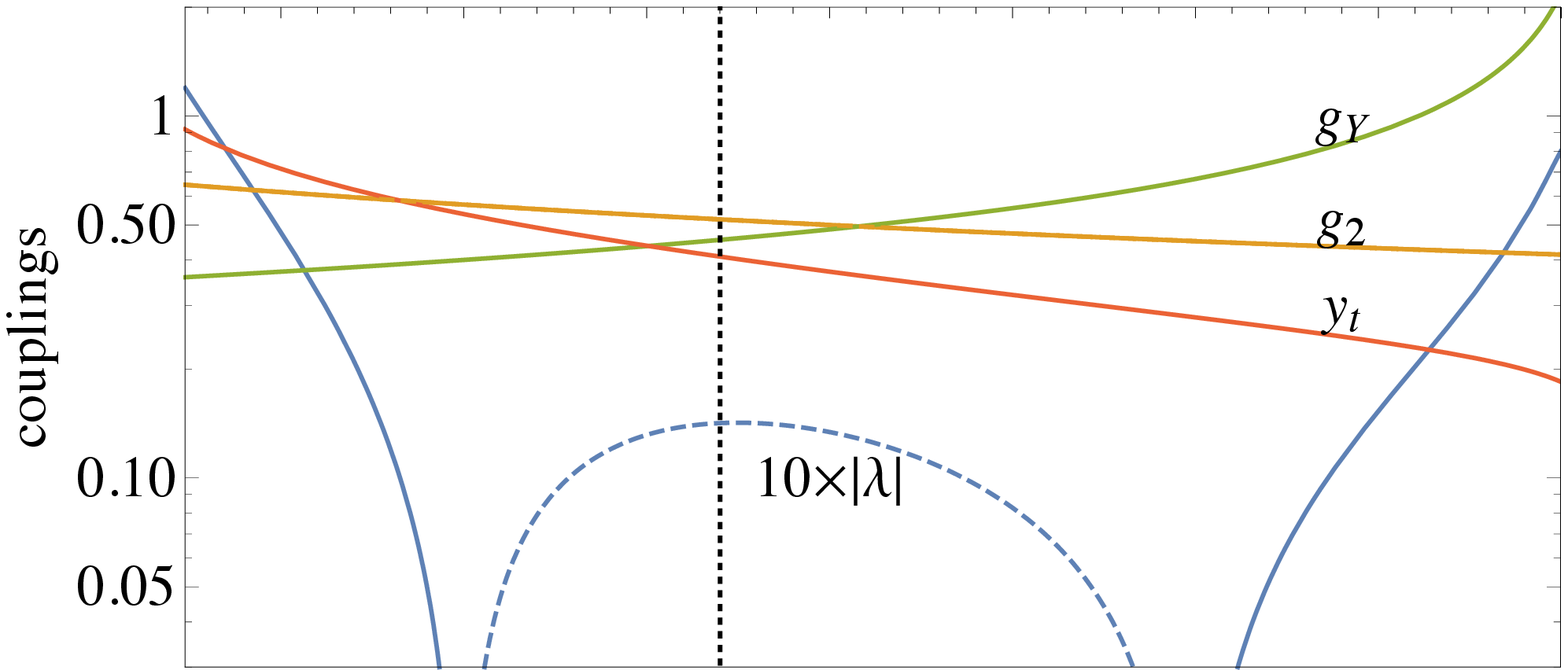}
   \includegraphics[width=0.625\linewidth]{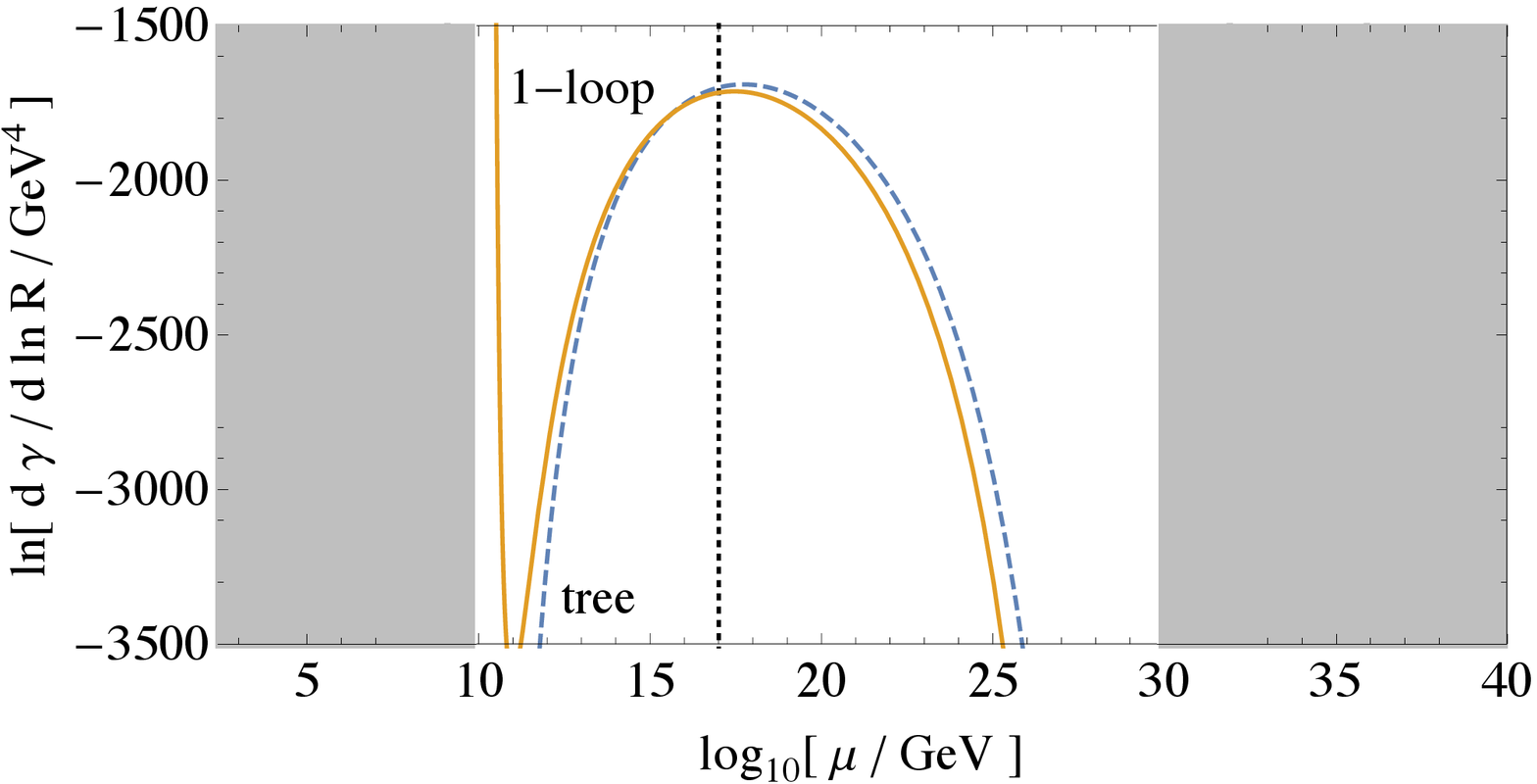}
 \end{center}
 \caption{{\it Top}: The RG evolution of the SM coupling constants as
   functions of $\mu$ (in units of GeV).  The dashed line indicates
   $\lambda<0$.  The black dotted line shows the scale where
   $\bar\phi_C=M_{\rm Pl}$, taking $\mu=1/R$. The horizontal axis is
   common with the bottom panel. {\it Bottom}: The integrand of the
   decay rate with $\mu=1/R$, taking the central values for the SM
   parameters. In the shaded region, $\lambda$ is positive and the
   integrand is zero. }
 \label{fig_rge}
\end{figure}

First, we show the RG evolution of the SM coupling constants in fig.\
\ref{fig_rge}. We use mainly three-loop $\beta$-functions summarized
in \cite{Buttazzo:2013uya} and the central values for the SM
parameters. The black dotted line indicates where $\bar\phi_C$ reaches
the Planck scale $M_{\rm Pl}\simeq 2.4\times 10^{18}\ {\rm GeV}$.  We
also show the running above the Planck scale, assuming there are no
significant corrections from gravity.  For $10^{10}\ {\rm GeV}\lesssim
\mu \lesssim10^{30}\ {\rm GeV}$, $\lambda$ becomes negative; for such
a region, we use a dashed line to indicate $\lambda<0$.

In order to understand the $\mu$ dependence of $\lambda$, let us show
one-loop RG equations of $\lambda$ and $y_t$ (although, in our numerical
calculation, we use RG equations including two- and three-loop effects
and the contribution from the bottom and the tau Yukawa couplings):
\begin{align}
  \left.
    16\pi^2\frac{d\lambda}{d\ln\mu}
  \right|_{\rm one\mathhyphen loop}
  & = 12\lambda
 \left(
  2\lambda+y_t^2-\frac{g_Y^2+g_2^2}{4}-\frac{g_2^2}{2}
 \right)-6y_t^4+6
 \left(
  \frac{g_Y^2+g_2^2}{4}
 \right)^2+12
 \left(
  \frac{g_2^2}{4}
 \right)^2, 
 \\
 \left.
   16\pi^2\frac{dy_t}{d\ln\mu}
 \right|_{\rm one\mathhyphen loop}
  & = y_t
 \left(
  \frac{9}{2}y_t^2-8g_3^2-\frac{9}{4}g_2^2-\frac{17}{12}g_Y^2
 \right).
\end{align}
At a low energy scale, the term proportional to $y_t^4$ drives
$\lambda$ to a negative value. As the scale increases, $y_t$ decreases
while $g_Y$ increases, which brings $\lambda$ back to a positive
value. Notice that $\lambda$ is bounded from below in the SM.

\begin{figure}[t]
  \begin{center}
    \hspace{.37ex}\includegraphics[width=0.614\linewidth]{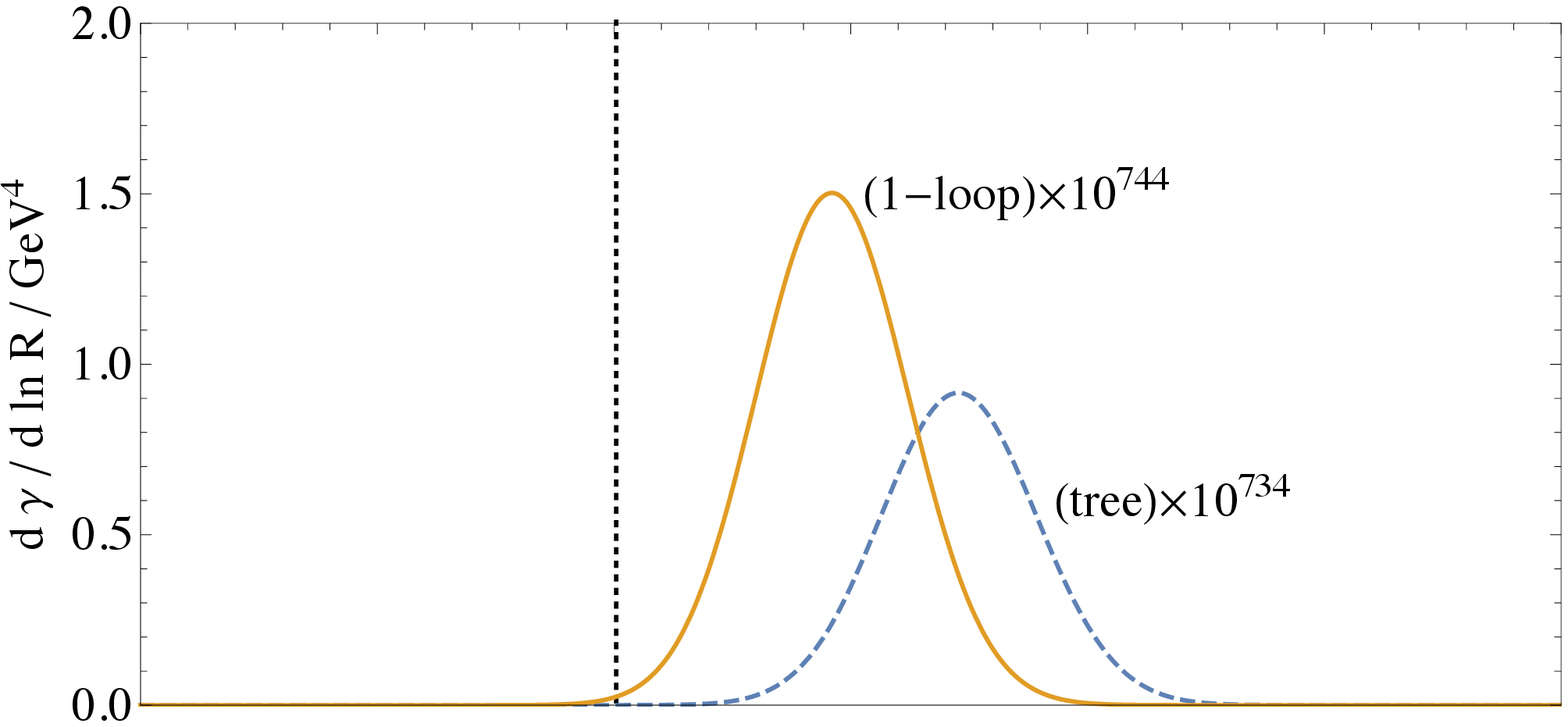}
    \includegraphics[width=0.6485\linewidth]{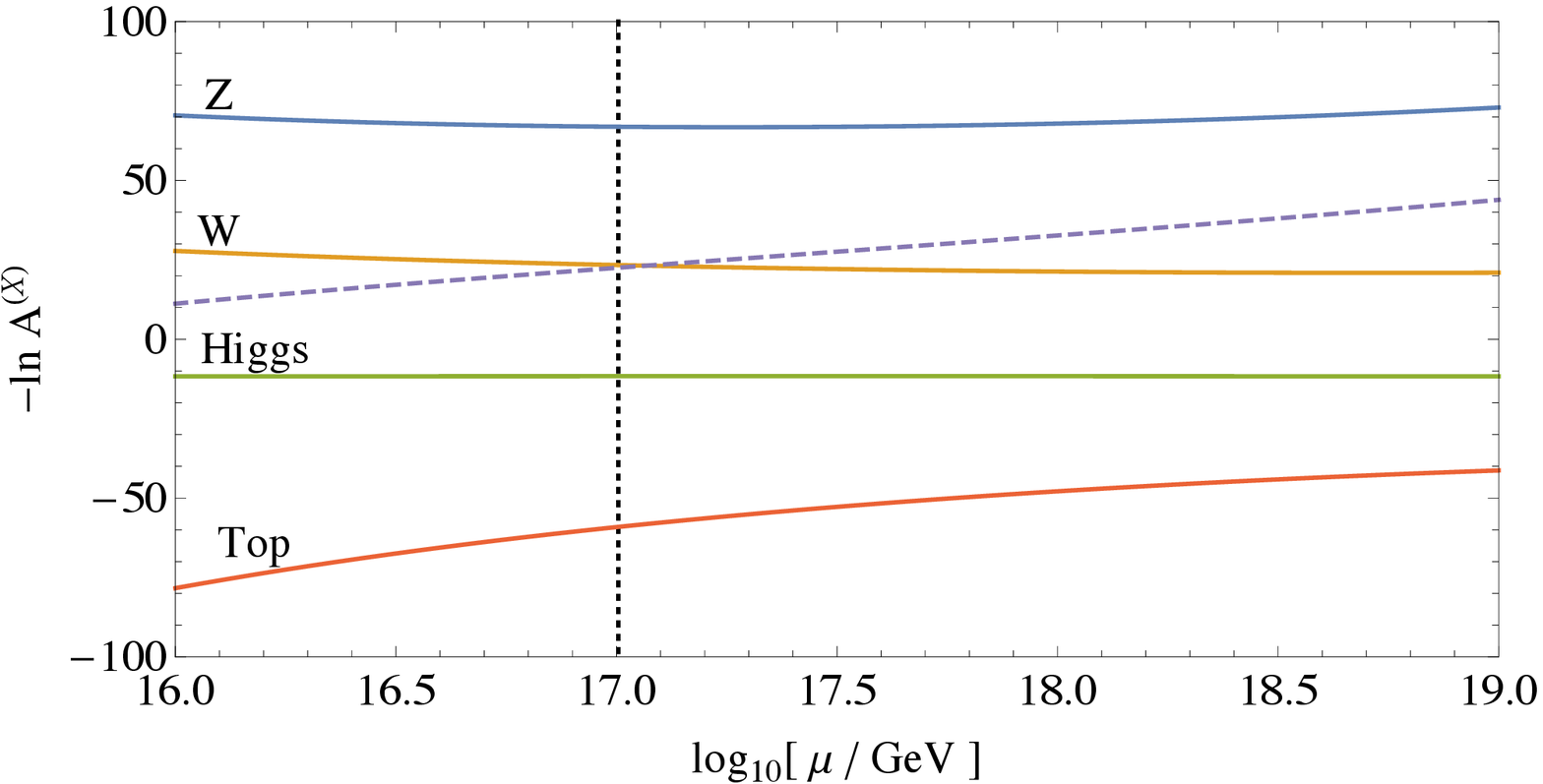}
  \end{center}
  \caption{{\it Top}: The integrand of the decay rate with the central
 values of SM parameters. The solid line corresponds to a result at the
 one-loop level and the dashed one corresponds to that at the tree
 level. The horizontal axis is common with the bottom figure. We show
 $\bar\phi_C = M_{\rm Pl}$ with the vertical dotted line.  {\it Bottom}:
 The size of each quantum correction. The dashed line corresponds to
 $\delta\mathcal B^{(1)}_{\rm eff}$.}  \label{fig_dgamma}
\end{figure}

We show the integrand of $\gamma$ in the bottom panel of fig.\
\ref{fig_rge}, together with that of
\begin{align}
 \gamma_{\rm tree} = \int d \ln R \frac{1}{R^4} e^{-\mathcal B}.
 \label{eq_treelevel}
\end{align}
They are also shown in a linear scale in the top panel of fig.\
\ref{fig_dgamma}.

There are some remarks on the integral over $R$.
\begin{itemize}
\item As indicated by the top panel of fig.\ \ref{fig_dgamma}, the
  integral is dominated by the interval $10^{17}~{\rm
    GeV}\lesssim1/R<10^{18}~{\rm GeV}$, corresponding to $10^{18}~{\rm
    GeV}\lesssim\bar\phi_C<10^{19}~{\rm GeV}$, which is close to the
  Planck scale.  We may formally perform the $R$-integration up to the
  scale where $\lambda$ becomes positive again; the result of such an
  analysis is denoted as $\gamma_\infty$.  Otherwise, we may stop the
  integration at $\bar{\phi}_C\sim M_{\rm Pl}$, expecting that the SM
  breaks down at the Planck scale due to an effect of quantum gravity;
  we also perform such a calculation terminating the integral at
  $\bar{\phi}_C=M_{\rm Pl}$, assuming that the bounce solution is
  unaffected by the effect of quantum gravity.  The result is denoted
  as $\gamma_{\rm Pl}$.

\item As one can see in the bottom panel of fig.\ref{fig_rge}, there
  is an artificial divergence of the integrand at $1/R\simeq10^{10}\
  {\rm GeV}$.  This is due to a breakdown of perturbative expansion
  owing to a too small $|\lambda|$, which makes the one-loop effect
  larger than the tree-level one.  We expect that the effect of such a
  bounce configuration is unimportant because the bounce action for
  such a small $|\lambda|$ suppresses the decay rate significantly.
  Thus, we exclude such a region from the region of integration.  In
  our numerical calculation, we require $\left|\frac{\delta\mathcal
      B_{\rm eff}^{(1)}}{\mathcal B}\right| <0.8$ and
  $\left|\frac{[\ln \mathcal A^{(X)}]_{\rm \overline{MS}}}{\mathcal
      B}\right| <0.8$ for each $X$, where $\delta \mathcal B_{\rm
    eff}^{(1)}$ is the one-loop contribution to $\mathcal{B}_{\rm
    eff}$, and $[\ln \mathcal A^{(X)}]_{\rm \overline{MS}}$ is a
  contribution from particle $X$; the region that does not satisfy
  these conditions is excluded from the integration.\footnote
  {The numerical result is insensitive to the cut parameter, $0.8$, as
    far as only the region where the perturbation breaks down is
    removed from the integration.  In the SM, with the central values
    of the couplings, the numerical result is not affected much even
    if we change the number from $0.04$ to $1.2$.  }
\end{itemize}

By numerically integrating over $R$, we obtain\footnote
{In our previous analysis \cite{Chigusa:2017dux}, we used a different
  renormalization scale, {\it i.e.}, $\mu=\bar\phi_C$ instead of
  $\mu=R^{-1}$.  With $\mu=\bar\phi_C$, the decay rate becomes
  $\log_{10}\left[\gamma\times{\rm Gyr~Gpc^3}\right]=-570$ for the
  best-fit values of the SM parameters.  Difference between this
  result and that in \cite{Chigusa:2017dux} is due to the correction
  of an error in Eq.\ (29) of \cite{Chigusa:2017dux} (see Eq.\
  \eqref{A_BG(A,NG)}).  The uncertainty related to the choice of $\mu$
  should be regarded as theoretical uncertainty.}
\begin{align}
  \log_{10}\left[\gamma_{\rm Pl}\times{\rm Gyr~Gpc^3}\right]
  & = -582^{+40~+184~+144~+2}_{-45~-329~-218~-1}, \\
  \log_{10}\left[\gamma_{\infty}\times{\rm Gyr~Gpc^3}\right]
  & = -580^{+40~+183~+145~+2}_{-44~-328~-218~-1},
\end{align}
where the 1st, 2nd, 3rd, and 4th errors are due to the Higgs mass, the
top mass, the strong coupling constant, and the renormalization scale,
respectively.  (In order to estimate the uncertainty due to the choice
of the renormalization scale, we vary the renormalization scale from
$1/2R$ to $2/R$.)  Currently, the largest error comes from the
uncertainty of the top mass.  With a better understanding of the top
quark mass at future LHC experiment \cite{Kharchilava:1999yj,
  Hill:2005zy, Biswas:2010sa, Agashe:2012bn, Alioli:2013mxa,
  Kawabataa:2014osa, Ravasio:2018lzi}, or even with at future $e^+e^-$
colliders \cite{Horiguchi:2013wra}, more accurate determination of the
decay rate will become possible.  One can see that the predicted decay
rate per unit volume is extremely small, in particular, compared with
$H_0^{-4}\simeq 10^3\ {\rm Gyr~Gpc^3}$ (with $H_0$ being the expansion
rate of the present universe).  Such a small decay rate is harmless
for realizing the present universe observed.\footnote
{Cosmologically, the Higgs field may evolve into the unstable region
  due to the dynamics during and after inflation
  \cite{Espinosa:2007qp, Lebedev:2012sy, Kobakhidze:2013tn,
    Enqvist:2014bua, Herranen:2014cua, Kobakhidze:2014xda,
    Kamada:2014ufa, Herranen:2015ima, Hook:2014uia, Espinosa:2015qea,
    Ema:2016kpf, Enqvist:2016mqj, Kearney:2015vba, Kohri:2016wof,
    East:2016anr, Ema:2017loe, Postma:2017hbk, Joti:2017fwe,
    Fairbairn:2014zia, Shkerin:2015exa, Ema:2017rkk}.  We do not
  consider such cases.}

\begin{figure}[t]
  \begin{center}
    \includegraphics[width=0.5\linewidth]{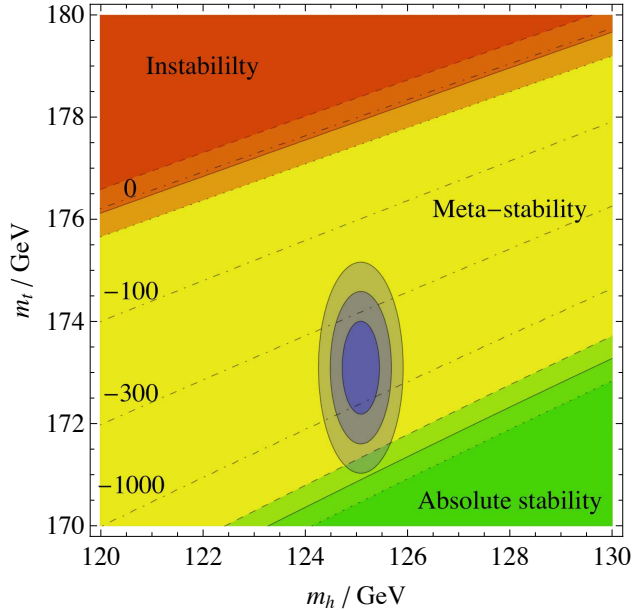}
  \end{center}
  \caption{The stability of the EW vacuum in the SM with a cut-off of
    the integration at $\bar\phi_C=M_{\rm Pl}$. The red region is
    unstable, the yellow region is meta-stable, and the green region
    is absolutely stable. The dashed, solid, and dotted lines
    correspond to $\alpha_s = 0.1192$, $0.1181$, and $0.1170$,
    respectively. The black dot-dashed lines indicate
    $\log_{10}\left[\gamma\times{\rm Gyr~Gpc^3}\right]=0$, $-100$, $-300$,
    and $-1000$ with the central value of $\alpha_s$. The blue circles
    indicate $68$, $95$, and $99$ \% C.L. constraints on the Higgs
    mass vs.\ top mass plane assuming that their errors are
    independently Gaussian. }
    \label{fig_SMconst}
\end{figure}

In fig.\ \ref{fig_SMconst}, we show the decay rate in $m_h$ vs.\ $m_t$
plane.  In the red region, $\gamma$ becomes larger than $H_0^4$, which
we call unstable region.  In the yellow region, the EW vacuum is
metastable, meaning that $0<\gamma<H_0^4$.  In the green region, the
EW vacuum is absolutely stable because $\lambda$ is always positive.
The dashed, solid, and dotted lines correspond to $\alpha_s =
0.1192,~0.1181$, and $0.1170$, respectively. The black dot-dashed
contours show $\log_{10}\left[\gamma\times{\rm Gyr~Gpc^3}\right]=0,-100$,
$-300$, and $-1000$ with the central value of $\alpha_s$.  We also
show $68$, $95$, and $99$ \% C.L. constraints on the Higgs mass vs.\
top mass plane assuming that their errors are independently Gaussian
distributed.  In fig.\ \ref{fig_SMconst}, we terminate the integral at
$\bar\phi_C = M_{\rm Pl}$, but it does not change the figure as far as
the cut-off is not so far from the Planck scale.\footnote
{Even with a lower cut-off such as $\bar\phi_C<0.1M_{\rm Pl}$, the
  result does not change significantly. It reduces $\log_{10}\gamma$ by
  about 20.}
The value of $\bar\phi_C$ at the maximum of the integrand ranges from
$10^{18}$ GeV to $10^{20}$ GeV.

It is well known that, currently, our universe is (almost) dominated
by the dark energy.  If it is a cosmological constant, then our
universe will eventually become de Sitter space with the expansion
rate of about $56.3\ {\rm km/sec/Mpc}$ \cite{Ade:2015xua}.  Then,
based on $\gamma_{\rm Pl}$, the phase transition rate within the Hubble
volume of such a universe is estimated to be $10^{-580}\ {\rm
  Gyr}^{-1}$, which may be regarded as a decay rate of the EW vacuum
in the SM.

For comparison, we also perform a ``tree-level'' calculation of the
decay rate using eq.~\eqref{eq_treelevel}.  The results are
$\log_{10}\left[\gamma_{\rm Pl}^{\rm (tree)}\times{\rm Gyr~Gpc^3}\right]
= -575$ and $\log_{10}\left[\gamma_{\infty}^{\rm (tree)}\times{\rm
Gyr~Gpc^3}\right] = -570$.  Thus, the difference between $\gamma$ and
$\gamma^{\rm (tree)}$ turns out to be rather small.  This is a
consequence of an accidental cancellation among the contributions of
several fields.  In the bottom panel of fig.\ \ref{fig_dgamma}, we show
individual quantum corrections separately, as well as the total one-loop
contribution.  We can see that the large quantum correction from the top
quark is cancelled by those from the gauge bosons.  We have also checked
that the unstable region on the $m_h$ vs.\ $m_t$ plane shifts upward by
$\Delta m_t\simeq0.2\ {\rm GeV}$ if we use $\gamma^{\rm (tree)}$.

\section{Models with Extra Fermions}
\label{sec_extra}
\setcounter{equation}{0}

So far, we assumed that the SM is valid up to the Planck or some
higher scale.  However, the decay rate of the EW vacuum may be
affected if there exist extra particles.  In particular, extra
fermions coupled to the Higgs boson may destabilize the EW vacuum
because the new Yukawa couplings tend to drive $\lambda$ to a negative
value through RG effects \cite{Gogoladze:2008ak, He:2012ub,
  Rodejohann:2012px, Chakrabortty:2012np, Chao:2012mx, Masina:2012tz,
  Khan:2012zw, Dev:2013ff, Kobakhidze:2013pya, Datta:2013mta,
  Chakrabortty:2013zja, Xiao:2014kba, Bambhaniya:2014hla,
  Salvio:2015cja, Lindner:2015qva, Rose:2015fua, Haba:2016zbu,
  Bambhaniya:2016rbb, Garg:2017iva}.  Consequently, the decay rate of
the EW vacuum becomes larger than that in the SM.  Potential
candidates of such fermions include vector-like fermions as well as
right-handed neutrinos for the seesaw mechanism
\cite{Minkowski:1977sc, Yanagida:1979as, GellMann:1980vs}.

In this section, we consider several models with such extra fermions.
We perform the RG analysis of the runnings of coupling constants with
the effects of the extra fermions.  We include two-loop effects of the
extra fermions into the $\beta$-functions, which can be calculated
using the result in \cite{Machacek:1983tz, Machacek:1983fi,
  Machacek:1984zw, Luo:2002ti}.  We also take account of one-loop
threshold corrections due to the extra fermions, which are summarized
in Appendix \ref{apx_threshold}.\footnote
{If we use the two-loop $\beta$-functions instead of three-loop ones
  in the SM calculation, the difference of
  $\log_{10}\left[\gamma\times{\rm Gyr~Gpc^3}\right]$ is around
  40. Thus, the systematic error of neglecting three-loop effects of
  the extra fermions is expected to be similar.}
For the integration over $R$, we follow the procedure in the SM case,
as well as the following treatments:
\begin{itemize}
\item We terminate the integration if any of the coupling constants
  (in particular, Yukawa coupling constants of extra particles)
  exceeds $\sqrt{4\pi}$.
\item In order to maintain the classical scale invariance at a good
  accuracy, we require $1/R>10M_{\rm ex}$, where $M_{\rm ex}$ is the
  mass scale of the new particles.
\end{itemize}

\subsection{Vector-like fermions}

Here, we consider two examples of vector-like fermions, one is colored
vector-like fermions and the other is non-colored ones.  We consider
the case where the extra fermions have a Yukawa coupling with the SM
Higgs boson.  (We assume that the mixing between the extra fermions
and the SM fermions is negligible.)

We first consider colored vector-like fermions, having the same SM gauge
quantum numbers as the left-handed quark doublet and the right-handed
down quark, as well as their vector-like partners; we add $Q$ $({\bf 3},
{\bf 2},\frac{1}{6})$, $\overline{Q}$ $(\bar{\bf 3}, {\bf
2},-\frac{1}{6})$, $D$ $({\bf 3}, {\bf 1},-\frac{1}{3})$, and
$\overline{D}$ $(\bar{\bf 3}, {\bf 1},\frac{1}{3})$.  (In the
parenthesis, we show the quantum numbers of $SU(3)_C$, $SU(2)_L$, and
$U(1)_Y$.) The Yukawa terms in the Lagrangian are given by
\begin{align}
  \mathcal L_{\rm Yukawa} = \mathcal L^{({\rm SM})}_{\rm Yukawa}
  +Y_{\overline{D}} \Phi^* Q \overline{D}
  +Y_D \Phi\overline{Q} D,
\end{align}
where $\mathcal L_{\rm SM}^{\rm (Yukawa)}$ is the SM part.  We also
add the following mass terms:
\begin{align}
  \mathcal L_{\rm mass} = 
  M_{Q}\overline{Q} Q
  +M_{D}\overline{D} D.
\end{align}
For simplicity, we assume $M_Q=M_D$.  We take the new Yukawa coupling
constants and mass parameters real and positive.  We also take
\begin{align}
  Y_D (\mu=M_D) = Y_{\overline{D}} (\mu=M_D) \equiv y_D.
\end{align}
As we have mentioned before, the scale dependence of the new Yukawa
coupling constants is evaluated by using two-loop RG equations and
one-loop threshold corrections (see Appendix \ref{apx_threshold}).

The calculation of the decay rate is parallel to the SM case, and the
decay rate is given in the following form:
\begin{align}
 \gamma = \int d\ln R \frac{1}{R^4}
 \left[
  \mathcal A'^{(h)}\mathcal A^{(t)}\mathcal A^{(A_\mu,\varphi)}
  \mathcal A^{(Q,\overline{D})}
  \mathcal A^{(\overline{Q},D)}
  e^{-\mathcal B}
 \right]_{\msbar,~\mu=1/R},
\end{align}
where $\mathcal A^{(Q,\overline{D})}$ and $\mathcal
A^{(\overline{Q},D)}$ are effects of the extra fermions on the
prefactor:
\begin{align}
 \left[
  \ln \mathcal A^{(Q,\overline{D})}
 \right]_{\msbar}
  & = \left.3
 \left[
  \ln \mathcal A^{(\psi)}
 \right]_{\msbar}
 \right|_{y\to Y_{\overline{D}}}, \\
 \left[
  \ln \mathcal A^{(\overline{Q},D)}
 \right]_{\msbar}
  & = \left.3
 \left[
  \ln \mathcal A^{(\psi)}
 \right]_{\msbar}
 \right|_{y\to Y_D}.
\end{align}

\begin{figure}[t]
  \begin{minipage}{0.49\linewidth}
    \begin{center}
      \includegraphics[width=\linewidth]{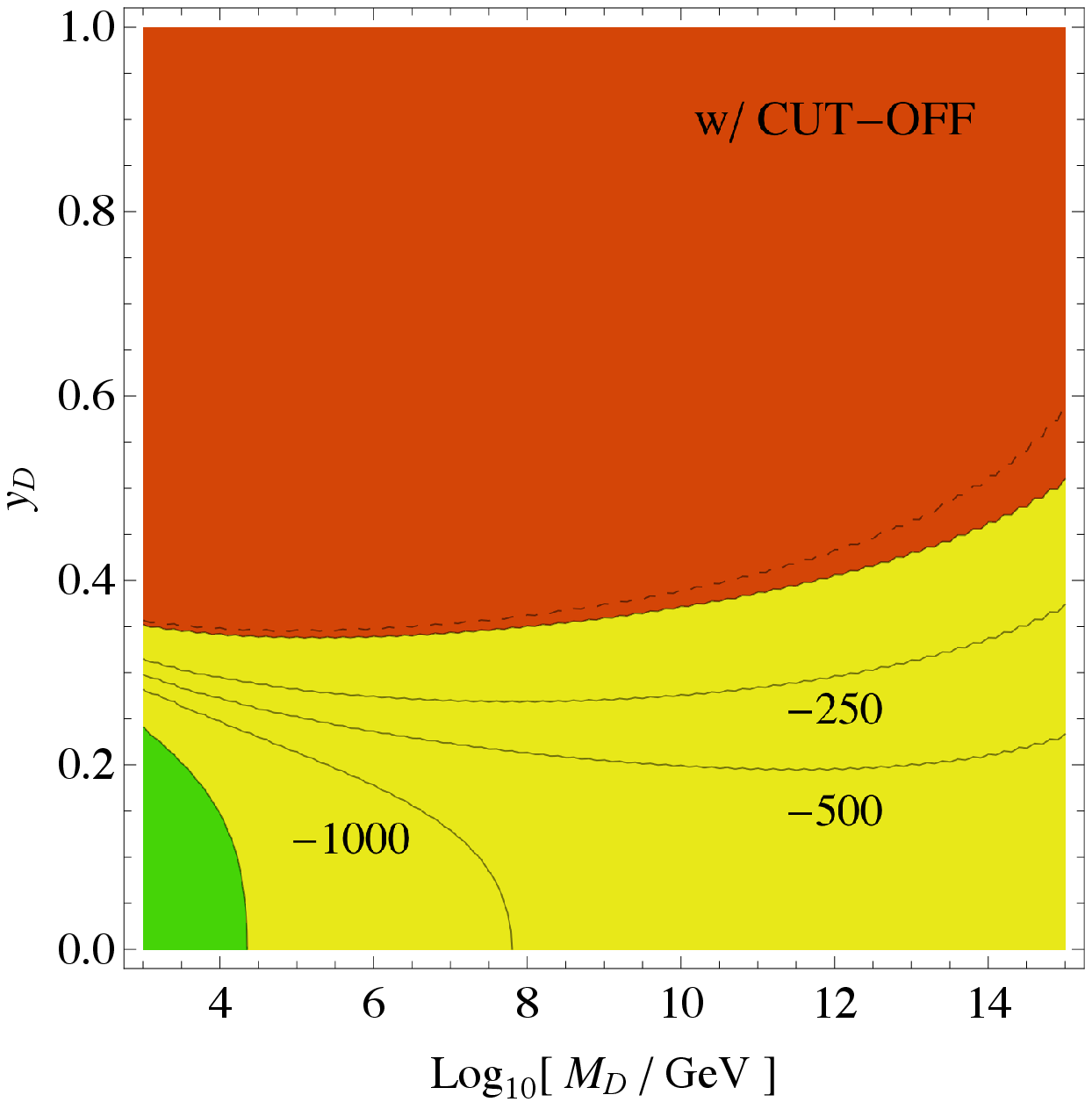}
    \end{center}
  \end{minipage}
  \begin{minipage}{0.49\linewidth}
    \begin{center}
      \includegraphics[width=\linewidth]{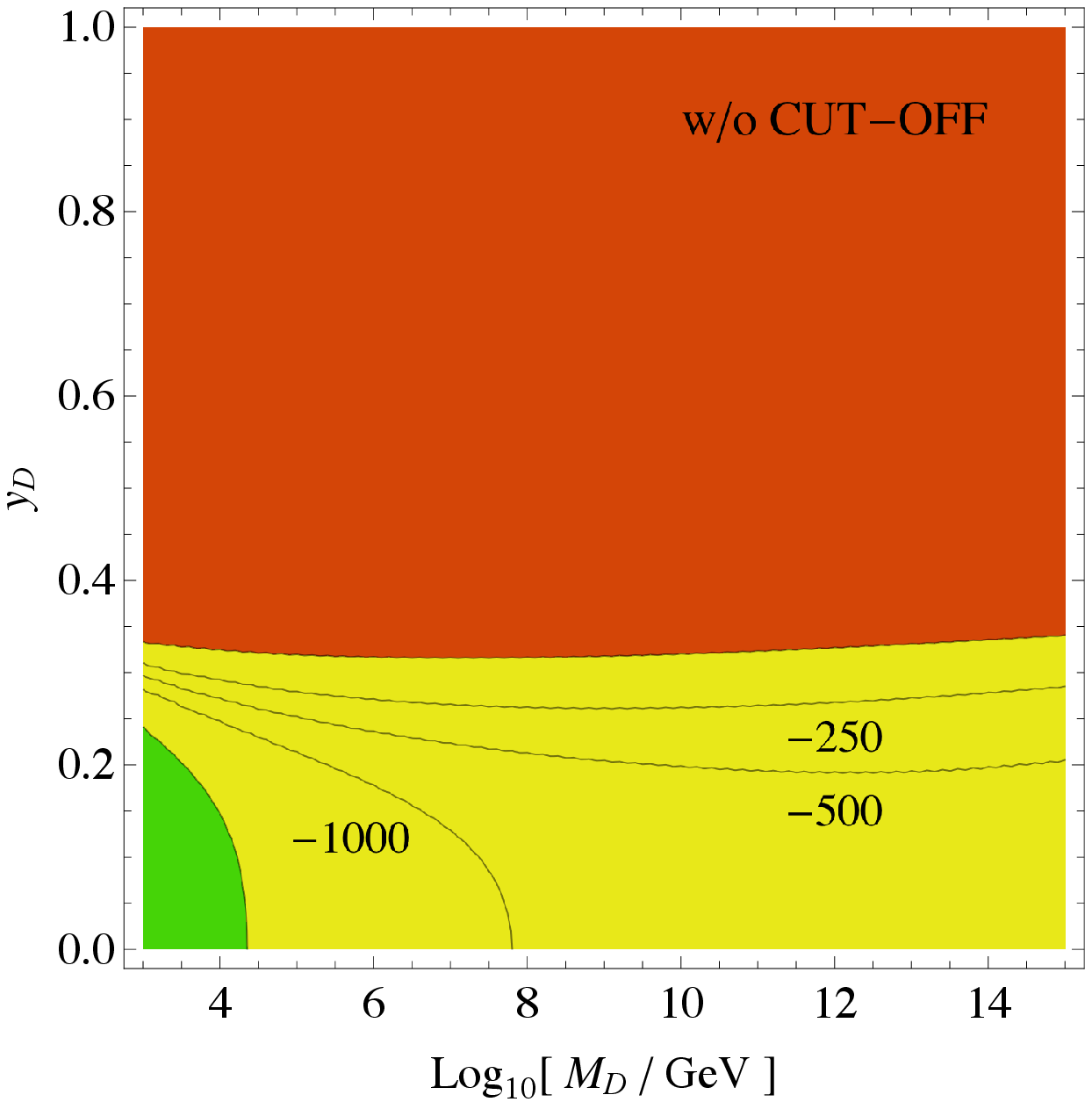}
    \end{center}
  \end{minipage}
  \caption{The decay rate of the EW vacuum for vector-like quarks with
    (left) and without (right) the cut-off at the Planck scale.  The
    red region is unstable, the yellow region is meta-stable, and the
    green region is absolutely stable.  The solid lines show
    $\log_{10}[\gamma~{\rm Gyr}~{\rm Gpc}^3]=-250,-500$, and
    $-1000$. The dashed line corresponds to the constraint when we
    stop the integration at $\bar\phi_C=0.1M_{\rm Pl}$. }
 \label{fig_VLQ}
\end{figure}

\begin{figure}[t]
 \begin{minipage}{0.49\linewidth}
  \begin{center}
   \includegraphics[width=\linewidth]{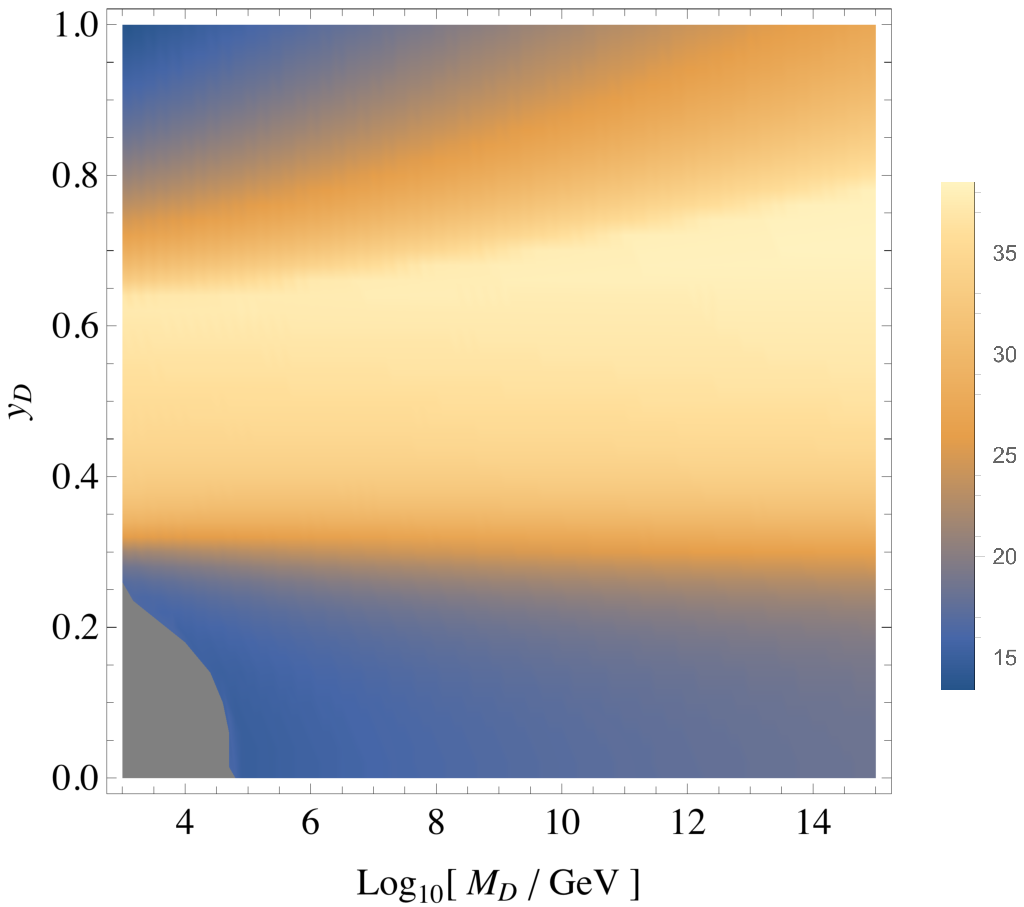}
  \end{center}
 \end{minipage}
 \begin{minipage}{0.49\linewidth}
  \begin{center}
   \includegraphics[width=\linewidth]{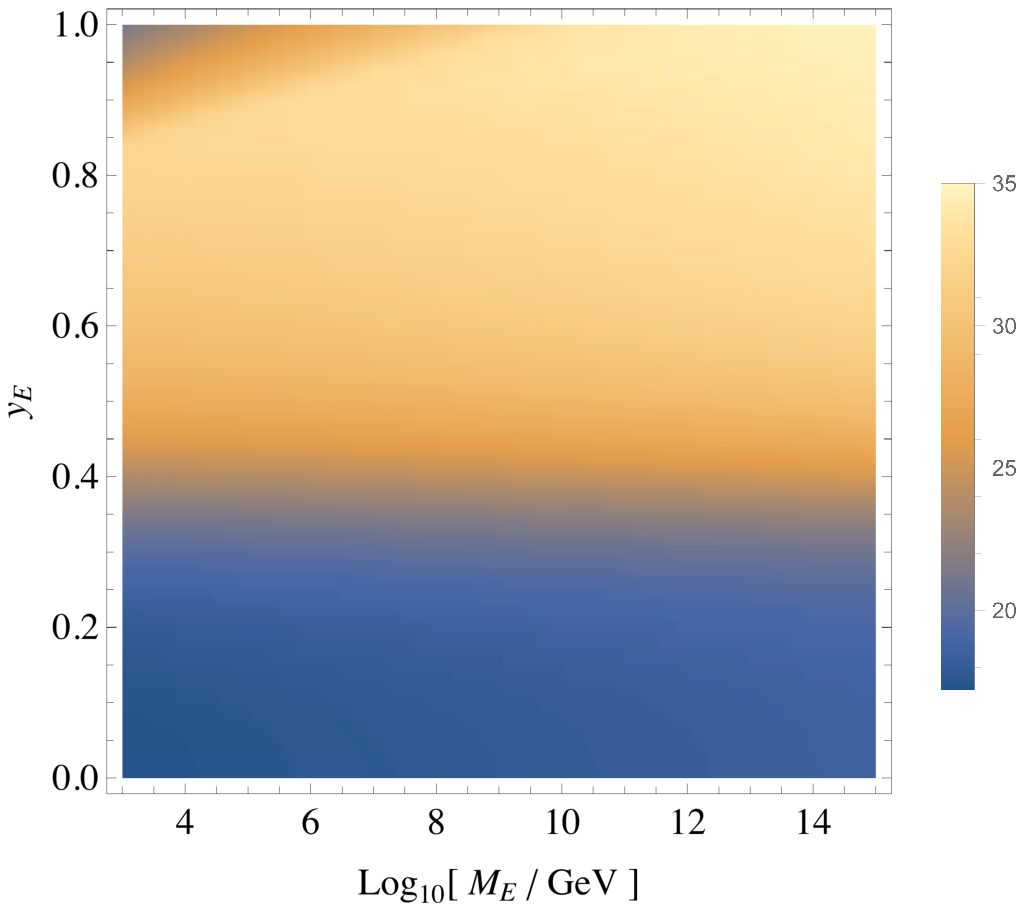}
  \end{center}
 \end{minipage}
 \caption{The value of $\bar\phi_C$ at the maximum of the integrand of
   $\gamma$, $\bar\phi_C^{\max}$. The left panel is for vector-like
   quarks and the right is for vector-like leptons. The numbers in the
   legends indicate $\log_{10}[\bar\phi_C^{\max}/{\rm GeV}]$. In the
   gray region, the EW vacuum is absolutely stable. }
   \label{fig_cutoffVLF}
\end{figure}

In fig.\ \ref{fig_VLQ}, we show the contours of the constant decay rate
on $M_D$ vs.\ $y_D$ plane.  Here, we use the central values of the SM
parameters. The meanings of the shading colors are the same as in the SM
case. The left and the right panels show the results with and without
imposing the condition $\bar{\phi}_C<M_{\rm Pl}$ in integrating over
$R$, respectively. As we can see, the effect of such a cut-off is
significant. This is because $\bar\phi_C$ at the maximum of the
integrand, $\bar\phi_C^{\max}$, can become much larger than the
Planck scale in the case with extra fermions; we show
$\bar\phi_C^{\max}$ for the case with vector-like colored fermions in
the left panel of fig.\ \ref{fig_cutoffVLF}.  (In the upper-left corner
of the figure, the value of $\bar\phi_C^{\max}$ becomes smaller; this is
because, in such a region, the Yukawa coupling constants become
non-perturbative at a lower RG scale, which gives an upper bound on
$\bar{\phi}_C$ in the integration over $R$.).  To see the cut-off
dependence of the decay rate, we show the constraint with terminating
the integration at $\bar\phi_C=0.1M_{\rm Pl}$ in the left panel (dashed
line).  In addition, when $M_D$ and $y_D$ are small, we have a region of
absolute stability. This is because the addition of colored particles
makes the strong coupling constant larger than the SM case. It rapidly
drives $y_t$ to a small value, which makes $\lambda$ always positive.
Requiring that the lifetime of the EW vacuum should be longer than the
age of the universe, we obtain $y_D\lesssim0.35-0.5$ for $10^3~{\rm
GeV}\lesssim M_D\lesssim10^{15}~{\rm GeV}$.

The second example is non-colored extra fermions, having the same SM quantum
numbers as leptons.
We introduce
$L$ $({\bf 1}, {\bf 2},-\frac{1}{2})$, 
$\overline{L}$ $({\bf 1}, {\bf 2},\frac{1}{2})$, 
$E$ $({\bf 1}, {\bf 1},-1)$, and
$\overline{E}$ $({\bf 1}, {\bf 1},1)$, and the 
Yukawa and mass terms in the Lagrangian are 
given by
\begin{align}
 \mathcal L_{\rm Yukawa} &= \mathcal L^{({\rm SM})}_{\rm Yukawa}
 +Y_{\overline{E}}\Phi^* L\overline{E}
 +Y_E\Phi\overline{L}E,\\
 \mathcal{L}_{\rm mass} &=
 M_L\overline{L}L
 +M_E\overline{E}E,
\end{align}
respectively.  For simplicity, we 
take $M_L = M_E$ and 
adopt the following renormalization
conditions
\begin{align}
  Y_E (\mu=M_E)
  = Y_{\overline{E}} (\mu=M_E)
  \equiv y_E.
\end{align}

The decay rate of the EW vacuum is given by
\begin{align}
  \gamma = \int d\ln R\frac{1}{R^4}
  \left[
    \mathcal A'^{(h)}\mathcal A^{(t)}\mathcal A^{(A_\mu,\varphi)}
    \mathcal A^{(L,\overline{E})}
    \mathcal A^{(\overline{L},E)}
    e^{-\mathcal B}
  \right]_{\msbar,~\mu=1/R},
\end{align}
where
\begin{align}
  \left[
    \ln \mathcal A^{(L,\overline{E})}
  \right]_{\msbar}
  & = \left.
    \left[
      \ln \mathcal A^{(\psi)}
    \right]_{\msbar}
  \right|_{y\to Y_{\overline{E}}}, \\
  \left[
    \ln \mathcal A^{(\overline{L},E)}
  \right]_{\msbar}
  & = \left.
    \left[
      \ln \mathcal A^{(\psi)}
    \right]_{\msbar}
  \right|_{y\to Y_E}.
\end{align}

\begin{figure}[t]
 \begin{minipage}{0.49\linewidth}
  \begin{center}
   \includegraphics[width=\linewidth]{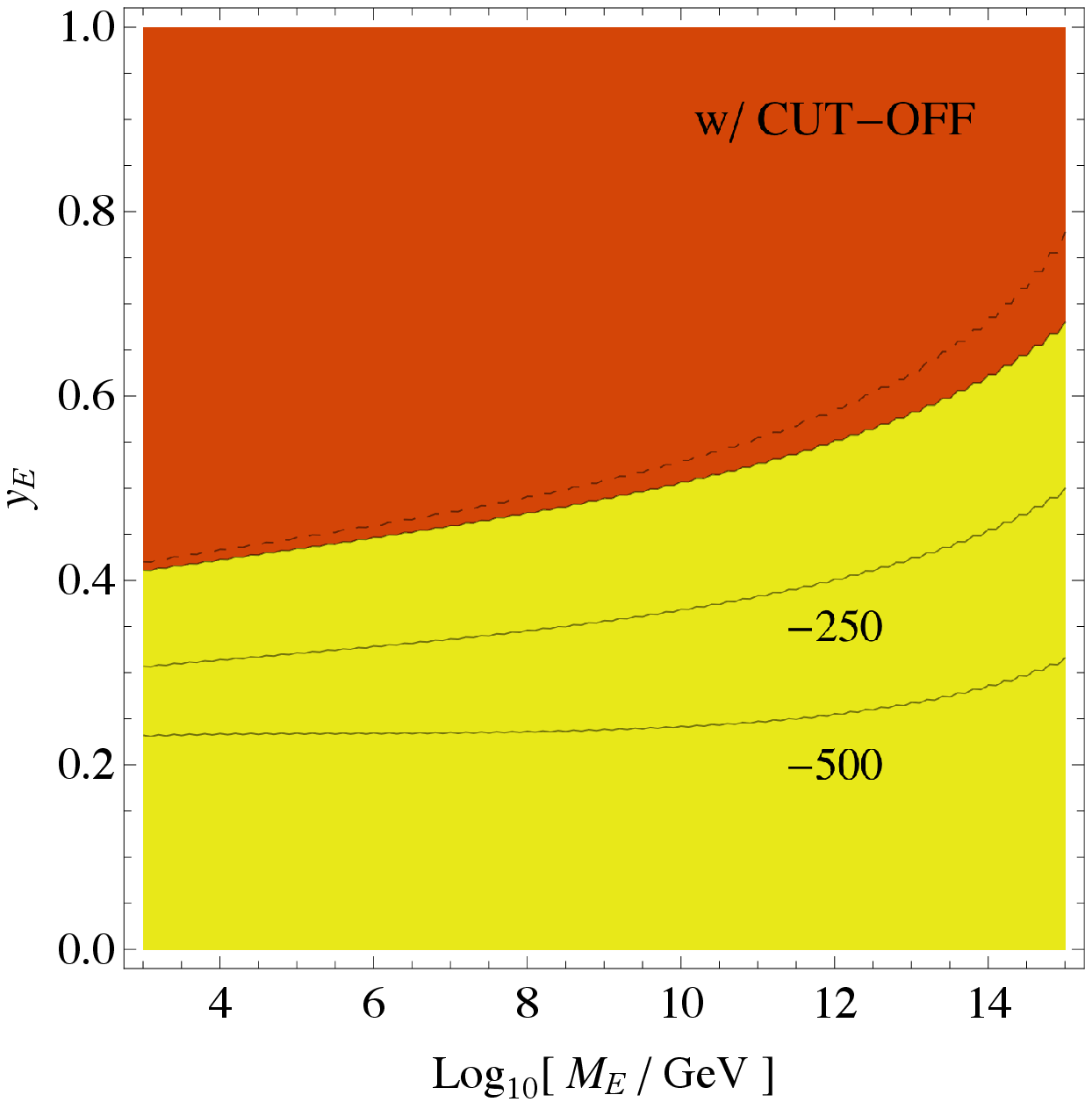}
  \end{center}
 \end{minipage}
 \begin{minipage}{0.49\linewidth}
  \begin{center}
   \includegraphics[width=\linewidth]{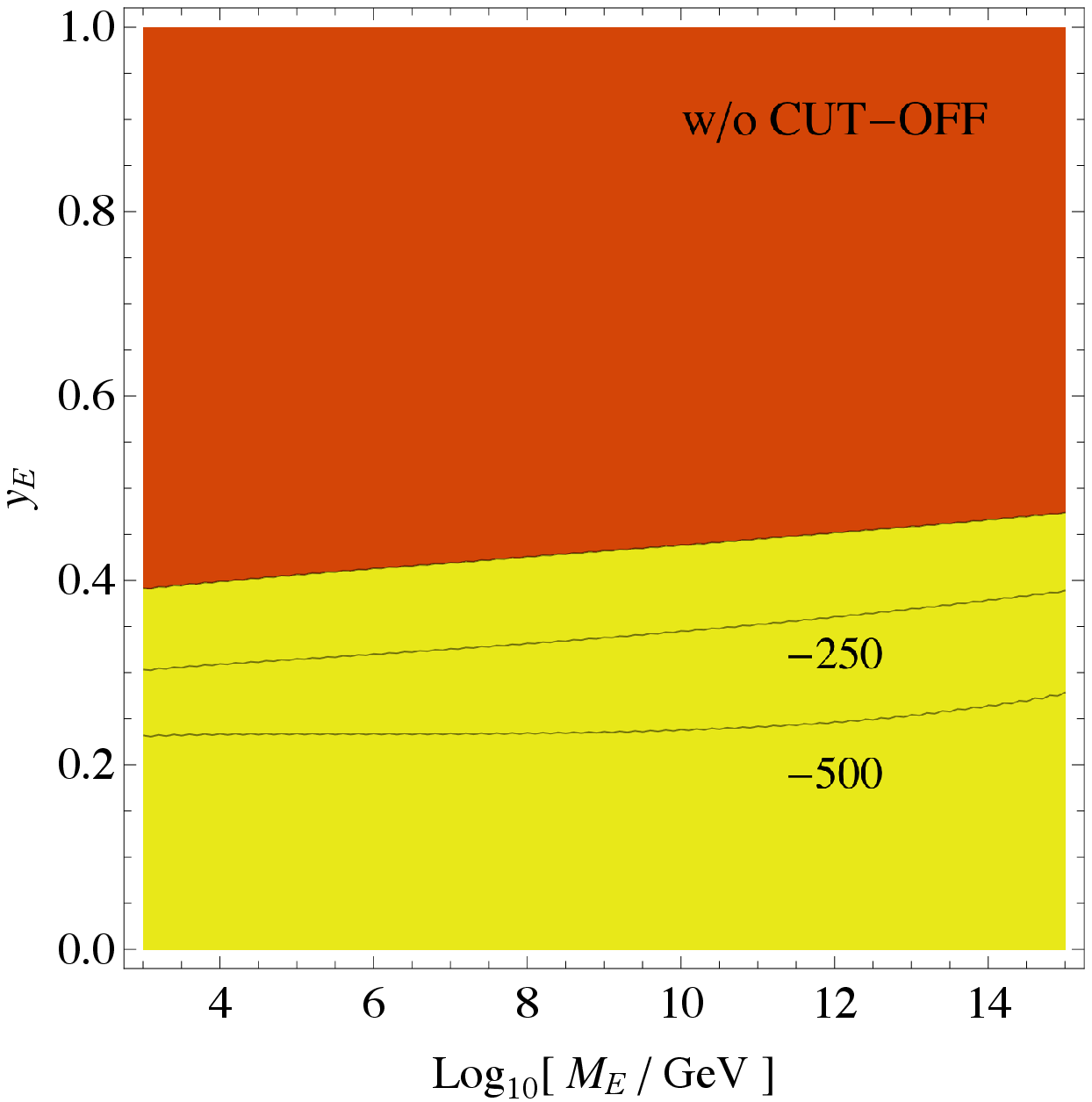}
  \end{center}
 \end{minipage}
 \caption{The same figure as fig.\ \ref{fig_VLQ} but for vector-like
   leptons.  }  \label{fig_VLL}
\end{figure}

In fig.\ \ref{fig_VLL}, we show the contours of constant decay
rate. Since the extra fermions are not colored, we do not have a
region of absolute stability. We observe a larger effect of the
cut-off at the Planck scale. This is because $\bar\phi_C^{\max}$ is
typically large in a wider parameter space, as indicated in the right
panel of fig.\ \ref{fig_cutoffVLF}.

Requiring that the lifetime of the EW vacuum should be longer than the
age of the universe, we obtain $y_E\lesssim0.4-0.7$ for $10^3~{\rm
GeV}\lesssim M_E\lesssim10^{15}~{\rm GeV}$. The constraint becomes
significantly weaker for larger $M_E$ owing to the cut-off at the Planck
scale.

\subsection{Right-handed neutrino}

Next, we consider the case with right-handed neutrinos, which is
responsible for the active neutrino masses via the seesaw mechanism
\cite{Minkowski:1977sc, Yanagida:1979as, GellMann:1980vs}.  For
simplicity, we concentrate on the case where only one mass eigenstate
of the right-handed neutrinos, denoted as $N$, strongly couples to the
Higgs boson (as well as to the third generation lepton doublet).
Then, the Yukawa and mass terms in the Lagrangian are
\begin{align}
  \mathcal L_{\rm Yukawa}  &= \mathcal L^{({\rm SM})}_{\rm Yukawa}
  +Y_N\Phi^* L \overline{N},\\
  \mathcal{L}_{\rm mass} &= \frac{1}{2}M_N\overline{N}N,
\end{align}
where, in this subsection, $L$ denotes one of the lepton doublets in
the SM. We define
\begin{align}
  Y_N (\mu=M_N) \equiv y_N.
\end{align}

Assuming that, for simplicity, the neutrino Yukawa matrix is diagonal
in the mass-basis of right-handed neutrinos, the following effective
operator shows up by integrating out $N$:
\begin{align}
 \Delta\mathcal L = \frac{C}{4}(\Phi L)^2,
\end{align}
with
\begin{align}
  C(M_N) = -2\frac{y_N^2}{M_N}.
\end{align}
One of the active neutrino masses is related to the value of $C$ at
the EW scale as
\begin{align}
  m_\nu = C(m_t)\frac{v^2}{4},
\end{align}
with $v\simeq 246\ {\rm GeV}$ being the vacuum expectation value of the
Higgs boson.  In our numerical calculation, we use the following
one-loop RG equation to estimate the neutrino mass \cite{Chankowski:1993tx}:
\begin{align}
  16\pi^2\frac{d}{d\ln\mu}C = \left(4\lambda+6y_t^2-3g_2^2\right)C.
\end{align}

\begin{figure}[t]
 \begin{minipage}{0.49\linewidth}
  \begin{center}
   \includegraphics[width=\linewidth]{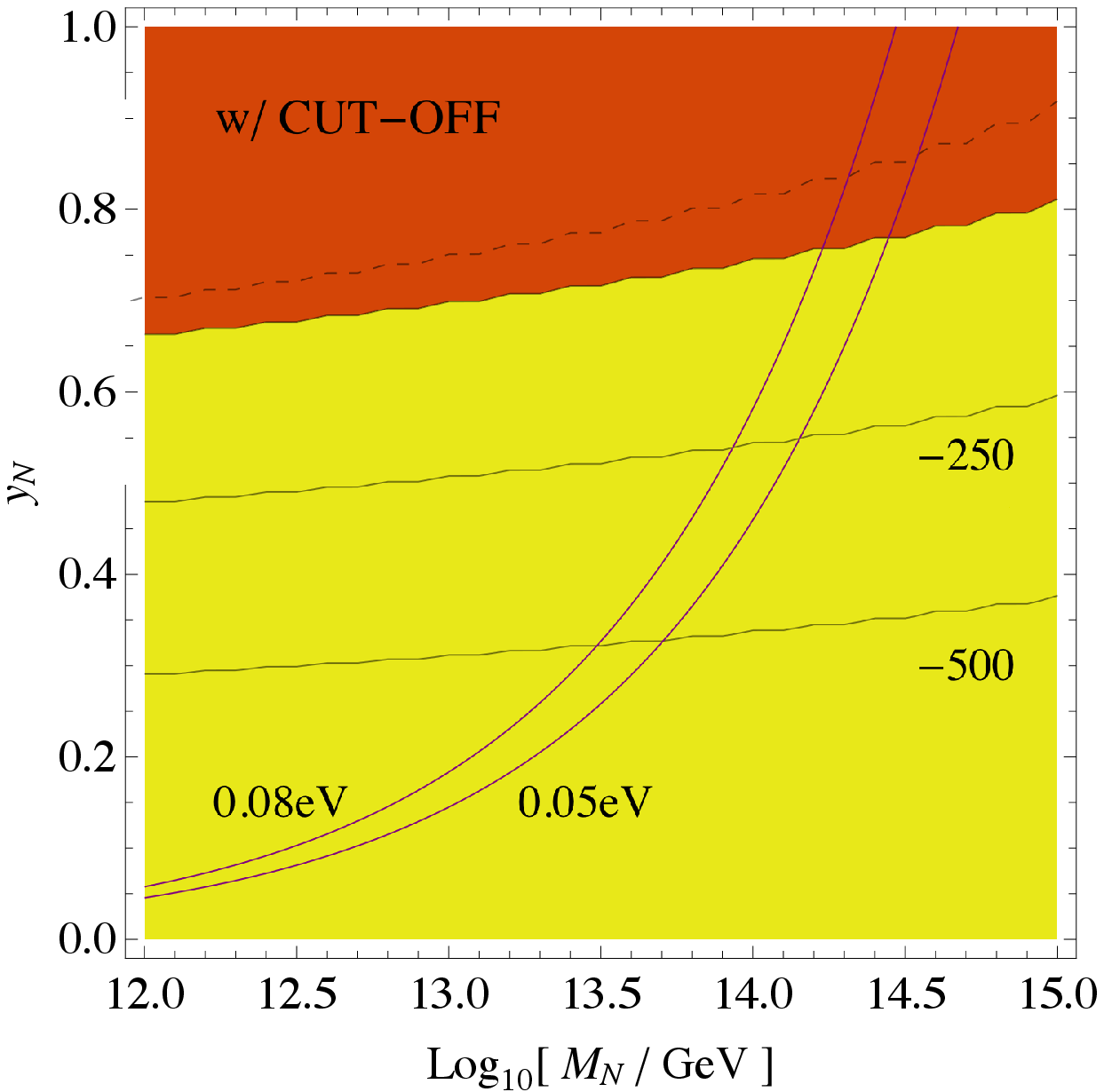}
  \end{center}
 \end{minipage}
 \begin{minipage}{0.49\linewidth}
  \begin{center}
   \includegraphics[width=\linewidth]{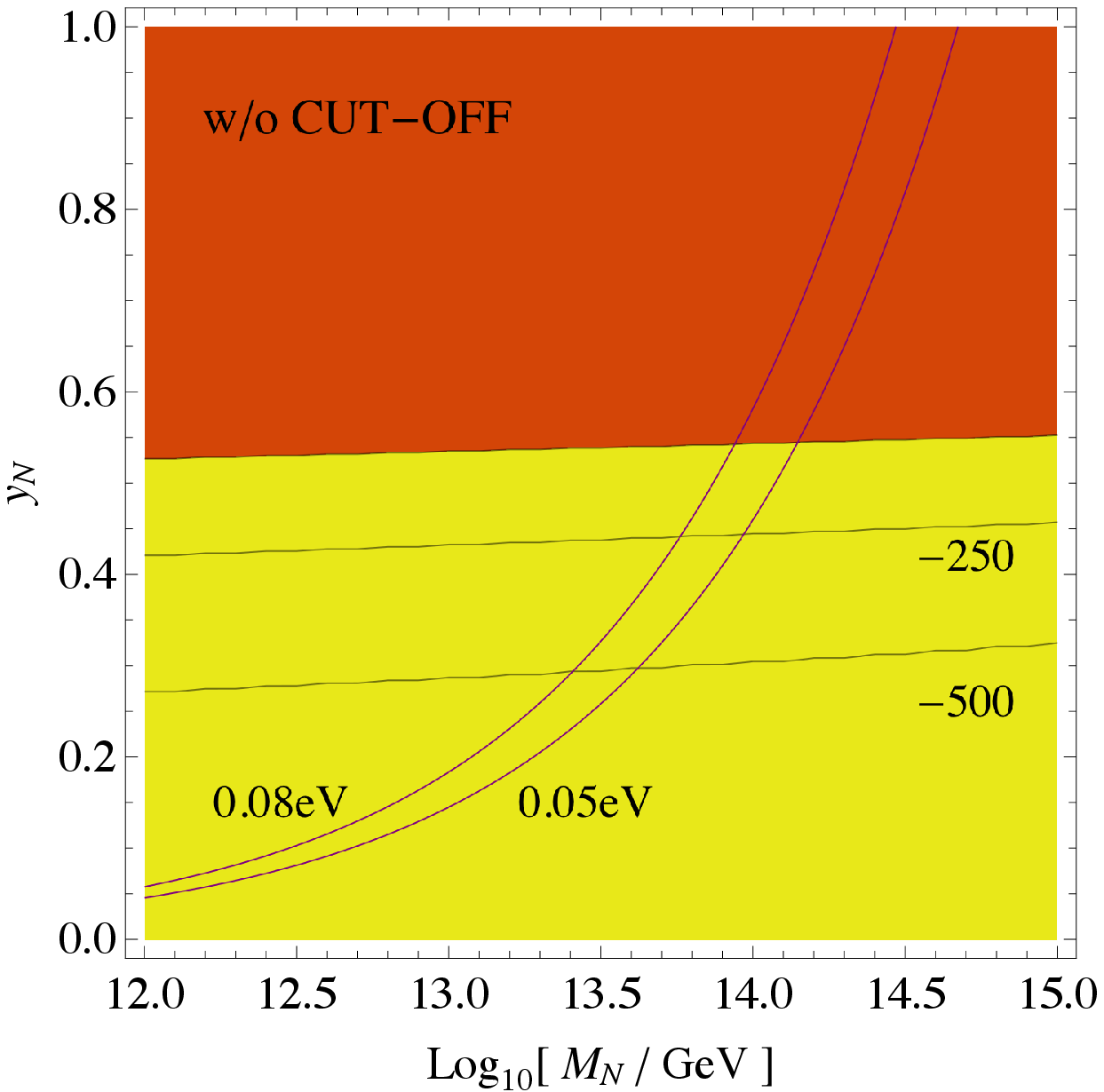}
  \end{center}
 \end{minipage}
 \caption{The same figure as fig.\ \ref{fig_VLQ} but for a right-handed
  neutrino. We also show lines indicating $m_\nu=0.05$ eV and
  $m_\nu=0.08$ eV with purple solid lines.}  \label{fig_RN}
\end{figure}

\begin{figure}[t]
 \begin{center}
  \includegraphics[width=0.5\linewidth]{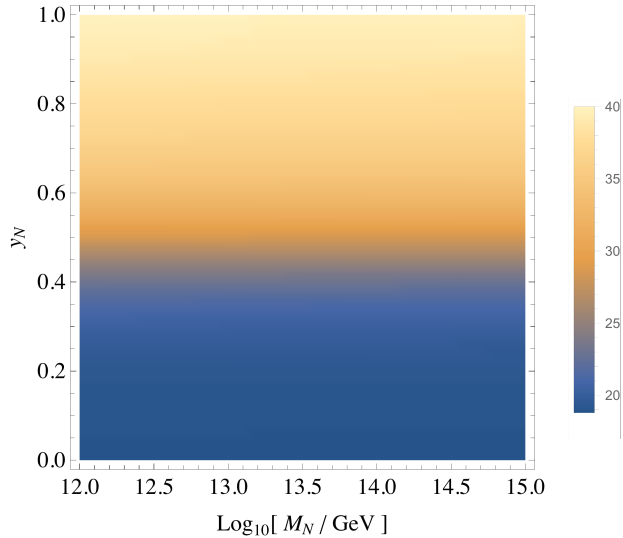}
 \end{center}
 \caption{The same figure as fig.\ \ref{fig_cutoffVLF} but for a
   right-handed neutrino. }  \label{fig_cutoffRHN}
\end{figure}

In the SM with right-handed neutrinos, the decay rate of the EW vacuum
is evaluated with
\begin{align}
 \gamma = \int d\ln R\frac{1}{R^4}
 \left[
  \mathcal A'^{(h)}\mathcal A^{(t)}\mathcal A^{(A_\mu,\varphi)}
  \mathcal A^{(L,N)}e^{-\mathcal B}
 \right]_{\msbar,~\mu=1/R},
\end{align}
where
\begin{align}
 \left[
  \ln \mathcal A^{(L,N)}
 \right]_{\msbar}
  & = \left.
 \left[
  \ln \mathcal A^{(\psi)}
 \right]_{\msbar}
 \right|_{y\to Y_{N}}.
\end{align}
In fig.\ \ref{fig_RN}, we show the contour plots of the decay
rate. Since it does not have any SM charges, the decay rate goes to
the SM value when $y_N$ goes to zero.  The effect of the cut-off at
the Planck scale is again large, which is because of a large
$\bar\phi_C^{\max}$ as shown in fig.\ \ref{fig_cutoffRHN}.  The purple
solid lines show the left-handed neutrino mass.  Requiring that the
decay rate should be smaller than the age of the universe, we obtain
$y_N\lesssim0.65-0.8$ for $10^{12}~{\rm GeV}\lesssim
M_N\lesssim10^{15}~{\rm GeV}$.

\section{Conclusion}
\label{sec_conclusion}
\setcounter{equation}{0}

In this paper, we have calculated the decay rate of the EW vacuum in the
framework of the SM and also in various models with extra fermions.  We
included the complete one-loop corrections as well as large logarithmic
terms in higher loop corrections.  We used a recently developed
technique to calculate functional determinants in the gauge sector,
which not only gives a prescription to perform a gauge invariant
calculation of the decay rate but also allows us to calculate the
functional determinants analytically.  In addition, in calculating the
decay rate of the EW vacuum, zero modes show up in association with the
dilatational and gauge symmetries.  We have properly taken into account
their effects, which was not possible in previous calculations.  We have
given an analytic formula of the decay rate of the EW vacuum, which is
also applicable to models that exhibit classical scale invariance at a
high energy scale.

The decay rate of the EW vacuum is sensitive to the coupling constants
in the SM and their RG behavior.  We have used three-loop RG equations
for the study of the RG behavior of the SM couplings.  The result is
used for the precise calculation of the decay rate of the EW vacuum.
The decay rate of the EW vacuum is estimated to be
\begin{align}
  \log_{10}\left[\gamma_{\rm Pl}\times{\rm Gyr~Gpc^3}\right]
  & = -582^{+40~+184~+144~+2}_{-45~-329~-218~-1},
\end{align}
where the errors come from the Higgs mass, the top mass, the strong
coupling constant, and the renormalization scale, respectively.  Here,
only the bounce configurations with its amplitude smaller than the
Planck scale is taken into account; for the decay rate of the EW vacuum,
such a cut-off of the bounce amplitude does not significantly affect the
result as far as it is not so far from the Planck scale. Since
$H_0^{-4}\simeq10^3~{\rm Gyr~Gpc^3}$, the lifetime is long enough
compared with the age of the universe.

We have also considered models with extra fermions. Since they
typically make the EW vacuum more unstable, the constraints on their
masses and couplings are phenomenologically important.  We have
analyzed the decay rate for the extensions of the SM with (i)
vector-like quarks, (ii) vector-like leptons, and (iii) a right-handed
neutrino.  We have obtained a constraint on the parameter space for
each model, requiring that the lifetime be long enough. The results
constrain the Yukawa couplings that are larger than about $0.3-0.5$ if
we do not consider the cut-off at the Planck scale. The effect of the
cut-off was found to be rather large and the constraints on the Yukawa
couplings become weaker, at most, by $0.3$ after including the
cut-off.

\section*{Acknowledgements}
The authors are grateful to A.~J.~Andreassen for helpful
correspondence.  This work was supported by the Grant-in-Aid for
Scientific Research C (No.26400239), and Innovative Areas
(No.16H06490). The work of S.C. was also supported in part by the
Program for Learning Graduate Schools, MEXT, Japan.

\vspace{3mm}\noindent {\it Note added}: The analytical and numerical
results given in this paper are consistent with those given in the
latest version of \cite{Andreassen:2017rzq}; in the earlier version,
(i) the calculation of the $J=0$ component of the gauge and NG
contributions, (ii) the subtraction of the divergence, and (iii) the
calculation of the volume of $SU(2)$ group, were not properly
performed, and have been corrected in the recent revision.  The
differences between our numerical results and those in
\cite{Andreassen:2017rzq} come mainly from the difference of the
threshold corrections to the $\msbar$ top Yukawa coupling constant,
which is regarded as a theoretical uncertainty.  In addition, there is
a difference in the treatment of the integration over the bounce size,
although it has little effect on the numerical results.

\appendix
\section{Functional Determinant}
\label{apx_determinantJ}
\setcounter{equation}{0}

In this Appendix, we present analytic formulae for various functional
determinants.  For simplicity, we consider the case with $U(1)$ gauge
interaction. The charge of the scalar field that is responsible for
the instability, $\Phi$, is set to be $+1$. Application of our results to
the case of general gauge groups is straightforward.  We are
interested in the case where the Lagrangian has (classical) scale
invariance; the potential of $\Phi$ is given in eq.\ \eqref{V(Phi)},
and the bounce solution is obtained as eq.\ \eqref{bounce}.

\subsection{Scalar contribution}

We first consider a real scalar field, $\sigma$, which couples to
$\Phi$ as
\begin{align}
  V=\kappa\sigma^2\Phi^\dagger\Phi,
\end{align}
where $\kappa$ is a positive coupling constant.  The contribution to the
prefactor is given by
\begin{align}
  \ln\mathcal A^{(\sigma)}
  = -\frac{1}{2}\ln\frac{\det[-\partial^2+\kappa\bar\phi^2]}{\det[-\partial^2]}.
\end{align}

First, we expand $\sigma$ into partial waves;
\begin{align}
  \sigma(r,\Omega) = 
  \alpha^{(\sigma)}_{J,m_A,m_B}(r)
  Y_{J,m_A,m_B}(\Omega).
\end{align}
Here and hereafter, $\alpha^{(X)}_{J,m_A,m_B}$ denotes the radial
mode function of $X$.  For notational simplicity, the summations over
$J$, $m_A$, and $m_B$ are implicit.

Since the fluctuation operator for partial waves does not depend on
$m_A$ and $m_B$, we have $(2J+1)^2$ degeneracy for each $J$.  Summing up
all the contributions, 
\begin{align}
 \ln\mathcal A^{(\sigma)}
 = -\frac{1}{2}\sum_{J=0}^{\infty}(2J+1)^2
 \ln\frac{\det[-\Delta_J+\kappa\bar\phi^2]}{\det[-\Delta_J]},
\end{align}
where
\begin{align}
 \Delta_J = \partial_r^2+\frac{3}{r}\partial_r-\frac{L^2}{r^2}.
\end{align}
Then, using eq.~\eqref{eq_theorem}, we have
\begin{align}
 \frac{\det[-\Delta_J+\kappa\bar\phi^2]}{\det[-\Delta_J]}
 = \lim_{r\to\infty}\frac{f^{(\sigma)}(r)}{r^{2J}},
\end{align}
where the function $f^{(\sigma)}$ satisfies
\begin{align}
 (-\Delta_J+\kappa\bar\phi^2)f^{(\sigma)} = 0,
\end{align}
and $\lim_{r\to0}f^{(\sigma)}(r)/r^{2J} = 1$.
The solution of the above differential equation is given by
\begin{align}
  f^{(\sigma)} = r^{2J}
  \left(
    1+\frac{|\lambda|}{8}\bar\phi_C^2r^2
  \right)^{1+z_\kappa}{}_2F_1
  \left(
    1+z_\kappa,2(J+1)+z_\kappa;2(J+1);-\frac{|\lambda|}{8}\bar\phi_C^2r^2
  \right),
\end{align}
where ${}_2F_1(a,b;c;z)$ is the hypergeometric function and
\begin{align}
  z_\kappa \equiv -\frac{1}{2}
  \left(
    1-\sqrt{1-8\frac{\kappa}{|\lambda|}}
  \right).
\end{align}
Taking the limit of $r\to\infty$, we have
\begin{align}
 \frac{\det[-\Delta_J+\kappa\bar\phi^2]}{\det[-\Delta_J]}
 = \frac{\Gamma(2J+1)\Gamma(2J+2)}{\Gamma(2J+1-z_\kappa)\Gamma(2J+2+z_\kappa)}.
 \label{eq_scalarDeterminant}
\end{align}

\subsection{Higgs contribution}

Using the bounce solution given in eq.\ \eqref{BounceSolution}, the
Higgs mode fluctuation is parametrized as
\begin{align}
  \Phi = \frac{1}{\sqrt{2}}
  \left[
    \bar\phi(r)
    + \alpha^{(h)}_{J,m_A,m_B}(r) Y_{J,m_A,m_B}(\Omega)
  \right].
\end{align}
Then, the Higgs contribution to the prefactor is given by
\begin{align}
  \ln\mathcal A^{(h)}
  & = -\frac{1}{2}\sum_{J=0}^\infty(2J+1)^2\ln\frac{\det[-\Delta_J-3|\lambda|\bar\phi^2]}{\det[-\Delta_J]}.
\end{align}
The functional determinant of the Higgs mode can be obtained with the
same procedure as the case of the scalar contribution, taking
$\kappa\to-3|\lambda|$:
\begin{align}
  \frac{\det[-\Delta_J-3|\lambda|\bar\phi^2]}{\det[-\Delta_J]}
  = \frac{2J(2J-1)}{(2J+3)(2J+2)}.
\end{align}
As we can see, the above ratio vanishes for $J = 0$ and $J = 1/2$,
which are due to the scale invariance and the translational
invariance, respectively.

\subsection{Fermion contribution}

Let us consider chiral fermions $\psi_L$ and $\psi_R$ with the
following Yukawa term
\begin{align}
  \mathcal L^{\rm (Yukawa)} = y \Phi \bar{\psi}_L \psi_R + {\rm h.c.}
\end{align}

The contribution to the prefactor is given by
\begin{align}
  \ln\mathcal A^{(\psi)}
  = &\, \ln\frac{\det
    \left[
      \cancel\partial+\frac{y}{\sqrt{2}}\bar\phi
    \right]}{\det[\cancel\partial]}
  \nonumber
  \\
  = &\, \frac{1}{2}\ln\frac{\det
  \left[
   \left(
    \cancel\partial+\frac{y}{\sqrt{2}}\bar\phi
   \right)
   \left(-\cancel\partial+\frac{y}{\sqrt{2}}\bar\phi
   \right)
  \right]}{\det[-\cancel\partial\cancel\partial]}.\nonumber \\
\end{align}
Taking the basis given in \cite{Avan:1985eg}, we expand it as
\begin{align}
  \ln\mathcal A^{(\psi)}
  & = \sum_{J=0}^{\infty} (2J+1)(2J+2)
  \ln\frac{\det\mathcal M^{(\psi)}_J}{\det\widehat{\mathcal M}^{(\psi)}_J},
\end{align}
where
\begin{align}
  \mathcal M^{(\psi)}_J =
  \begin{pmatrix}
    -\Delta_J+\frac{y^2}{2}\bar\phi^2
    & \frac{y}{\sqrt{2}}\bar\phi'
    \\
    \frac{y}{\sqrt{2}}\bar\phi'
    & -\Delta_{J+1/2}+\frac{y^2}{2}\bar\phi^2
  \end{pmatrix},
\end{align}
with $\bar\phi'=d\bar\phi/dr$, and $\widehat{\mathcal M}^{(\psi)}_J$
is obtained from $\mathcal M^{(\psi)}_J$ by replacing $\bar\phi\to0$
and $\bar\phi'\to0$.

Using eq.~\eqref{eq_theorem}, we have
\begin{align}
 \frac{\det\mathcal M^{(\psi)}_J}{\det\widehat{\mathcal M}^{(\psi)}_J}
 =
 \left(
  \lim_{r\to\infty}\frac{\det[\Psi_1~\Psi_2]}{r^{4J+1}}
 \right)
 \left(
  \lim_{r\to0}\frac{\det[\Psi_1~\Psi_2]}{r^{4J+1}}
 \right)^{-1},
\end{align}
where
\begin{align}
 \mathcal M^{(\psi)}_J\Psi_i(r) = 0.
 \label{Mpsi*psi=0}
\end{align}

Solutions of eq.\ \eqref{Mpsi*psi=0} can be expressed by using two
functions $\chi^{(\psi)}$ and $\eta^{(\psi)}$ as
\begin{align}
  \Psi_i =
  \begin{pmatrix}
    \frac{\sqrt{2}}{y\bar\phi}\frac{1}{r^{2J+3}}\partial_rr^{2J+3}\chi^{(\psi)}_i \\
    \chi^{(\psi)}_i
  \end{pmatrix} +
  \begin{pmatrix}
    -\frac{2}{y^2\bar\phi^2}\frac{1}{r^{2J+3}}\partial_rr^{2J+3}\eta^{(\psi)}_i \\
    0
  \end{pmatrix},
\end{align}
where $i=1$ and $2$ are for two independent solutions of eq.\
\eqref{Mpsi*psi=0}, and $\chi^{(\psi)}$ and $\eta^{(\psi)}$ obey
\begin{align}
  \left[
    -\Delta_{J+1/2}
    +\frac{\bar\phi'}{\bar\phi}\frac{1}{r^{2J+3}}\partial_rr^{2J+3}
    +\frac{y^2}{2}\bar\phi^2
  \right]\chi^{(\psi)}_i
  & =\frac{\sqrt{2}\bar\phi'}{y\bar\phi^2}
  \frac{1}{r^{2J+3}}\partial_rr^{2J+3}\eta^{(\psi)}_i,
  \\
  \left[
    -\Delta_{J+1/2}
    +\frac{\bar\phi'}{\bar\phi}\frac{1}{r^{2J+3}}\partial_rr^{2J+3}
    +\frac{y^2}{2}\bar\phi^2
  \right]\eta^{(\psi)}_i
  & =0.
\end{align}

For the first solution, we take
\begin{align}
 \chi^{(\psi)}_1(r) = f^{(\psi)}(r),~\eta^{(\psi)}_1(r) = 0,
\end{align}
where the function $f^{(\psi)}$ satisfies 
\begin{align}
  \left[
    -\Delta_{J+1/2}
    +\frac{\bar\phi'}{\bar\phi}\frac{1}{r^{2J+3}}\partial_rr^{2J+3}
    +\frac{y^2}{2}\bar\phi^2
  \right]f^{(\psi)} = 0,
\end{align}
and 
\begin{align}
  \lim_{r\to0}\frac{f^{(\psi)}(r)}{r^{2J+1}} = 1.
\end{align}
The analytic formula of $f^{(\psi)}$ is given by
\begin{align}
 f^{(\psi)}(r) = r^{2J+1}
 \left(
  1+\frac{|\lambda|}{8}\bar\phi_C^2r^2
 \right)^{-i\frac{y}{\sqrt{|\lambda|}}}{}_2F_1
 \left(
  2J+2-i\frac{y}{\sqrt{|\lambda|}},1-i\frac{y}{\sqrt{|\lambda|}};
  2J+3;-\frac{|\lambda|}{8}\bar\phi_C^2r^2
 \right),
\end{align}
and $f^{(\psi)}$ behaves as
\begin{align}
  \lim_{r\to\infty}\frac{f^{(\psi)}(r)}{r^{2J-1}}
  = \frac{8}{|\lambda|\bar\phi_C^2}
  \frac{\Gamma(2J+1)\Gamma(2J+3)}{\Gamma
    \left(
      2J+2-i\frac{y}{\sqrt{\lambda}}
    \right)
    \Gamma
    \left(
   2J+2+i\frac{y}{\sqrt{\lambda}}
 \right)}.
\end{align}

For the second solution, we take
\begin{align}
 \eta^{(\psi)}_2(r) = f^{(\psi)}(r).
\end{align}
Then, $\chi^{(\psi)}_2$ behaves as
\begin{align}
 \lim_{r\to0}\frac{\chi^{(\psi)}_2(r)}{r^{2J+3}}
  & = -\frac{\sqrt{2}\bar\phi''(0)}{y\bar\phi_C^2}\frac{J+1}{2J+3}, \\
 \lim_{r\to\infty}\frac{\chi^{(\psi)}_2(r)}{r^{2J+1}}
  & = \frac{\sqrt{2}}{y\bar\phi_C}
 \frac{[\Gamma(2J+2)]^2}{\Gamma
  \left(
   2J+2-i\frac{y}{\sqrt{\lambda}}
  \right)\Gamma
  \left(
   2J+2+i\frac{y}{\sqrt{\lambda}}
  \right)}.
\end{align}
Using these solutions, we get
\begin{align}
 \lim_{r\to0}\frac{\det[\Psi_1~\Psi_2]}{r^{4J+1}}
  & = \frac{8(J+1)}{y^2\bar\phi_C^2}, \\
  \lim_{r\to\infty}\frac{\det[\Psi_1~\Psi_2]}{r^{4J+1}}
  & = \frac{8(J+1)}{y^2\bar\phi^2}
 \left[
  \frac{[\Gamma(2J+2)]^2}{\Gamma
   \left(
    2J+2-i\frac{y}{\sqrt{\lambda}}
   \right)\Gamma
   \left(
    2J+2+i\frac{y}{\sqrt{\lambda}}
   \right)}
 \right]^2,
\end{align}
and hence
\begin{align}
  \frac{\det\mathcal M^{(\psi)}_J}{\det\widehat{\mathcal M}^{(\psi)}_J}
  =
  \left[
    \frac{[\Gamma(2J+2)]^2}{\Gamma
      \left(
        2J+2-i\frac{y}{\sqrt{\lambda}}
      \right)\Gamma
      \left(
        2J+2+i\frac{y}{\sqrt{\lambda}}
      \right)}
  \right]^2.
  \label{eq_FermionDeterminant}
\end{align}

\subsection{Gauge contribution}

We consider the contributions from gauge bosons, NG bosons, and
Faddeev-Popov ghosts.  The Lagrangian is given in the following form:
\begin{align}
  \mathcal L = \frac{1}{4} F_{\mu\nu} F_{\mu\nu}
  + \left[
    \left(
      \partial_\mu - i g A_\mu 
    \right) \Phi
  \right]^\dagger
  \left(
    \partial_\mu - i g A_\mu 
  \right) \Phi
  +V(\Phi)+\mathcal L_{\rm GF}+\mathcal L_{\rm FP},
\end{align}
where $F_{\mu\nu}$ is the field strength tensor, and
\begin{align}
  \mathcal L_{\rm GF} & = \frac{1}{2\xi}\mathcal F^2, \\
  \mathcal L_{\rm FP} & = \bar{c}(-\partial_\mu \partial_\mu)c,
\end{align}
with
\begin{align}
  \mathcal F = \partial_\mu A_\mu.
  \label{gaugefixingfn}
\end{align}

Since Faddeev-Popov ghosts do not directly couple to the Higgs field
with our choice of the gauge fixing function, we have
\begin{align}
  \ln \mathcal A^{(c,\bar{c})} = 0.
\end{align}

At the one-loop level, $\ln\mathcal A^{(A_\mu,\varphi)}$ is given by
\begin{align}
  \ln\mathcal A^{(A_\mu,\varphi)}
  & = 
  - \frac{1}{2}
  \ln
  \frac{\det\mathcal M^{(A_\mu,\varphi)}}
  {\det\widehat{\mathcal M}^{(A_\mu,\varphi)}},
\end{align}
where
\begin{align}
 \mathcal M^{(A_\mu,\varphi)} =
 \begin{pmatrix}
  -\partial^2\delta_{\mu\nu}+
  \left(
   1-\frac{1}{\xi}
  \right)\partial_\mu\partial_\nu+g^2\bar\phi^2
   & g(\partial_\nu\bar\phi)-g\bar\phi\partial_\nu \\
  2g(\partial_\mu\bar\phi)+g\bar\phi\partial_\mu
   & -\partial^2+ V_{\varphi\varphi}
 \end{pmatrix},
 \label{eq_gaugeFluct}
\end{align}
with $V_{\varphi\varphi} = d^2V/d\varphi^2$.

For the partial wave expansions, we use the following basis;
\begin{align}
  A_\mu(r,\Omega)
  = &\,
  \alpha^{(S)}_{J,m_A,m_B}(r)\frac{x_\mu}{r} Y_{J,m_A,m_B}(\Omega)
  +\alpha^{(L)}_{J,m_A,m_B}(r)\frac{r}{L}\partial_\mu Y_{J,m_A,m_B}(\Omega)
  \nonumber \\ &\,
  +\alpha^{(T1)}_{J,m_A,m_B}(r) i \varepsilon_{\mu\nu\rho\sigma}
  V_\nu^{(1)}L_{\rho\sigma}Y_{J,m_A,m_B}(\Omega)
  \nonumber \\ &\,
  +\alpha^{(T2)}_{J,m_A,m_B}(r)i\varepsilon_{\mu\nu\rho\sigma}
  V_\nu^{(2)}L_{\rho\sigma}Y_{J,m_A,m_B}(\Omega),
  \\
  \varphi(r,\Omega)
  = &\,
  \alpha^{(\varphi)}_{J,m_A,m_B}(r)Y_{J,m_A,m_B}(\Omega),
  \label{eq_gaugeBasis}
\end{align}
where $V_\nu^{(i)}$'s are arbitrary independent vectors and
$\varepsilon_{\mu\nu\rho\sigma}$ is a fully anti-symmetric tensor.
Then, we have
\begin{align}
  \ln\mathcal A^{(A_\mu,\varphi)}
  &= -\frac{1}{2}\ln\frac{\det\mathcal M^{(S,\varphi)}_0}{\det\widehat{\mathcal M}^{(S,\varphi)}_0}
  -\frac{1}{2}\sum_{J = 1/2}^{\infty}(2J+1)^2
  \left[
    \ln\frac{\det\mathcal M^{(S,L,\varphi)}_J}{\det\widehat{\mathcal M}^{(S,L,\varphi)}_J}
    +2\ln\frac{\det\mathcal M^{(T)}_J}{\det\widehat{\mathcal M}^{(T)}_J}
  \right],
\end{align}
where
\begin{align}
  \mathcal M^{(S,\varphi)}_0  = &\,
  \begin{pmatrix}
    \frac{1}{\xi}\left(-\Delta_{1/2}+\xi g^2\bar\phi^2\right)
    & g\bar\phi'-g\bar\phi\partial_r \\
    2g\bar\phi'+g\bar\phi\frac{1}{r^3}\partial_rr^3
    & -\Delta_0+V_{\varphi\varphi}
  \end{pmatrix},
  \\
  \mathcal M^{(S,L,\varphi)}_J = &\,
  \begin{pmatrix}
    -\Delta_J+\frac{3}{r^2}+g^2\bar\phi^2
    & -\frac{2L}{r^2}
    & g\bar\phi'-g\bar\phi\partial_r
    \\
    -\frac{2L}{r^2}
    & -\Delta_J-\frac{1}{r^2}+g^2\bar\phi^2 & -\frac{L}{r}g\bar\phi
    \\
    2g\bar\phi'+g\bar\phi\frac{1}{r^3}\partial_rr^3
    & -\frac{L}{r}g\bar\phi
    & -\Delta_J+V_{\varphi\varphi}
  \end{pmatrix}
  \nonumber \\ &\,
  +\left(1-\frac{1}{\xi}\right)
  \begin{pmatrix}
    \Delta_{1/2}
    & -L\partial_r\frac{1}{r}
    & 0                       \\
    \frac{L}{r^4}\partial_rr^3
    & -\frac{L^2}{r^2}
    & 0                       \\
    0
    & 0
    & 0
  \end{pmatrix},
\end{align}
and
\begin{align}
 \mathcal M^{(T)}_J = -\Delta_J+g^2\bar\phi^2.
\end{align}

Three independent solutions of $\mathcal M^{(S,\varphi)}_J\Psi_i=0$
(with $i=1-3$) can be constructed from the functions, $\chi_i$,
$\eta_i$, and $\zeta_i$, as \cite{Endo:2017gal,Endo:2017tsz}\footnote
{The function $\eta$ in the present analysis corresponds to $\eta/L$
  in \cite{Endo:2017gal,Endo:2017tsz}; such a rescaling makes the
  formulae simpler.}
\begin{align}
 \Psi_i\equiv
 \begin{pmatrix}
  \Psi_i^{\rm (top)} \\
  \Psi_i^{\rm (mid)} \\
  \Psi_i^{\rm (bot)}
 \end{pmatrix}=
 \begin{pmatrix}
  \partial_r\chi_i  \\
  \frac{L}{r}\chi_i \\
  g\bar\phi\chi_i
 \end{pmatrix}
 +
 \begin{pmatrix}
  \frac{L}{r}\frac{1}{g^2\bar\phi^2}\eta_i                \\
  \frac{1}{g^2\bar\phi^2}\frac{1}{r^2}\partial_rr^2\eta_i \\
  0
 \end{pmatrix}
 +
 \begin{pmatrix}
  -\frac{2\bar\phi'}{g^2\bar\phi^3}\zeta_i \\
  0                                          \\
  \frac{1}{g\bar\phi}\zeta_i
 \end{pmatrix},
\end{align}
where $\chi_i$, $\eta_i$, and $\zeta_i$ obey
\begin{align}
  & \Delta_J\chi_i = \frac{L}{r}\frac{2\bar\phi'}{g^2\bar\phi^3}\eta_i
 +\frac{1}{r^3}\partial_rr^3\frac{2\bar\phi'}{g^2\bar\phi^3}\zeta_i
 -\xi\zeta_i,                                                            \\
  &
 \left(
  \Delta_J-2\frac{\bar\phi'}{\bar\phi}\frac{1}{r^2}\partial_rr^2-g^2\bar\phi^2
 \right)\eta_i
 = -2\frac{L}{r}\frac{\bar\phi'}{\bar\phi}\zeta_i,                       \\
  & \Delta_J\zeta_i = 0.
\end{align}
We also note useful relations:
\begin{align}
  \frac{1}{r^3}\partial_rr^3\Psi^{\rm (top)}_i
  & = \frac{L}{r}\Psi_i^{\rm (mid)}-\xi\zeta_i,
  \label{eq_relationPositiveJ1}                  \\
  \frac{1}{r}\partial_rr\Psi^{\rm (mid)}_i
  & = \frac{L}{r}\Psi^{\rm (top)}_i+\eta_i,
  \label{eq_relationPositiveJ2}                  \\
  \Psi^{\rm (bot)}_i
  & = \frac{rg\bar\phi}{L}\Psi^{\rm (mid)}_i
  -\frac{1}{g\bar\phi L}\frac{1}{r}\partial_rr^2\eta_i
  +\frac{1}{g\bar\phi}\zeta_i.
  \label{eq_relationPositiveJ3}
\end{align}

The first solution is obtained by setting $\zeta_1=0$ and $\eta_1=0$
as
\begin{align}
 \chi_1 = r^{2J},
\end{align}
resulting in
\begin{align}
 \Psi_1 =
 \begin{pmatrix}
  2Jr^{2J-1} \\
  Lr^{2J-1}  \\
  g\bar\phi r^{2J}
 \end{pmatrix}.
\end{align}

For the second solution, we set $\zeta_2 = 0$ and
\begin{align}
 \eta_2 = f^{(\eta)},
\end{align}
where the function $f^{(\eta)}$ satisfies
\begin{align}
  \left(
    \Delta_J
    -2\frac{\bar\phi'}{\bar\phi}\frac{1}{r^2}\partial_rr^2
    -g^2\bar\phi^2
  \right)f^{(\eta)} = 0,
\end{align}
and
\begin{align}
 \lim_{r\to0}\frac{f^{(\eta)}(r)}{r^{2J}} = 1.
\end{align}
We can find an analytic formula of $f^{(\eta)}$ as
\begin{align}
  f^{(\eta)} = r^{2J}
  \left(
    1+\frac{|\lambda|}{8}\bar\phi_C^2r^2
  \right)^{z_g}{}_2F_1
  \left(
    1+z_g,2(J+1)+z_g;2(J+1);-\frac{|\lambda|}{8}\bar\phi_C^2r^2
  \right),
\end{align}
with
\begin{align}
  z_g = -\frac{1}{2}\left(
    1-\sqrt{1-8\frac{g^2}{|\lambda|}}
  \right).
\end{align}
Then, we get
\begin{align}
  \lim_{r\to\infty}\frac{f^{(\eta)}(r)}{r^{2J-2}}
  & = \frac{8}{|\lambda|\bar\phi_C^2}
  \frac{\Gamma(2J+1)\Gamma(2J+2)}{\Gamma(2J+1-z_g)\Gamma(2J+2+z_g)}.
\end{align}
Using eqs.~\eqref{eq_relationPositiveJ1},
\eqref{eq_relationPositiveJ2}, and \eqref{eq_relationPositiveJ3}, we
obtain
\begin{align}
  \lim_{r\to0}\Psi_2      & =
  \begin{pmatrix}
    \frac{Lr^{2J+1}}{8(J+1)}   \\
    \frac{J+2}{4(J+1)}r^{2J+1} \\
    -\frac{2}{g\bar\phi_C}\frac{J+1}{L}r^{2J}
  \end{pmatrix}, \\
  \lim_{r\to\infty}\Psi_2 & =
  \frac{\Gamma(2J+1)\Gamma(2J+2)}{\Gamma(2J+1-z_g)\Gamma(2J+2+z_g)}
  \begin{pmatrix}
    \frac{4L}{(2J+1)|\lambda|\bar\phi^2}r^{2J-1}\ln r       \\
    \frac{8(J+1)}{(2J+1)|\lambda|\bar\phi_C^2}r^{2J-1}\ln r \\
    -\frac{2J}{gL\bar\phi_C}r^{2J}
  \end{pmatrix}.
\end{align}

The last solution can be obtained with
\begin{align}
  \zeta_3 = r^{2J}.
\end{align}
The asymptotic form of $\eta_3$ is given by
\begin{align}
  \lim_{r\to0}\frac{\eta_3(r)}{r^{2J+2}} = &\,
  \frac{L|\lambda|\bar\phi_C^2}{16(J+1)}, \\
  \lim_{r\to\infty}\frac{\eta_3(r)}{r^{2J}} = &\,
  \frac{L}{2(J+1)}.
\end{align}
Using eqs.~\eqref{eq_relationPositiveJ1}, \eqref{eq_relationPositiveJ2},
and \eqref{eq_relationPositiveJ3}, we have
\begin{align}
  \lim_{r\to0}\Psi_3      & =
 \begin{pmatrix}
   -\frac{\xi}{4}r^{2J+1}   \\
   -\frac{J\xi}{2L}r^{2J+1} \\
   \frac{1}{g\bar\phi_C}r^{2J}
 \end{pmatrix}, \\
 \lim_{r\to\infty}\Psi_3 & =
 \begin{pmatrix}
   \frac{J-\xi(J+1)}{4(J+1)}r^{2J+1}         \\
   \frac{J[(J+2)-\xi(J+1)]}{2L(J+1)}r^{2J+1} \\
   \frac{g}{|\lambda|\bar\phi_C}\frac{(J+2)-\xi(J+1)}{(J+1)^2}r^{2J}
 \end{pmatrix}.
\end{align}

We also need three independent solutions around the false vacuum.  We
take
\begin{align}
  (\hat\Psi_1~\hat\Psi_2~\hat\Psi_3)=
  \begin{pmatrix}
    2Jr^{2J-1} & \frac{(J+1)\xi-J}{2L^2}r^{2J+1}        & 0      \\
    Lr^{2J-1}  & \frac{(J+1)\xi-(J+2)}{4L(J+1)}r^{2J+1} & 0      \\
    0          & 0                                      & r^{2J}
  \end{pmatrix}.
\end{align}

Then, using eq.\ \eqref{eq_theorem} with combining the above
expressions, we obtain the ratio of the functional determinants for
$S$, $L$, and NG modes as
\begin{align}
  \frac{\det\mathcal M^{(S,L,\varphi)}_J}
  {\det\widehat{\mathcal M}^{(S,L,\varphi)}_J}
  = &\,\frac{J}{J+1}
  \frac{\Gamma(2J+1)\Gamma(2J+2)}{\Gamma(2J+1-z_g)\Gamma(2J+2+z_g)}.
\end{align}
The functional determinant for $T$ modes can be obtained by using the
result for the scalar contribution.  With the replacement $\kappa\to
g^2$ in eq.~\eqref{eq_scalarDeterminant}:
\begin{align}
 \frac{\det\mathcal M_J^{(T)}}{\det\widehat{\mathcal M}_J^{(T)}}
 = \frac{\Gamma(2J+1)\Gamma(2J+2)}{\Gamma(2J+1-z_g)\Gamma(2J+2+z_g)}.
\end{align}

For $J=0$, the solutions of $\mathcal M_0^{(S,\varphi)}\Psi = 0$ can
be decomposed as \cite{Endo:2017gal,Endo:2017tsz}
\begin{align}
  \Psi_i\equiv
  \begin{pmatrix}
    \Psi_i^{\rm (top)} \\
    \Psi_i^{\rm (bot)}
  \end{pmatrix} =
  \begin{pmatrix}
    \partial_r\chi_i \\
    g\bar\phi\chi_i
  \end{pmatrix}
  +
  \begin{pmatrix}
    -\frac{2\bar\phi'}{g^2\bar\phi^3}\zeta_i \\
    \frac{1}{g\bar\phi}\zeta_i
  \end{pmatrix},
\end{align}
with $i=1$ and $2$ for two independent solutions, where the functions
$\chi_i$ and $\zeta_i$ satisfy
\begin{align}
  & \Delta_0\chi_i =
  \frac{1}{r^3}\partial_rr^3\frac{2\bar\phi'}{g^2\bar\phi^3}\zeta_i
  -\xi\zeta_i,
  \\
  & \Delta_0\zeta_i =
  0.
  \label{eq_gauge0zeta}
\end{align}
Notice that there are useful relations:
\begin{align}
  \frac{1}{r^3}\partial_rr^3\Psi_i^{\rm (top)}
  & = -\xi\zeta_i,
  \label{eq_relationZero1}
  \\
  \partial_r\frac{1}{g\bar\phi}\Psi_i^{\rm (bot)}
  & = \Psi_i^{\rm (top)}+\frac{1}{g^2\bar\phi^2}\partial_r\zeta_i.
  \label{eq_relationZero2}
\end{align}
The first solution is given by
\begin{align}
  \Psi_1 =
  \begin{pmatrix}
    0 \\
    \frac{\bar\phi}{\bar\phi_C}
  \end{pmatrix}.
  \label{eq_gauge0psi1}
\end{align}
The second solution can be obtained with
\begin{align}
 \zeta_2 = 1,
\end{align}
giving rise to 
\begin{align}
  \Psi_2 =
  \begin{pmatrix}
    -\xi\frac{r}{4} \\
    -\xi g\bar\phi\frac{r^2}{8}
  \end{pmatrix}.
\end{align}
Then, using eq.\ \eqref{eq_theorem}, we obrtain
\begin{align}
  \frac{\det\mathcal M^{(S,\varphi)}_0}{\det\widehat{\mathcal M}^{(S,\varphi)}_0}
  = 0.\label{eq_gaugezero}
\end{align}
The vanishing of the above ratio is due to the existence of zero
mode.  The treatment of the zero mode is discussed in Appendix
\ref{apx_zeromode}.

\section{Zero Modes}
\label{apx_zeromode}
\setcounter{equation}{0}

In this Appendix, we discuss the zero modes associated with the
dilatation, translation, and global transformation.

\subsection{Dilatational zero mode}

In $J=0$ mode of the Higgs fluctuation, there exists a dilatational
zero mode. The dilatational transformation is parametrized by
$\bar\phi_C$; with $\bar\phi_C\rightarrow\bar\phi_C+\delta\bar\phi_C$,
such a change can be absorbed by
\begin{align}
  h \to h + \delta\bar\phi_C\tilde{\fnc}_DY_{0,0,0}+\cdots,
  \label{h(dilatation)}
\end{align}
where we neglect higher order terms in $\delta\bar\phi_C$, and
\begin{align}
 \tilde{\fnc}_D(r) = \sqrt{2\pi^2}\frac{d\bar\phi}{d\bar\phi_C}.
\end{align}
The second term in the right-hand side of eq.\ \eqref{h(dilatation)}
is nothing but the change of the amplitude of the dilatational zero
mode; one can easily check
${\cal M}^{(h)}_0\tilde{\fnc}_D=0$.

In order to translate the path integral over the dilatational zero
mode to the integration over $\bar\phi_C$, we calculate
\begin{align}
  \frac{\det[-\Delta_0-3|\lambda|\bar\phi^2+\nu \rho_D(r)]}{\det[-\Delta_0]},
\end{align}
with\footnote
{A prescription used in \cite{Chigusa:2017dux} is consistent with our
  argument, where $\rho_D$ is a constant.}
\begin{align}
  \rho_D(r) = \frac{15}{16\pi}|\lambda|^2\bar\phi_C^2\bar\phi(r)^2.
\end{align}
Notice that the condition \eqref{rho-normalization} is satisfied with
the above choice of $\rho_D$, {\it i.e.},
\begin{align}
  \int dr r^3 \tilde{\fnc}_D^2 (r) \rho_D (r) = 2 \pi.
\end{align}
The ratio of the determinant can be obtained from
eq.~\eqref{eq_scalarDeterminant} by replacing
$\kappa\to-3|\lambda|+\frac{15}{16\pi}|\lambda|^2\bar\phi_C^2\nu$. Expanding
the result with respect to $\nu$, we get
\begin{align}
 \frac{\det[-\Delta_0-3|\lambda|\bar\phi^2+\nu \rho_D(r)]}{\det[-\Delta_0]}
 = -\nu\frac{|\lambda|}{16\pi}\bar\phi_C^2+\mathcal O(\nu^2).
\end{align}
Thus, we obtain
\begin{align}
  \left|
    \frac{\det[-\Delta_0-3|\lambda|\bar\phi^2]}{\det[-\Delta_0]}
  \right|^{-1/2}
  \to \int d\bar\phi_C\sqrt{\frac{16\pi}{|\lambda|}}\frac{1}{\bar\phi_C}
  = 
  \int d\ln R \sqrt{\frac{16\pi}{|\lambda|}}.
\end{align}
Here, we take an absolute value since there is a negative mode
\cite{Coleman:1977py,Callan:1977pt}.

\subsection{Translational zero mode}

In $J = 1/2$ mode of the Higgs fluctuation, translational zero modes
exist.  The translation is parametrized by the center of the bounce.
The shift of the center of the bounce to $a_\mu$ can be absorbed by the
transformation of the Higgs mode as
\begin{align}
 h \to h + a_\mu \tilde{\fnc}_T Y_{1/2,\mu} + \cdots,
\end{align}
where
\begin{align}
  \tilde{\fnc}_T (r) = -\frac{\pi}{\sqrt{2}}\frac{d\bar\phi}{dr},
\end{align}
and 
\begin{align}
  Y_{1/2,\mu}(\hat{x}) = \frac{\sqrt{2}}{\pi}\hat{x}_\mu,
\end{align}
with $\hat{x}_\mu = x_\mu/|x|$.  Notice that $Y_{1/2,\mu}$ is given by
a linear combination of $Y_{1/2,m_A,m_B}$ with the same
normalization as $Y_{1/2,m_A,m_B}$.  Thus, as noted in
\cite{Callan:1977pt}, the path integral over the translational zero
modes can be converted to the integration over the spacetime volume.

In order to take care of the translational zero mode, we calculate
\begin{align}
 \frac{\det[-\Delta_{1/2}-3|\lambda|\bar\phi^2+\rho_T(r)\nu]}
 {\det[-\Delta_{1/2}]},
\end{align}
with
\begin{align}
 \rho_T(r) = \frac{3|\lambda|}{4\pi}.
\end{align}
One can see that
\begin{align}
  \int dr r^3 \tilde{\fnc}_T^2 (r) \rho_T(r) = 2 \pi,
\end{align}
which is consistent with eq.\ \eqref{rho-normalization}.

Following the argument in Subsection \ref{subsec:zeromodes}, we have
\begin{align}
 \left(
  \frac{\det[-\Delta_{1/2}-3|\lambda|\bar\phi^2]}{\det[-\Delta_{1/2}]}
 \right)^{-2}\to\int d^4a
 \lim_{r\rightarrow\infty}
 \left(
  \frac{\check f^{(h)}_{1/2}}{r}
 \right)^{-2},
\end{align}
where
\begin{align}
 [-\Delta_{1/2}-3|\lambda|\bar\phi^2]\check f^{(h)}_{1/2} = -\rho_Tf_{1/2}^{(h)},
\end{align}
with
\begin{align}
 f_{1/2}^{(h)} = -\frac{4}{|\lambda|\bar\phi_C^3}\frac{d\bar\phi}{dr}.
\end{align}
The function $\check f^{(h)}_{1/2}$ behaves as
\begin{align}
 \lim_{r\to\infty}\frac{\check f^{(h)}_{1/2}}{r} = \frac{1}{4\pi\bar\phi_C^2}.
\end{align}
Thus, we obtain
\begin{align}
  \left(
    \frac{\det[-\Delta_{1/2}-3|\lambda|\bar\phi^2]}{\det[-\Delta_{1/2}]}
  \right)^{-2}
  \to
  \left(
    \frac{32\pi}{|\lambda|}
  \right)^2
  \left(
    \frac{|\lambda|}{8}\bar\phi_C^2
  \right)^2\mathcal V_{\rm 4D} =
  \left(
    \frac{32\pi}{|\lambda|}
  \right)^2
  \frac{\mathcal V_{\rm 4D}}{R^4},
\end{align}
where $\mathcal V_{\rm 4D}$ is the spacetime volume.

\subsection{Gauge zero mode}

In $J = 0$ mode of the gauge field, we have a gauge zero mode.  For
the case of the $U(1)$ gauge symmetry, the bounce solution is
parameterized as eq.\ \eqref{bounce} with the parameter $\theta$.  The
path integral over the gauge zero mode can be understood as the
integration over the variable $\theta$, as we see below.

The effect of the shift $\theta\rightarrow\theta+\delta\theta$
can be absorbed by the transformation of the NG mode as
\begin{align}
  \varphi \to 
  \varphi + \delta \theta \tilde{\fnc}_G Y_{0,0,0} + \cdots,
\end{align}
where
\begin{align}
  \tilde{\fnc}_G = \sqrt{2\pi^2} \bar\phi.
\end{align}
Using the equation of motion of the bounce solution, the following
relation holds:
\begin{align}
  \mathcal M^{(S,\varphi)}_0 
  \begin{pmatrix}
    0 \\ \tilde \fnc_G 
  \end{pmatrix} = 0.
\end{align}

In order to deal with gauge zero mode, we calculate
\begin{align}
  \frac{\det
    \left[
      \mathcal M_0^{(S,\varphi)}+\nu\rho_G(r)
    \right]}{\det\widehat{\mathcal M}_0^{(S,\varphi)}},
  \label{detG(nu)}
\end{align}
with
\begin{align}
  \rho_G(r) = \frac{|\lambda|^2}{16\pi}\bar\phi_C\bar\phi.
\end{align}
Notice that the following relation holds:
\begin{align}
  \int dr r^3 \tilde{\fnc}_G^2 (r) \rho_G (r) = 2 \pi.
\end{align}

For the evaluaton of the ratio \eqref{detG(nu)} at the leading order
in $\nu$, we introduce the function $\check\Psi_1$ satisfying
\begin{align}
  \mathcal M^{(S,\varphi)}_0\check\Psi_1(r) = -\rho_G(r)\Psi_1(r).
\end{align}
with which
\begin{align}
  \left(
    \frac{\det\mathcal M^{(S,\varphi)}_0}{\det\widehat{\mathcal M}^{(S,\varphi)}_0}
  \right)^{-1/2}
  \to
  \int d\theta
  \left(
    \lim_{r\to\infty}\frac{\det[\check\Psi_1(r)~\Psi_2(r)]}{r}
  \right)^{-1/2}
  \left(
    \lim_{r\to0}\frac{\det[\Psi_1(r)~\Psi_2(r)]}{r}
  \right)^{1/2}.
\end{align}
We can decompose $\check\Psi_1$ as
\begin{align}
  \check\Psi_1
  =
  \begin{pmatrix}
    \partial_r \check\chi_1 \\
    g\bar\phi \check\chi_1
  \end{pmatrix}
  +
  \begin{pmatrix}
    -\frac{2\bar\phi'}{g^2\bar\phi^3} \check\zeta_1 \\
    \frac{1}{g\bar\phi} \check\zeta_1
  \end{pmatrix},
\end{align}
where $\check\chi_1$ and $\check\zeta_1$ satisfies
\begin{align}
  & \Delta_0 \check\chi_1 =
  \frac{1}{r^3}\partial_rr^3\frac{2\bar\phi'}{g^2\bar\phi^3}
  \check\zeta_1
  -\xi \check\zeta_1,
  \\
  & \Delta_0 \check\zeta_1 = g \rho_G \frac{\bar\phi^2}{\bar\phi_C}.
\end{align}
We can solve $\check\zeta_1$ as
\begin{align}
  \check\zeta_1(r) = -\frac{|\lambda|g}{16\pi}\bar\phi(r),
\end{align}
based on which $\check\Psi_1$ should behave as
\begin{align}
  \lim_{r\to\infty}\check\Psi_1 =
  \begin{pmatrix}
    \frac{\xi g}{4\pi \bar\phi_C r} \\
    \frac{|\lambda|}{16\pi}
  \end{pmatrix}.
\end{align}
Thus, we obtain
\begin{align}
  \left(
    \frac{\det\mathcal M^{(S,\varphi)}_0}{\det\widehat{\mathcal M}^{(S,\varphi)}_0}
  \right)^{-1/2}
  \to
  \int d\theta \sqrt{\frac{16\pi}{|\lambda|}}.
\end{align}

\section{Infinite Sum over Angular Momentum}
\label{apx_infiniteSum}
\setcounter{equation}{0}

In this Appendix, we perform various infinite sums appearing in the
calculation of functional determinants.

We first evaluate the following sum:
\begin{align}
  I_\sigma(z) =
 \sum_{J=0}^\infty\frac{(2J+1)^2}{(1+\varepsilon_\sigma)^{2J}}
 \ln\frac{\Gamma(2J+1)\Gamma(2J+2)}{\Gamma(2J+1-z)\Gamma(2J+2+z)}, 
\end{align}
which can be used for the calculation of the scalar contribution to
the prefactor.  Notice that $J=0,1/2,1,\cdots$ is half-integer.  In addition, here,
$z$ is a complex number satisfying
\begin{align}
  -2<\Re(z)<1.
  \label{eq_rez}
\end{align}
For $\varepsilon_\sigma>0$, the sum converges thanks to the factor
$1/(1+\varepsilon_\sigma)^{2J}$.

First, we rewrite the log-gamma functions with integrals of digamma
functions as
\begin{align}
 I_\sigma(z) = \sum_{J=0}^{\infty}\frac{(2J+1)^2}{(1+\varepsilon_\sigma)^{2J}}
 \int_\infty^{2J+1}dx
 \left[
  \psi_{\Gamma}(x)-\psi_{\Gamma}(x-z)+\psi_{\Gamma}(x+1)-\psi_{\Gamma}(x+1+z)
 \right],
\end{align}
where $\psi_{\Gamma}(z)$ is the digamma function.
Then, we use the following relation:
\begin{align}
 \psi_{\Gamma}(x)-\psi_{\Gamma}(y) = \int_0^1\frac{u^{y-1}-u^{x-1}}{1-u}du,
\end{align}
which is valid for $\Re(x)>0$ and $\Re(y)>0$, and we obtain
\begin{align}
 I_\sigma(z) = \sum_{J=0}^{\infty}\frac{(2J+1)^2}{(1+\varepsilon_\sigma)^{2J}}
 \int_\infty^{2J+1}dx\int_0^1du\frac{u^{x-z-1}(1-u^z)(1-u^{z+1})}{1-u}.
\end{align}
Notice that this is verified only in region \eqref{eq_rez}.
Then, we interchange the two integrals,\footnote
{This is justified when
  \begin{align*}
    \int_\infty^{2J+1}dx\int_0^1du
    \left|
      \frac{u^{x-z-1}(1-u^z)(1-u^{z+1})}{1-u}
    \right|<\infty.
  \end{align*}
}
and integrate over $x$ first. Consequently, it becomes
\begin{align}
 I_\sigma(z) = \sum_{J=0}^{\infty}\frac{(2J+1)^2}{(1+\varepsilon_\sigma)^{2J}}
 \int_0^1du\frac{u^{2J}(1-u^z)(1-u^{1+z})}{u^z(1-u)\ln u}.
\end{align}
Notice that the integration over $u$ is convergent.

Since we have regularized the sum, the result should be finite. Thus,
we can take the sum first:
\begin{align}
 I_\sigma(z) = (1+\varepsilon_\sigma)^2\int_0^1du
 \frac{(1-u^z)(1-u^{1+z})(1+u+\varepsilon_\sigma)}{u^z(1-u)(1+\varepsilon_\sigma-u)^3\ln u}.
\end{align}
We can see that new poles appear at $u = 1+\varepsilon_\sigma$, but the
integral is still convergent for positive $\varepsilon_\sigma$.  Then, using
\begin{align}
 \frac{1-u^z}{\ln u} = -z\int_0^1dtu^{zt},
\end{align}
we obtain
\begin{align}
 I_\sigma(z) = -z(1+\varepsilon_\sigma)^2\int_0^1du\int_0^1dt
 \frac{(1-u^{1+z})(1+u+\varepsilon_\sigma)}{u^{z(1-t)}(1-u)(1+\varepsilon_\sigma-u)^3}.
\end{align}

Interchanging the integrals and integrating over $u$, we get
\begin{align}
  I_\sigma(z) = &\,
  z \int_0^1 dt
  \Bigg[
  \frac{(1+\varepsilon_\sigma)(z-1+tz\varepsilon_\sigma-\varepsilon_\sigma)}{\varepsilon_\sigma^2}
  +\frac{(1+\varepsilon_\sigma)^2(2+\varepsilon_\sigma)[\psi_{\Gamma}(1+tz-z)
    -\psi_{\Gamma}(1+tz)]}{\varepsilon_\sigma^3}
  \nonumber \\ &\,
  +\frac{2+(3+2z-2tz)\varepsilon_\sigma
    +(1+z-tz)^2\varepsilon_\sigma^2}{(1+tz-z)\varepsilon_\sigma^3}{}_2F_1
  \left(
    1,1+tz-z;2+tz-z;\frac{1}{1+\varepsilon_\sigma}
  \right)
  \nonumber \\ &\,
  -\frac{(1+\varepsilon_\sigma)
    (2+\varepsilon_\sigma-2tz\varepsilon_\sigma+t^2z^2\varepsilon_\sigma^2)}{(1+tz)\varepsilon_\sigma^3}{}_2F_1
  \left(
    1,1+tz;2+tz;\frac{1}{1+\varepsilon_\sigma}
  \right)
  \Bigg],
\end{align}
where ${}_2F_1(a,b;c;z)$ is the hypergeometric function.  Since we do
not need higher order terms in $\varepsilon_\sigma$, we expand $I_\sigma$ as
\begin{align}
  I_\sigma(z) = &\,
  \int_0^1dt
  \Bigg\{
  -\frac{1}{\varepsilon_\sigma^2}z(z+1)-\frac{1}{2\varepsilon_\sigma}z(z+1)(4+z-2tz)
  \nonumber\\ &\,
  +\frac{\ln\varepsilon_\sigma}{6}z^2(1+z)(1+2z-6tz+6t^2z)
  \nonumber\\ &\,
  -z+z^2
  \left[
    -\frac{53}{36}+t+\frac{1}{6}(1-t)H(tz-z)+\frac{1}{6}tH(tz)
  \right]
  \nonumber\\ &\,
  +\frac{z^3}{12}[-7+14t+6(1-t)^2H(tz-z)+6t^2H(tz)]
  \nonumber\\ &\,
  +\frac{z^4}{9}[-1+3(1-t)t+3(1-t)^3H(tz-z)+3t^3H(tz)]
  +\mathcal O(\varepsilon_\sigma)
 \Bigg\},
\end{align}
where $H(z)$ is the harmonic number.

After the final integral, we obtain
\begin{align}
  I_\sigma(z) =
  -\frac{1}{\varepsilon_\sigma^2}z(z+1)
  -\frac{2}{\varepsilon_\sigma}z(z+1)
  +\frac{1}{6}z^2(z+1)^2\ln\varepsilon_\sigma
  +\mathcal S_\sigma(z)+\mathcal{O}(\epsilon_\sigma),
\end{align}
where
\begin{align}
  \mathcal S_\sigma(z) = &\,
  \frac{1}{6}z(1+z)(1+2z)
  \left[
    \ln\Gamma(1+z)-\ln\Gamma(1-z)
  \right]
  \nonumber \\ &\,
  -
  \left(
    z+z^2+\frac{1}{6}
  \right)
  \left[
    \psi_{\Gamma}^{(-2)}(1+z)+\psi_{\Gamma}^{(-2)}(1-z)
  \right]
  \nonumber \\ &\,
  +(1+2z)
  \left[
    \psi_{\Gamma}^{(-3)}(1+z)-\psi_{\Gamma}^{(-3)}(1-z)
  \right]
  \nonumber \\ &\,
  -2
  \left[
    \psi_{\Gamma}^{(-4)}(1+z)+\psi_{\Gamma}^{(-4)}(1-z)
  \right]
  \nonumber \\ &\,
  +\frac{1}{6}\gamma_Ez^2(z+1)^2-z-\frac{35}{36}z^2-\frac{z^4}{18}
  \nonumber \\ &\,
  +\frac{1}{2}\ln 2\pi+2\ln A_G+\frac{\zeta(3)}{2\pi^2}.
  \label{fnS_B}
\end{align}

We can repeat a similar calculation for
\begin{align}
  I_\psi(z) = \sum_{J=0}^\infty\frac{(2J+1)(2J+2)}{(1+\varepsilon_\psi)^{2J}}\ln
  \left[
    \frac{[\Gamma(2J+2)]^2}{\Gamma(2J+2+z)\Gamma(2J+2-z)}
  \right]^2,
\end{align}
which can be used for the calculation of the fermionic contribution to
the prefactor.  Here, $-2<\Re(z)<2$.  The result is
\begin{align}
  I_\psi(z) = 
  -\frac{2}{\varepsilon_\psi^2}z^2-\frac{5}{\varepsilon_\psi}z^2
  +\frac{1}{3}z^2(z^2-2)\ln\varepsilon_\psi+\mathcal S_\psi(z)+\mathcal{O}(\epsilon_\psi),
\end{align}
where
\begin{align}
  \mathcal S_\psi(z) = &\,
  \frac{2}{3}z(z^2-1)
  \left[
    \ln\Gamma(1+z)-\ln\Gamma(1-z)
  \right]
  \nonumber \\ &\,
  -2
  \left(
    z^2-\frac{1}{3}
  \right)\left[\psi_{\Gamma}^{(-2)}(1+z)+\psi_{\Gamma}^{(-2)}(1-z)
  \right]
  \nonumber \\ &\,
  +4z[\psi_{\Gamma}^{(-3)}(1+z)-\psi_{\Gamma}^{(-3)}(1-z)]
  \nonumber \\ &\,
  -4[\psi_{\Gamma}^{(-4)}(1+z)+\psi_{\Gamma}^{(-4)}(1-z)]
  \nonumber \\ &\,
  +\frac{1}{3}z^2(z^2-2)\gamma_E-\frac{z^2}{9}(z^2+31)
  \nonumber \\ &\,
  +4\ln A_G+\frac{\zeta(3)}{\pi^2}.
  \label{fnS_F}
\end{align}

Next, we consider
\begin{align}
 I_h = \sum_{J=1}^{\infty}\frac{(2J+1)^2}{(1+\varepsilon_h)^{2J}}
 \ln\frac{2J(2J-1)}{(2J+3)(2J+2)},
\end{align}
which is for the calculation of the Higgs contribution.  Using a
similar technique as in the previous cases, we obtain
\begin{align}
  I_h = -\sum_{J=1}^{\infty}\frac{(2J+1)^2}{(1+\varepsilon_h)^{2J}}\int_0^3dx
  \left[
    \frac{1}{x+2J}+\frac{1}{x+2J-1}
  \right].
\end{align}
Then, performing the sum first, $I_h$ is obtained as
\begin{align}
  I_h = -\frac{6}{\varepsilon_h^2}-\frac{12}{\varepsilon_h}+6\ln\varepsilon_h
  +\frac{3}{2}+6\gamma_E+12\ln A_G+9\ln2+5\ln3+\mathcal O(\varepsilon_h).
\end{align}

Finally, for the gauge and NG contributions, let us consider
\begin{align}
  I_A = 
  \sum_{J=1/2}^{\infty}\frac{(2J+1)^2}{(1+\varepsilon_A)^{2J}}\ln\frac{2J}{2J+2}.
\end{align}
Similary to $I_h$, it is expressed as
\begin{align}
  I_A = -\sum_{J=1/2}^{\infty}
  \frac{(2J+1)^2}{(1+\varepsilon_A)^{2J}}\int_0^2dx\frac{1}{x+2J},
\end{align}
which results in
\begin{align}
  I_A = 
  -\frac{2}{\varepsilon_A^2}-\frac{4}{\varepsilon_A}+\frac{2}{3}\ln\varepsilon_A
  -\frac{3}{2}+\frac{2}{3}\gamma_E+\ln2+4\ln A_G+\mathcal{O}(\epsilon_A).
\end{align}

\section{Renormalization with $\msbar$-scheme}
\label{apx_divpart}
\setcounter{equation}{0}

In this Appendix, we relate the regularization based on the angular
momentum expansion, which we call ``angular momentum regularization,''
and the dimensional regularization.  In particular, we derive the
relation between $\eps{X}$, which shows up in the angular momentum
regularization, and $\epsD$, which is for dimensional regularization.

\subsection{Scalar field}

Let us start with the scalar contribution.  For this purpose, let us
consider $\left[\ln\mathcal A^{(\sigma)}\right]_{{\rm div}}$, defined
in \eqref{eq_deltaMExpansion}.

First, we calculate $\left[\ln\mathcal A^{(\sigma)}\right]_{{\rm
    div}}$ with angular momentum regularization with using
$\eps{\sigma}$ as a regularization parameter.  The expansion of
eq.~\eqref{eq_expansionEachJ} is exactly the same as that with respect
to $\kappa$. Thus, we get
\begin{align}
  \left[
    \ln\frac{\det[-\Delta_J+\kappa\bar\phi^2]}{\det[-\Delta_J]}
  \right]_{\mathcal O(\kappa^2)}
  =\frac{2\kappa}{|\lambda|}\frac{1}{2J+1}+
  \left(
    \frac{2\kappa}{|\lambda|}
  \right)^2
  \left[
    \frac{1}{2(2J+1)^2}+\frac{1}{2J+1}-\psi^{(1)}(2J+1)
  \right],
\end{align}
where $\psi^{(n)}(z)$ is the polygamma function.  Summing over $J$, we get
\begin{align}
 \left[
  \ln\mathcal A^{(\sigma)}
 \right]_{{\rm div},\eps{\sigma}}
  & = 
  -\frac{\kappa}{|\lambda|}
  \left(
   \frac{1}{\eps{\sigma}^2}+\frac{2}{\eps{\sigma}}+\frac{\kappa}{3|\lambda|}\ln\eps{\sigma}
  \right)-\frac{\kappa}{|\lambda|}
  \left(
   1+\frac{\kappa}{18|\lambda|}
  \right)
 +\mathcal O(\eps{\sigma}).
   \label{Asigma(AM)}
\end{align}
As is expected, it has the same divergence as $\left[\ln\mathcal
  A^{(\sigma)}\right]_{\eps{\sigma}}$ does.

Next, we directly calculate $\left[\ln\mathcal
  A^{(\sigma)}\right]_{\rm div}$ by using the dimensional
regularization. Using the ordinary Feynman rules, we obtain
\begin{align}
  \left[
    \ln\mathcal A^{(\sigma)}
  \right]_{{\rm div},\epsD}
  = &\,
  -\frac{1}{2}
  \Bigg[
  \kappa\int\frac{d^D k}{(2\pi)^D}\frac{1}{k^2}F[\bar\phi^2](0)
  \nonumber \\ &\,
  -\frac{\kappa^2}{2}
  \int\frac{d^D k}{(2\pi)^D}\frac{d^D k'}{(2\pi)^D}
  \frac{1}{k^2}F[\bar\phi^2](k-k')\frac{1}{k'^2}F[\bar\phi^2](k'-k)
  \Bigg],
\end{align}
where $F[\cdots]$ is the Fourier transform of the argument.  With
performing the integration, we obtain
\begin{align}
  \left[
    \ln\mathcal A^{(\sigma)}
  \right]_{{\rm div},\epsD}
  = &\,
 \frac{\kappa^2}{3|\lambda|^2}
 \left(
 \frac{1}{2\epsD}+\frac{5}{6}+\gamma_E+\ln\frac{\mu R}{2}
 \right).
  \label{Asigma(DR)}
\end{align}

Comparing eqs.\ \eqref{Asigma(AM)} and \eqref{Asigma(DR)}, we obtain
the relation between the two regularizations as
\begin{align}
  \left(
    \frac{1}{\eps{\sigma}^2}+\frac{2}{\eps{\sigma}}
    +\frac{\kappa}{3|\lambda|}\ln\eps{\sigma}\right)
  \to -1-\frac{\kappa}{3|\lambda|}
  \left[
    \frac{1}{2\epsD}+1+\gamma_E+\ln\frac{\mu R}{2}
  \right].
  \label{eq_ScalarRegulatorRel}
\end{align}

\subsection{Higgs}

For the Higgs case, the relation between $\eps{h}$ and $\epsD$ can be
obtained from eq.~\eqref{eq_ScalarRegulatorRel} with the replacement
of $\kappa\to-3|\lambda|$:
\begin{align}
 \frac{1}{\eps{h}^2}+\frac{2}{\eps{h}}
 -\ln\eps{h}\to\frac{1}{2\epsD}+\gamma_E+\ln\frac{\mu R}{2}.
\end{align}
Notice that the zero modes have nothing to do with the UV divergence.

\subsection{Fermion}

Next, we derive the relation for the fermionic contribution. As
discussed in \cite{Andreassen:2017rzq}, it is convenient to expand
with respect to $y$.  The expansion in eq.~\eqref{eq_deltaMExpansion}
is equivalent to
\begin{align}
  \left[\ln\mathcal A^{(\psi)}\right]_{{\rm div},\eps{\psi}}
  & = \sum_{J=0}^{\infty} \frac{(2J+1)(2J+2)}{(1+\eps{\psi})^{2J}}
  \nonumber \\
  & \hspace{3ex}
  \left\{
    \left[
      \ln\frac{\det\mathcal M^{(\psi)}_J}{\det\widehat{\mathcal M}^{(\psi)}_J}
    \right]_{\mathcal O(y^2)}+
    \left[
      \ln\frac{\det\mathcal M^{(\psi)}_{J,{\rm diag}}}{\det\widehat{\mathcal M}^{(\psi)}_{J,{\rm diag}}}
    \right]_{\mathcal O(y^4)}-
    \left[
      \ln\frac{\det\mathcal M^{(\psi)}_{J,{\rm diag}}}{\det\widehat{\mathcal M}^{(\psi)}_{J,{\rm diag}}}
    \right]_{\mathcal O(y^2)}
  \right\},
\end{align}
where
\begin{align}
 \mathcal M^{(\psi)}_{J,{\rm diag}} =
 \begin{pmatrix}
  -\Delta_J+\frac{y^2}{2}\bar\phi^2 & 0                                       \\
  0                                 & -\Delta_{J+1/2}+\frac{y^2}{2}\bar\phi^2
 \end{pmatrix},
\end{align}
and $\widehat{\mathcal M}^{(\psi)}_J$ is obtained from ${\mathcal
  M}^{(\psi)}_J$ by taking $\bar\phi\to0$.  From
eq.~\eqref{eq_scalarDeterminant}, we have
\begin{align}
 \frac{\det\mathcal M^{(\psi)}_{J,{\rm diag}}}{\det\widehat{\mathcal M}^{(\psi)}_{J,{\rm diag}}}
 = \frac{\Gamma(2J+1)\Gamma(2J+2)\Gamma(2J+2)\Gamma(2J+3)}
 {\Gamma(2J+1-\tilde z_y)\Gamma(2J+2+\tilde z_y)\Gamma(2J+2-\tilde z_y)\Gamma(2J+3+\tilde z_y)},
\end{align}
where
\begin{align}
 \tilde{z}_y = -\frac{1}{2}\left(1-\sqrt{1-4\frac{y^2}{|\lambda|}}\right).
\end{align}
Then, together with eq.~\eqref{eq_FermionDeterminant}, we have
\begin{align}
 \left[\ln\mathcal A^{(\psi)}\right]_{{\rm div},\eps{\psi}}
  & = \frac{y^2}{|\lambda|}
 \left(
  \frac{2}{\eps{\psi}^2}+\frac{5}{\eps{\psi}}+\frac{1}{3}\frac{y^2+2|\lambda|}{|\lambda|}\ln\eps{\psi}
 \right)+\frac{31}{9}\frac{y^2}{|\lambda|}+\frac{1}{18}\frac{y^4}{|\lambda|^2}+\mathcal O(\eps{\psi}).
\end{align}
We can also calculate $\left[\ln\mathcal A^{(\psi)}\right]_{{\rm
    div}}$ with dimensional regularization:
\begin{align}
 \left[\ln\mathcal A^{(\psi)}\right]_{{\rm div},\epsD}
  & = -\frac{y^4}{3|\lambda|^2}
 \left(
  \frac{1}{2\epsD}+\frac{5}{6}+\gamma_E+\ln\frac{\mu R}{2}
 \right)-\frac{2y^2}{3|\lambda|}
 \left(
  \frac{1}{2\epsD}+\frac{13}{12}+\gamma_E+\ln\frac{\mu R}{2}
 \right).
\end{align}
Thus, we obtain
\begin{align}
  &
 \left(
  \frac{2}{\eps{\psi}^2}+\frac{5}{\eps{\psi}}+\frac{1}{3}\frac{y^2+2|\lambda|}{|\lambda|}\ln\eps{\psi}
 \right)\nonumber                            \\
  & \hspace{3ex} \to -\frac{y^2}{3|\lambda|}
 \left(
  \frac{1}{2\epsD}+1+\gamma_E+\ln\frac{\mu R}{2}
 \right)-\frac{2}{3}
 \left(
  \frac{1}{2\epsD}+\frac{25}{4}+\gamma_E+\ln\frac{\mu R}{2}
 \right).
\end{align}

\subsection{Gauge and NG fields}

Finally, we consider the gauge and NG contributions.  Although we may
use eq.~\eqref{eq_deltaMExpansion} to subtract the divergence, it is
more convenient to use the expression of the prefactor with a
different choice of the gauge fixing function from that in eq.\
\eqref{gaugefixingfn}.

Here, we use the result with the following choice of the gauge fixing
function:
\begin{align}
  \mathcal F_{\rm BG} = \partial_\mu A_\mu-\xi g\bar\phi\varphi,
\end{align}
which we call the background gauge. We may perform a calculation of
the prefactor $\mathcal A$ with this choice of the gauge fixing
function.  We have checked that, irrespective of the choice of $\xi$,
the $J>0$ contribution from the gauge and NG sectors agrees with the
one that we have obtained using the gauge fixing function in
\eqref{gaugefixingfn}.  We also comment here that, in the background
gauge, the treatment of the gauge zero mode is highly non-trivial.
However, such an issue is unimportant for the following discussion
because the zero mode does not affect the behavior of the divergence,
which we will discuss below.

For simplicity, we take $\xi = 1$ in the following.  Then, using the
same basis as eq.~\eqref{eq_gaugeBasis}, we define
\begin{align}
 \ln\mathcal A_{\rm BG}^{(A_\mu,\varphi)}
  & = -\frac{1}{2}
 \ln\frac{\det\mathcal M^{(A_\mu,\varphi)}_{\rm BG}}
 {\det\widehat{\mathcal M}^{(A_\mu,\varphi)}_{\rm BG}},
\end{align}
where the fluctuation operator in the background gauge is given by
\begin{align}
  \mathcal M^{(A_\mu,\varphi)}_{\rm BG} =
  \begin{pmatrix}
    -\partial^2\delta_{\mu\nu}+g^2\bar\phi^2 
    & 2g(\partial_\nu\bar\phi)
    \\
    2g(\partial_\mu\bar\phi)
    & -\partial^2 + V_{\varphi\varphi} + g^2\bar\phi^2
  \end{pmatrix}.
\end{align}
With the angular momentum decomposition, we have
\begin{align}
  \ln\mathcal A_{\rm BG}^{(A_\mu,\varphi)}
  & = -\frac{1}{2}
 \ln\frac{\det\mathcal M^{(S,\varphi)}_{0,\rm BG}}
 {\det\widehat{\mathcal M}^{(S,\varphi)}_{0,\rm BG}}
 -\frac{1}{2}\sum_{J=1/2}^{\infty}(2J+1)^2
 \left[
  \ln\frac{\det\mathcal M^{(S,L,\varphi)}_{J,\rm BG}}
  {\det\widehat{\mathcal M}^{(S,L,\varphi)}_{J,\rm BG}}
  +2\ln\frac{\det\mathcal M^{(T)}_{J,\rm BG}}
  {\det\widehat{\mathcal M}^{(T)}_{J,\rm BG}}
 \right],
\end{align}
where 
\begin{align}
  \mathcal M^{(S,L,\varphi)}_{J,\rm BG} & =
  \begin{pmatrix}
    -\Delta_J+\frac{3}{r^2}+g^2\bar\phi^2 & -\frac{2L}{r^2}                         & 2g\bar\phi'
    \\
    -\frac{2L}{r^2}
    & -\Delta_J-\frac{1}{r^2}+g^2\bar\phi^2 & 0
    \\
    2g\bar\phi'                           & 0                                       & -\Delta_J+V_{\varphi\varphi}+g^2\bar\phi^2
  \end{pmatrix},
\end{align}
for the partial waves with $J>0$,
and
\begin{align}
  \mathcal M^{(S,\varphi)}_{0,\rm BG} =
  \begin{pmatrix}
    -\Delta_{1/2}+g^2\bar\phi^2 
    & 
    2g\bar\phi'
    \\
    2g\bar\phi'
    &
    -\Delta_0+V_{\varphi\varphi}+g^2\bar\phi^2
  \end{pmatrix},
\end{align}
for $J=0$.
In addition,
\begin{align}
  \mathcal M^{(T)}_{J,\rm BG} =  -\Delta_J + g^2\bar\phi^2,
\end{align}
for the $T$ mode.
Remenber that the $T$ modes exist only for $J>0$.

In the background gauge, Faddeev-Popov ghosts also couple to the
bounce; the fluctuation operator of the ghosts is given by
\begin{align}
  \mathcal M^{(c,\bar{c})}_{\rm BG} & = -\partial^2+g^2\bar\phi^2,
\end{align}
and, after the angular momentum decomposition, we have
\begin{align}
  \mathcal M^{(c,\bar{c})}_{J,{\rm BG}} = -\Delta_J + g^2\bar\phi^2.
\end{align}
Then, we define
\begin{align}
  \ln\mathcal A_{\rm BG}^{(c,\bar{c})}
  = \sum_{J=0}^{\infty}(2J+1)^2
  \ln\frac{\det\mathcal M^{(c,\bar{c})}_{J,\rm BG}}
  {\det\widehat{\mathcal M}^{(c,\bar{c})}_{J,\rm BG}},
\end{align}
where the hatted fluctuation operators are defined through the
replacements of $\bar\phi\to0$ and $\bar\phi'\to0$.

One feature of the above gauge fixing is that, by a rotation using an
orthogonal transformation, ${\mathcal M}^{(S,L,\varphi)}_{J,\rm BG}$
becomes
\begin{align}
  B^t\mathcal M^{(S,L,\varphi)}_{J,\rm BG} B =
  \begin{pmatrix}
    -\Delta_{J-1/2}+g^2\bar\phi^2 & 0 & 2\sqrt{\frac{J}{2J+1}}g\bar\phi'
    \\
    0 & -\Delta_{J+1/2}+g^2\bar\phi^2  & -2\sqrt{\frac{J+1}{2J+1}}g\bar\phi'
    \\
    2\sqrt{\frac{J}{2J+1}}g\bar\phi' & -2\sqrt{\frac{J+1}{2J+1}}g\bar\phi'
    & -\Delta_J+m_a^2+g^2\bar\phi^2
  \end{pmatrix},
\end{align}
with 
\begin{align}
 B =
 \begin{pmatrix}
  \sqrt{\frac{J}{2J+1}}   & -\sqrt{\frac{J+1}{2J+1}} & 0 \\
  \sqrt{\frac{J+1}{2J+1}} & \sqrt{\frac{J}{2J+1}}    & 0 \\
  0                       & 0                        & 1
 \end{pmatrix}.
\end{align}
Notice that, in the new basis, the fluctuation operator around the
false vacuum, $\widehat{\mathcal M}^{(S,L,\varphi)}_{J,\rm BG}$,
becomes diagonal.  This makes the calculation of the functional
determinant easier.

Following \cite{Endo:2017gal}, we can show that
\begin{align}
  \ln\frac{\det\mathcal M^{(S,L,\varphi)}_{J}}
  {\det\widehat{\mathcal M}^{(S,L,\varphi)}_{J}}
  +2\ln\frac{\det\mathcal M^{(T)}_{J}}
  {\det\widehat{\mathcal M}^{(T)}_{J}}
  = 
    \ln\frac{\det\mathcal M^{(S,L,\varphi)}_{J,\rm BG}}
  {\det\widehat{\mathcal M}^{(S,L,\varphi)}_{J,\rm BG}}
  +2\ln\frac{\det\mathcal M^{(T)}_{J,\rm BG}}
  {\det\widehat{\mathcal M}^{(T)}_{J,\rm BG}}
  -2\ln\frac{\det\mathcal M^{(c,\bar{c})}_{J,\rm BG}}
  {\det\widehat{\mathcal M}^{(c,\bar{c})}_{J,\rm BG}}.
\end{align}
Thus, a divergent part can be subtracted from $\ln\mathcal
A'^{(A_\mu,\varphi)}$ as
\begin{align}
  \left[
    \ln\mathcal A'^{(A_\mu,\varphi)}
  \right]
  -
  \left[
    \ln\mathcal A^{(A_\mu,\varphi)}_{\rm BG}
  \right]_{{\rm div}}
  -
  \left[
    \ln\mathcal A_{\rm BG}^{(c,\bar{c})}
  \right]_{{\rm div}}
  = (\rm finite),
\end{align}
where $[\cdots]_{{\rm div}}$ is defined accordingly to
eq.~\eqref{eq_deltaMExpansion}.  The relation between the angular
momentum regularization and dimensional regularization for the gauge
and NG contributions can be understood by evaluating $[\ln\mathcal
A^{(A_\mu,\varphi)}_{\rm BG}]_{{\rm div}}+ [\ln\mathcal A_{\rm
  BG}^{(c,\bar{c})}]_{{\rm div}}$.

Let us evaluate the divergent part with $\eps{A}$ regularization.
Before proceeding, we define the following quantities:
\begin{align}
  K_J^{\rm (1)}(\cc)
  = &\,
  \left[
    \ln\frac{\det[-\Delta_J+\cc\bar\phi^2]}{\det[-\Delta_J]}
  \right]_{\mathcal O(\cc^2)},
  \\
  K_J^{\rm (2)}(\cc)
  = &\,
  \left[
  \ln\frac{\det
   \begin{pmatrix}
    -\Delta_J
     & \cc\bar\phi' \\
    \cc\bar\phi'
     & -\Delta_{J+1/2}
   \end{pmatrix}}{\det
   \begin{pmatrix}
    -\Delta_J
     & 0               \\
    0
     & -\Delta_{J+1/2}
   \end{pmatrix}}
 \right]_{\mathcal O(\cc^2)}.
\end{align}
Notice that $K_J^{\rm (2)}(\cc)$ can be also expressed as
\begin{align}
  K_J^{\rm (2)}(\cc)
  = &\,
  \left[
    \ln\frac{\det
      \begin{pmatrix}
        -\Delta_J+\cc^2\bar\phi^2
        & \cc\bar\phi'
        \\
        \cc\bar\phi'
        & -\Delta_{J+1/2}+\cc^2\bar\phi^2
      \end{pmatrix}}{\det
      \begin{pmatrix}
        -\Delta_J
        & 0               \\
        0
        & -\Delta_{J+1/2}
      \end{pmatrix}}
  \right]_{\mathcal O(\cc^2)}
  \nonumber \\ &\,
  - 
  \left[
    \ln\frac{\det[-\Delta_J+\cc^2\bar\phi^2]}{\det[-\Delta_J]}
  \right]_{\mathcal O(\cc^2)}-
  \left[
    \ln\frac{\det[-\Delta_{J+1/2}+\cc^2\bar\phi^2]}{\det[-\Delta_{J+1/2}]}
  \right]_{\mathcal O(\cc^2)}.
\end{align}
In addition, $K_J^{\rm (1)}(\cc)$ and $K_J^{\rm (2)}(\cc)$ can be
analytically calculated as
\begin{align}
  K_J^{\rm (1)}(\cc)
  = \frac{2\cc}{|\lambda|}\frac{1}{2J+1}+
  \left(
    \frac{2\cc}{|\lambda|}
  \right)^2
  \left[
    \frac{1}{2(2J+1)^2}+\frac{1}{2J+1}-\psi^{(1)}(2J+1)
  \right],
\end{align}
and
\begin{align}
 K_J^{\rm (2)}(\cc)
  & = \frac{\cc^2}{|\lambda|}
 \left[
  4\psi_\Gamma^{(1)}(2J+2)-\frac{2}{2J+1}-\frac{1}{J+1}
 \right].
\end{align}
Then,
\begin{align}
  \left[
    \ln\mathcal A^{(A_\mu,\varphi)}_{\rm BG}
  \right]_{{\rm div},\eps{A}}+
  \left[
    \ln\mathcal A_{\rm BG}^{(c,\bar{c})}
  \right]_{{\rm div},\eps{A}}
  = &\,
  -\frac{1}{2}
  \left[
    \ln\frac{\det \mathcal M^{(S,\varphi)}_{0,\rm BG}}{\det \widehat{\mathcal M}^{(S,\varphi)}_{0,\rm BG}}
  \right]_{\mathcal O(\delta \mathcal M^2)}
  +
  \left[
    \ln\frac{\det \mathcal M^{(c,\bar{c})}_{0,\rm BG}}{\det \widehat{\mathcal M}^{(c,\bar{c})}_{0,\rm BG}}
  \right]_{\mathcal O(\delta \mathcal M^2)}
  \nonumber \\ &\,
  -\frac{1}{2}
  \sum_{J=1/2}^{\infty}\frac{(2J+1)^2}{(1+\eps{A})^{2J}}
  \left[
    \ln\frac{\det \mathcal M^{(S,L,\varphi)}_{J,\rm BG}}{\det \widehat{\mathcal M}^{(S,L,\varphi)}_{J,\rm BG}}
  \right]_{\mathcal O(\delta \mathcal M^2)},
\end{align}
where
\begin{align}
  \left[
    \ln\frac{\det \mathcal M^{(S,\varphi)}_{0,\rm BG}}{\det \widehat{\mathcal M}^{(S,\varphi)}_{0,\rm BG}}
  \right]_{\mathcal O(\delta \mathcal M^2)}
  & = K_{1/2}^{\rm (1)}(g^2)+K_{0}^{\rm (1)}(g^2-|\lambda|)+K_0^{\rm (2)}(2g), \\
  \left[
    \ln\frac{\det \mathcal M^{(S,L,\varphi)}_{J,\rm BG}}{\det \widehat{\mathcal M}^{(S,L,\varphi)}_{J,\rm BG}}
  \right]_{\mathcal O(\delta \mathcal M^2)}
  & = K_{J+1/2}^{\rm (1)}(g^2)+K_{J-1/2}^{\rm (1)}(g^2)
  +K_{J}^{\rm (1)}(g^2-|\lambda|)\nonumber                                               \\
  & \hspace{3ex}+K_J^{\rm (2)}\left(-2\sqrt{\frac{J+1}{2J+1}}g\right)
  +K_{J-1/2}^{\rm (2)}\left(2\sqrt{\frac{J}{2J+1}}g\right),                               \\
  \left[
    \ln\frac{\det \mathcal M^{(c,\bar{c})}_{0,\rm BG}}{\det \widehat{\mathcal M}^{(c,\bar{c})}_{0,\rm BG}}
  \right]_{\mathcal O(\delta \mathcal M^2)}
  & = K_{0}^{\rm (1)}(g^2).
\end{align}
Notice that $V_{\varphi\varphi}=-|\lambda|\bar\phi^2$, and that the
Faddeev-Popov and $T$-mode contributions cancel out for $J>0$.
Summing over $J$, we obtain
\begin{align}
  \left[
    \ln\mathcal A^{(A_\mu,\varphi)}_{\rm BG}
  \right]_{{\rm div},\eps{A}}+
  \left[
    \ln\mathcal A_{\rm BG}^{(c,\bar{c})}
  \right]_{{\rm div},\eps{A}}
  = &\,
  -\left(
    \frac{3g^2}{|\lambda|}-1
  \right)
  \left(
    \frac{1}{\eps{A}^2}+\frac{2}{\eps{A}}
  \right)-
  \left(
    \frac{1}{3}+\frac{g^4}{|\lambda|^2}
  \right)\ln\eps{A}
  \nonumber \\ &\,
  +\frac{17}{18}+\frac{g^2}{3|\lambda|}+\frac{g^4}{6|\lambda|^2}(59-6\pi^2)
  +\mathcal O(\eps{A}).
\end{align}

With the dimensional regularization, it becomes
\begin{align}
   \left[
  \ln\mathcal A^{(A_\mu,\varphi)}_{\rm BG}
 \right]_{{\rm div},\epsD}+
 \left[
  \ln\mathcal A_{\rm BG}^{(c,\bar{c})}
 \right]_{{\rm div},\epsD}
  =&\,
  \left(
   \frac{1}{3}+\frac{2g^2}{|\lambda|}+\frac{g^4}{|\lambda|^2}
  \right)
  \left(
   \frac{1}{2\epsD}+\gamma_E+\ln\frac{\mu R}{2}
  \right)\nonumber\\&\,
  +\frac{5}{18}+\frac{7g^2}{3|\lambda|}+\frac{g^4}{2|\lambda|^2}.
  \label{A_BG(A,NG)}
\end{align}

Thus, we obtain
\begin{align}
    & \left(
      \frac{3g^2}{|\lambda|}-1
    \right)
    \left(
      \frac{1}{\eps{A}^2}+\frac{2}{\eps{A}}
    \right)+
    \left(
      \frac{1}{3}+\frac{g^4}{|\lambda|^2}
    \right)\ln\eps{A}
    \nonumber \\
    & \hspace{3ex}\to
    -
    \left(
      \frac{1}{3}+\frac{2g^2}{|\lambda|}+\frac{g^4}{|\lambda|^2}
    \right)
    \left(
      \frac{1}{2\epsD}+1+\gamma_E+\ln\frac{\mu R}{2}
    \right)
    +1+\frac{g^4}{|\lambda|^2}
    \left(
      \frac{31}{3}-\pi^2
    \right).
\end{align}

\section{Vacuum decay with global symmetry}
\setcounter{equation}{0}
\label{apx_global}

For completeness, we discuss the case where the field that is
responsible for the decay transforms under a global symmetry, although
it is not the case of the SM Higgs field. In such a case, we need to
take into account quantum corrections from the associated NG bosons. Similarly to
the gauge contributions, the NG fluctuation operator has zero modes in
association with the breaking of the global symmetry.

Let us consider a $U(1)$ symmetry, for simplicity. The contribution from
the NG boson, $\varphi$, is given by
\begin{align}
 \ln \mathcal A^{(\varphi)}&=
 -\frac{1}{2}\frac{\det
 \left[
  -\partial^2-|\lambda|\bar\phi^2
 \right]}{\det
 \left[
  -\partial^2
 \right]}\nonumber\\
 &=-\frac{1}{2}\sum_{J=0}^{\infty}(2J+1)^2\ln\frac{J}{J+1}.
\end{align}
As we can see, there is a zero mode for $J=0$.  Since it is obtained in
the limit of $g\to0$ in eq.~\eqref{detM^gauge}, $\left[\ln \mathcal
A^{(\varphi)}\right]_{\rm \overline{MS}}$ is given by
eqs.~\eqref{A(A,NG)} and \eqref{A'(A,NG)} with taking $g=z_g=0$.

\section{Numerical Recipe}
\label{apx_recipe}
\setcounter{equation}{0}

In this Appendix, we give fitting formulae of the prefactors at the
one-loop level.  Contrary to the analytic formulae including various
special functions with complex arguments, which may be inconvenient
for numerical calculations, the fitting formulae give a simple
procedure to perform a numerical calculation of the decay rate with
saving computational time.  Compared to the analytic expressions, the
errors of the fitting formulae are $0.05\%$ or smaller.  A C++ package
based on our fitting formulae for the study of the \verb|EL|ectroweak
\verb|VA|cuum \verb|S|tability, \verb|ELVAS|, can be found at
\cite{ELVAS}.

\begin{itemize}
 \item Higgs

       \begin{align}
        -\left[\ln\mathcal A'^{(h)}\right]_{\msbar} = -0.99192944327027 + 2.5\ln|\lambda| -
        3\ln\mu R.
       \end{align}
 \item Scalar

       Let $x = \kappa/|\lambda|$.
       For $x<0.7$,
       \begin{align}
        -\left[
         \ln\mathcal A^{(\sigma)}
        \right]_{\msbar}
         & = -0.239133939224974 x^2 + 0.222222222222222 x^3\nonumber                 \\
         & \hspace{3ex} - 0.134704602106396 x^4 + 0.102278606592866 x^5\nonumber     \\
         & \hspace{3ex} - 0.0839329261179402 x^6 + 0.0715956882048009 x^7\nonumber   \\
         & \hspace{3ex} - 0.0625481711576628 x^8 + 0.0555697470602515 x^9\nonumber   \\
         & \hspace{3ex} - 0.0500042455037409 x^{10} - 0.333333333333333 x^2\ln\mu R.
       \end{align}
       For $x>0.7$,
       \begin{align}
        -\left[\ln\mathcal A^{(\sigma)}\right]_{\msbar}
         & = -0.0261559272783723 + 0.0000886704923163256/x^4\nonumber                   \\
         & \hspace{3ex} + 0.0000962000962000962/x^3 + 0.000198412698412698/x^2\nonumber \\
         & \hspace{3ex} + 0.00105820105820106/x + 0.111111111111111 x\nonumber          \\
         & \hspace{3ex} - 0.181204187497805 x^2 + (-0.0055555555555556\nonumber         \\
         & \hspace{3ex} + 0.166666666666667 x^2)\ln x - 0.333333333333333 x^2 \ln\mu R.
       \end{align}
 \item Fermion

       Let $x=y^2/|\lambda|$.
       For $x<1.3$,
       \begin{align}
        -\left[\ln\mathcal A^{(\psi)}\right]_{\msbar}
         & = 0.64493454511661 x + 0.005114971505109 x^2\nonumber                                  \\
         & \hspace{3ex} - 0.0366953662258276 x^3 +  0.00476307962690785 x^4\nonumber              \\
         & \hspace{3ex} - 0.000845451274112082 x^5 + 0.000168244913551417 x^6\nonumber            \\
         & \hspace{3ex} - 0.0000353785958610453 x^7 +  7.67709260595572\times10^{-6} x^8\nonumber \\
         & \hspace{3ex} + (0.66666666666667 x +  0.333333333333333 x^2) \ln\mu R.
       \end{align}
       For $x>1.3$,
       \begin{align}
        -\left[\ln\mathcal A^{(\psi)}\right]_{\msbar}
         & = -0.227732960077634 + 0.00260942760942761/x^3\nonumber                  \\
         & \hspace{3ex} +  0.00271164021164021/x^2 + 0.00820105820105820/x\nonumber \\
         & \hspace{3ex} + 0.53790187962670 x +  0.296728717591129 x^2\nonumber      \\
         & \hspace{3ex} + (-0.06111111111111111 - 0.3333333333333333 x\nonumber     \\
         & \hspace{3ex} - 0.1666666666666666 x^2) \ln x\nonumber                    \\
         & \hspace{3ex} + (0.66666666666667 x + 0.333333333333333 x^2) \ln\mu R.
       \end{align}
 \item Gauge

       Let $x=g^2/|\lambda|$.
       For $x<1.4$,
       \begin{align}
        -\left[\ln\mathcal A'^{(A_\mu,\varphi)}\right]_{\msbar} = & 
         -0.96686103284373 - 1.76813696868318 x\nonumber                          \\
         & + 0.61593151565841 x^2 + 0.145084271024101 x^3\nonumber       \\
         & - 0.0241469799983579 x^4 + 0.00555917805602827 x^5\nonumber   \\
         & - 0.00145020891759152 x^6 + 0.000402580447036276 x^7\nonumber \\
         & - 0.000115821925959136 x^8 \nonumber         \\
         & + 0.5\ln|\lambda| + (-0.333333333333333 -  2 x -  x^2) \ln\mu R.
       \end{align}
       For $x>1.4$,
       \begin{align}
        -\left[\ln\mathcal A'^{(A_\mu,\varphi)}\right]_{\msbar} = &
         -27.0091748854198 + 0.000266011476948977/x^4\nonumber                               \\
         & + 0.000288600288600289/x^3 + 0.000595238095238095/x^2\nonumber           \\
         & + 0.00317460317460317/x + 1.56519636465016 x\nonumber                    \\
         & - 0.07988363024944 x^2 + (-3.54033527491510\times 10^{-6}/x^{5}\nonumber \\
         & - 0.0000404609745704583/x^{4} - 0.00051790047450187/x^{3}\nonumber       \\
         & - 0.0082864075920299/x^{2} - 0.265165042944955/x\nonumber                \\
         & + 4.24264068711929) \sqrt{x}\arcsin
        \left[
         \frac{s}{\sqrt{s^2+79164837199872 x^9}}
        \right]\nonumber                                                                         \\
         & + (-6.01666666666667 +  0.5 x^2)\ln x\nonumber                           \\
         & + 1.5\ln[3.14159265358979 (-98796.7402597403\nonumber                    \\
         & + 136316.571428571 x -136594.285714286 x^2\nonumber                      \\
         & + 92160 x^3 +7372800 x^4 + 6553600 x^5)]\nonumber                        \\
         & +  0.5 \ln|\lambda| + (-0.333333333333333 -  2 x - x^2) \ln\mu R,
       \end{align}
       where
       \begin{align}
        s = 7 + 80 x +1024 x^2 + 16384 x^3 +524288 x^4 -  8388608 x^5.
       \end{align}
\end{itemize}

\section{Threshold Corrections}
\label{apx_threshold}
\setcounter{equation}{0}

In this Appendix, we summarize the one-loop threshold corrections for
the coupling constants in the models with extra fermions discussed in
Section \ref{sec_extra}.  We parameterize the threshold corrections as
\begin{align}
  c^{\rm (below)} = c + \frac{1}{16\pi^2} \Delta_c,
\end{align}
where $c{\rm (below)}$ is a coupling constant below the matching scale,
while $c$ is that above the scale.  For the notational simplicity, we
only show the quantity $\Delta_c$ for each coupling constant.  Notice
that $\Delta_c$ depends on the extra fermion mass $M_X$ ($X=D,E,N$). In
our analysis, we take the matching scale to be equal to $M_X$.

\begin{itemize}
\item Model with vector-like quarks $Q$, $\overline{Q}$, $D$, and
  $\overline{D}$
\end{itemize}
\begin{align}
  \Delta_{g_1} &= - \frac{1}{5} g_1^2 \log \frac{\mu^2}{M_D^2},\\
  \Delta_{g_2} &= - g_2^2 \log \frac{\mu^2}{M_D^2}, \\
  \Delta_{g_3} &= - g_3^2 \log \frac{\mu^2}{M_D^2}, \\
  \Delta_{y_t} &= - 6 y_t y_D^2 \left[ \frac{1}{2} \log
    \frac{\mu^2}{M_D^2} - \frac{1}{3} \right],\\
  \Delta_{y_b} &= - 6 y_b y_D^2 \left[ \frac{1}{2} \log
    \frac{\mu^2}{M_D^2} - \frac{1}{3} \right],\\
  \Delta_{y_\tau} &= - 6 y_\tau y_D^2 \left[ \frac{1}{2} \log
    \frac{\mu^2}{M_D^2} - \frac{1}{3} \right],\\
  \Delta_{\lambda} &= - 24 \lambda y_D^2 \left[ \frac{1}{2} \log
    \frac{\mu^2}{M_D^2} - \frac{1}{3} \right] + 12 y_D^4 \left[
    \frac{1}{2} \log \frac{\mu^2}{M_D^2} - \frac{4}{3} \right].
\end{align}

\begin{itemize}
\item Model with vector-like leptons $L$, $\overline{L}$, $E$, and
  $\overline{E}$
\end{itemize}
\begin{align}
 \Delta_{g_1} &= - \frac{3}{5} g_1^2 \log \frac{\mu^2}{M_E^2},\\
 \Delta_{g_2} &= - \frac{1}{3} g_2^2 \log \frac{\mu^2}{M_E^2}, \\
 \Delta_{g_3} &= 0, \\
 \Delta_{y_t} &= - 2 y_t y_E^2 \left[ \frac{1}{2} \log
 \frac{\mu^2}{M_E^2} - \frac{1}{3} \right],\\
 \Delta_{y_b} &= - 2 y_b y_E^2 \left[ \frac{1}{2} \log
 \frac{\mu^2}{M_E^2} - \frac{1}{3} \right],\\
 \Delta_{y_\tau} &= - 2 y_\tau y_E^2 \left[ \frac{1}{2} \log
 \frac{\mu^2}{M_E^2} - \frac{1}{3} \right],\\
 \Delta_{\lambda} &= - 8 \lambda y_E^2 \left[ \frac{1}{2} \log
 \frac{\mu^2}{M_E^2} - \frac{1}{3} \right] + 4 y_E^4 \left[
 \frac{1}{2} \log \frac{\mu^2}{M_E^2} - \frac{4}{3} \right].
\end{align}

\begin{itemize}
\item Model with right-handed neutrino $N$\footnote
{For simplicity, we assume that $N$ couples only to third generation
  leptons.}
\end{itemize}
\begin{align}
 \Delta_{g_i} &= 0\ (i = 1,2,3),\\
 \Delta_{y_t} &= - y_t y_N^2 \left[ \frac{1}{2} \log
 \frac{\mu^2}{M_N^2} + \frac{1}{4} \right],\\
 \Delta_{y_b} &= - y_b y_N^2 \left[ \frac{1}{2} \log
 \frac{\mu^2}{M_N^2} + \frac{1}{4} \right],\\
 \Delta_{y_\tau} &= y_\tau y_N^2 \left[ \frac{1}{4} \log
 \frac{\mu^2}{M_N^2} + \frac{3}{8} \right],\\
 \Delta_{\lambda} &= - 4 \lambda y_N^2 \left[ \frac{1}{2} \log
 \frac{\mu^2}{M_N^2} + \frac{1}{4} \right] + 2 y_N^4 \left[
 \frac{1}{2} \log \frac{\mu^2}{M_N^2} - \frac{1}{2} \right].
\end{align}

\bibliographystyle{JHEP}
\bibliography{extramatter}

\providecommand{\href}[2]{#2}\begingroup\raggedright\begin{thebibliography}{100}

\bibitem{Cabibbo:1979ay}
N.~Cabibbo, L.~Maiani, G.~Parisi and R.~Petronzio, \emph{{Bounds on the
  Fermions and Higgs Boson Masses in Grand Unified Theories}},
  \href{https://doi.org/10.1016/0550-3213(79)90167-6}{\emph{Nucl. Phys.}
  {\bfseries B158} (1979) 295--305}.

\bibitem{Hung:1979dn}
P.~Q. Hung, \emph{{Vacuum Instability and New Constraints on Fermion Masses}},
  \href{https://doi.org/10.1103/PhysRevLett.42.873}{\emph{Phys. Rev. Lett.}
  {\bfseries 42} (1979) 873}.

\bibitem{Lindner:1988ww}
M.~Lindner, M.~Sher and H.~W. Zaglauer, \emph{{Probing Vacuum Stability Bounds
  at the Fermilab Collider}},
  \href{https://doi.org/10.1016/0370-2693(89)90540-6}{\emph{Phys. Lett.}
  {\bfseries B228} (1989) 139--143}.

\bibitem{Ford:1992mv}
C.~Ford, D.~R.~T. Jones, P.~W. Stephenson and M.~B. Einhorn, \emph{{The
  Effective potential and the renormalization group}},
  \href{https://doi.org/10.1016/0550-3213(93)90206-5}{\emph{Nucl. Phys.}
  {\bfseries B395} (1993) 17--34},
  [\href{https://arxiv.org/abs/hep-lat/9210033}{{\ttfamily hep-lat/9210033}}].

\bibitem{Casas:1994qy}
J.~A. Casas, J.~R. Espinosa and M.~Quiros, \emph{{Improved Higgs mass stability
  bound in the standard model and implications for supersymmetry}},
  \href{https://doi.org/10.1016/0370-2693(94)01404-Z}{\emph{Phys. Lett.}
  {\bfseries B342} (1995) 171--179},
  [\href{https://arxiv.org/abs/hep-ph/9409458}{{\ttfamily hep-ph/9409458}}].

\bibitem{Casas:1996aq}
J.~A. Casas, J.~R. Espinosa and M.~Quiros, \emph{{Standard model stability
  bounds for new physics within LHC reach}},
  \href{https://doi.org/10.1016/0370-2693(96)00682-X}{\emph{Phys. Lett.}
  {\bfseries B382} (1996) 374--382},
  [\href{https://arxiv.org/abs/hep-ph/9603227}{{\ttfamily hep-ph/9603227}}].

\bibitem{Einhorn:2007rv}
M.~B. Einhorn and D.~R.~T. Jones, \emph{{The Effective potential, the
  renormalisation group and vacuum stability}},
  \href{https://doi.org/10.1088/1126-6708/2007/04/051}{\emph{JHEP} {\bfseries
  04} (2007) 051}, [\href{https://arxiv.org/abs/hep-ph/0702295}{{\ttfamily
  hep-ph/0702295}}].

\bibitem{Ellis:2009tp}
J.~Ellis, J.~R. Espinosa, G.~F. Giudice, A.~Hoecker and A.~Riotto, \emph{{The
  Probable Fate of the Standard Model}},
  \href{https://doi.org/10.1016/j.physletb.2009.07.054}{\emph{Phys. Lett.}
  {\bfseries B679} (2009) 369--375},
  [\href{https://arxiv.org/abs/0906.0954}{{\ttfamily 0906.0954}}].

\bibitem{Degrassi:2012ry}
G.~Degrassi, S.~Di~Vita, J.~Elias-Miro, J.~R. Espinosa, G.~F. Giudice,
  G.~Isidori et~al., \emph{{Higgs mass and vacuum stability in the Standard
  Model at NNLO}}, \href{https://doi.org/10.1007/JHEP08(2012)098}{\emph{JHEP}
  {\bfseries 08} (2012) 098},
  [\href{https://arxiv.org/abs/1205.6497}{{\ttfamily 1205.6497}}].

\bibitem{Alekhin:2012py}
S.~Alekhin, A.~Djouadi and S.~Moch, \emph{{The top quark and Higgs boson masses
  and the stability of the electroweak vacuum}},
  \href{https://doi.org/10.1016/j.physletb.2012.08.024}{\emph{Phys. Lett.}
  {\bfseries B716} (2012) 214--219},
  [\href{https://arxiv.org/abs/1207.0980}{{\ttfamily 1207.0980}}].

\bibitem{Bezrukov:2012sa}
F.~Bezrukov, M.~{\relax Yu}. Kalmykov, B.~A. Kniehl and M.~Shaposhnikov,
  \emph{{Higgs Boson Mass and New Physics}},
  \href{https://doi.org/10.1007/JHEP10(2012)140}{\emph{JHEP} {\bfseries 10}
  (2012) 140}, [\href{https://arxiv.org/abs/1205.2893}{{\ttfamily 1205.2893}}].

\bibitem{Andreassen:2014gha}
A.~Andreassen, W.~Frost and M.~D. Schwartz, \emph{{Consistent Use of the
  Standard Model Effective Potential}},
  \href{https://doi.org/10.1103/PhysRevLett.113.241801}{\emph{Phys. Rev. Lett.}
  {\bfseries 113} (2014) 241801},
  [\href{https://arxiv.org/abs/1408.0292}{{\ttfamily 1408.0292}}].

\bibitem{DiLuzio:2014bua}
L.~Di~Luzio and L.~Mihaila, \emph{{On the gauge dependence of the Standard
  Model vacuum instability scale}},
  \href{https://doi.org/10.1007/JHEP06(2014)079}{\emph{JHEP} {\bfseries 06}
  (2014) 079}, [\href{https://arxiv.org/abs/1404.7450}{{\ttfamily 1404.7450}}].

\bibitem{Bednyakov:2015sca}
A.~V. Bednyakov, B.~A. Kniehl, A.~F. Pikelner and O.~L. Veretin,
  \emph{{Stability of the Electroweak Vacuum: Gauge Independence and Advanced
  Precision}},
  \href{https://doi.org/10.1103/PhysRevLett.115.201802}{\emph{Phys. Rev. Lett.}
  {\bfseries 115} (2015) 201802},
  [\href{https://arxiv.org/abs/1507.08833}{{\ttfamily 1507.08833}}].

\bibitem{Isidori:2001bm}
G.~Isidori, G.~Ridolfi and A.~Strumia, \emph{{On the metastability of the
  standard model vacuum}},
  \href{https://doi.org/10.1016/S0550-3213(01)00302-9}{\emph{Nucl. Phys.}
  {\bfseries B609} (2001) 387--409},
  [\href{https://arxiv.org/abs/hep-ph/0104016}{{\ttfamily hep-ph/0104016}}].

\bibitem{Arnold:1991cv}
P.~B. Arnold and S.~Vokos, \emph{{Instability of hot electroweak theory: bounds
  on m(H) and M(t)}},
  \href{https://doi.org/10.1103/PhysRevD.44.3620}{\emph{Phys. Rev.} {\bfseries
  D44} (1991) 3620--3627}.

\bibitem{DiLuzio:2015iua}
L.~Di~Luzio, G.~Isidori and G.~Ridolfi, \emph{{Stability of the electroweak
  ground state in the Standard Model and its extensions}},
  \href{https://doi.org/10.1016/j.physletb.2015.12.009}{\emph{Phys. Lett.}
  {\bfseries B753} (2016) 150--160},
  [\href{https://arxiv.org/abs/1509.05028}{{\ttfamily 1509.05028}}].

\bibitem{Espinosa:1995se}
J.~R. Espinosa and M.~Quiros, \emph{{Improved metastability bounds on the
  standard model Higgs mass}},
  \href{https://doi.org/10.1016/0370-2693(95)00572-3}{\emph{Phys. Lett.}
  {\bfseries B353} (1995) 257--266},
  [\href{https://arxiv.org/abs/hep-ph/9504241}{{\ttfamily hep-ph/9504241}}].

\bibitem{ArkaniHamed:2008ym}
N.~Arkani-Hamed, S.~Dubovsky, L.~Senatore and G.~Villadoro, \emph{{(No) Eternal
  Inflation and Precision Higgs Physics}},
  \href{https://doi.org/10.1088/1126-6708/2008/03/075}{\emph{JHEP} {\bfseries
  03} (2008) 075}, [\href{https://arxiv.org/abs/0801.2399}{{\ttfamily
  0801.2399}}].

\bibitem{EliasMiro:2011aa}
J.~Elias-Miro, J.~R. Espinosa, G.~F. Giudice, G.~Isidori, A.~Riotto and
  A.~Strumia, \emph{{Higgs mass implications on the stability of the
  electroweak vacuum}},
  \href{https://doi.org/10.1016/j.physletb.2012.02.013}{\emph{Phys. Lett.}
  {\bfseries B709} (2012) 222--228},
  [\href{https://arxiv.org/abs/1112.3022}{{\ttfamily 1112.3022}}].

\bibitem{Plascencia:2015pga}
A.~D. Plascencia and C.~Tamarit, \emph{{Convexity, gauge-dependence and
  tunneling rates}}, \href{https://doi.org/10.1007/JHEP10(2016)099}{\emph{JHEP}
  {\bfseries 10} (2016) 099},
  [\href{https://arxiv.org/abs/1510.07613}{{\ttfamily 1510.07613}}].

\bibitem{Espinosa:2016nld}
J.~R. Espinosa, M.~Garny, T.~Konstandin and A.~Riotto, \emph{{Gauge-Independent
  Scales Related to the Standard Model Vacuum Instability}},
  \href{https://doi.org/10.1103/PhysRevD.95.056004}{\emph{Phys. Rev.}
  {\bfseries D95} (2017) 056004},
  [\href{https://arxiv.org/abs/1608.06765}{{\ttfamily 1608.06765}}].

\bibitem{Lalak:2016zlv}
Z.~Lalak, M.~Lewicki and P.~Olszewski, \emph{{Gauge fixing and renormalization
  scale independence of tunneling rate in Abelian Higgs model and in the
  standard model}},
  \href{https://doi.org/10.1103/PhysRevD.94.085028}{\emph{Phys. Rev.}
  {\bfseries D94} (2016) 085028},
  [\href{https://arxiv.org/abs/1605.06713}{{\ttfamily 1605.06713}}].

\bibitem{Andreassen:2017rzq}
A.~Andreassen, W.~Frost and M.~D. Schwartz, \emph{{Scale Invariant Instantons
  and the Complete Lifetime of the Standard Model}},
  \href{https://arxiv.org/abs/1707.08124}{{\ttfamily 1707.08124}}.

\bibitem{Chigusa:2017dux}
S.~Chigusa, T.~Moroi and Y.~Shoji, \emph{{State-of-the-Art Calculation of the
  Decay Rate of Electroweak Vacuum in the Standard Model}},
  \href{https://doi.org/10.1103/PhysRevLett.119.211801}{\emph{Phys. Rev. Lett.}
  {\bfseries 119} (2017) 211801},
  [\href{https://arxiv.org/abs/1707.09301}{{\ttfamily 1707.09301}}].

\bibitem{Coleman:1977py}
S.~R. Coleman, \emph{{The Fate of the False Vacuum. 1. Semiclassical Theory}},
  \href{https://doi.org/10.1103/PhysRevD.15.2929,
  10.1103/PhysRevD.16.1248}{\emph{Phys. Rev.} {\bfseries D15} (1977)
  2929--2936}.

\bibitem{Callan:1977pt}
C.~G. Callan, Jr. and S.~R. Coleman, \emph{{The Fate of the False Vacuum. 2.
  First Quantum Corrections}},
  \href{https://doi.org/10.1103/PhysRevD.16.1762}{\emph{Phys. Rev.} {\bfseries
  D16} (1977) 1762--1768}.

\bibitem{Endo:2017gal}
M.~Endo, T.~Moroi, M.~M. Nojiri and Y.~Shoji, \emph{{On the Gauge Invariance of
  the Decay Rate of False Vacuum}},
  \href{https://doi.org/10.1016/j.physletb.2017.05.057}{\emph{Phys. Lett.}
  {\bfseries B771} (2017) 281--287},
  [\href{https://arxiv.org/abs/1703.09304}{{\ttfamily 1703.09304}}].

\bibitem{Endo:2017tsz}
M.~Endo, T.~Moroi, M.~M. Nojiri and Y.~Shoji, \emph{{False Vacuum Decay in
  Gauge Theory}}, \href{https://doi.org/10.1007/JHEP11(2017)074}{\emph{JHEP}
  {\bfseries 11} (2017) 074},
  [\href{https://arxiv.org/abs/1704.03492}{{\ttfamily 1704.03492}}].

\bibitem{ELVAS}
S.~Chigusa, T.~Moroi and Y.~Shoji, ``{ELVAS: c++ package for ELectroweak VAcuum
  Stability}.'' \url{https://github.com/YShoji-HEP/ELVAS/}.

\bibitem{Casas:1999cd}
J.~A. Casas, V.~Di~Clemente, A.~Ibarra and M.~Quiros, \emph{{Massive neutrinos
  and the Higgs mass window}},
  \href{https://doi.org/10.1103/PhysRevD.62.053005}{\emph{Phys. Rev.}
  {\bfseries D62} (2000) 053005},
  [\href{https://arxiv.org/abs/hep-ph/9904295}{{\ttfamily hep-ph/9904295}}].

\bibitem{Gogoladze:2008ak}
I.~Gogoladze, N.~Okada and Q.~Shafi, \emph{{Higgs Boson Mass Bounds in the
  Standard Model with Type III and Type I Seesaw}},
  \href{https://doi.org/10.1016/j.physletb.2008.08.023}{\emph{Phys. Lett.}
  {\bfseries B668} (2008) 121--125},
  [\href{https://arxiv.org/abs/0805.2129}{{\ttfamily 0805.2129}}].

\bibitem{He:2012ub}
B.~He, N.~Okada and Q.~Shafi, \emph{{125 GeV Higgs, type III seesaw and
  gauge-Higgs unification}},
  \href{https://doi.org/10.1016/j.physletb.2012.08.012}{\emph{Phys. Lett.}
  {\bfseries B716} (2012) 197--202},
  [\href{https://arxiv.org/abs/1205.4038}{{\ttfamily 1205.4038}}].

\bibitem{Rodejohann:2012px}
W.~Rodejohann and H.~Zhang, \emph{{Impact of massive neutrinos on the Higgs
  self-coupling and electroweak vacuum stability}},
  \href{https://doi.org/10.1007/JHEP06(2012)022}{\emph{JHEP} {\bfseries 06}
  (2012) 022}, [\href{https://arxiv.org/abs/1203.3825}{{\ttfamily 1203.3825}}].

\bibitem{Chakrabortty:2012np}
J.~Chakrabortty, M.~Das and S.~Mohanty, \emph{{Constraints on TeV scale
  Majorana neutrino phenomenology from the Vacuum Stability of the Higgs}},
  \href{https://doi.org/10.1142/S0217732313500326}{\emph{Mod. Phys. Lett.}
  {\bfseries A28} (2013) 1350032},
  [\href{https://arxiv.org/abs/1207.2027}{{\ttfamily 1207.2027}}].

\bibitem{Chao:2012mx}
W.~Chao, M.~Gonderinger and M.~J. Ramsey-Musolf, \emph{{Higgs Vacuum Stability,
  Neutrino Mass, and Dark Matter}},
  \href{https://doi.org/10.1103/PhysRevD.86.113017}{\emph{Phys. Rev.}
  {\bfseries D86} (2012) 113017},
  [\href{https://arxiv.org/abs/1210.0491}{{\ttfamily 1210.0491}}].

\bibitem{Masina:2012tz}
I.~Masina, \emph{{Higgs boson and top quark masses as tests of electroweak
  vacuum stability}},
  \href{https://doi.org/10.1103/PhysRevD.87.053001}{\emph{Phys. Rev.}
  {\bfseries D87} (2013) 053001},
  [\href{https://arxiv.org/abs/1209.0393}{{\ttfamily 1209.0393}}].

\bibitem{Khan:2012zw}
S.~Khan, S.~Goswami and S.~Roy, \emph{{Vacuum Stability constraints on the
  minimal singlet TeV Seesaw Model}},
  \href{https://doi.org/10.1103/PhysRevD.89.073021}{\emph{Phys. Rev.}
  {\bfseries D89} (2014) 073021},
  [\href{https://arxiv.org/abs/1212.3694}{{\ttfamily 1212.3694}}].

\bibitem{Dev:2013ff}
P.~S. Bhupal~Dev, D.~K. Ghosh, N.~Okada and I.~Saha, \emph{{125 GeV Higgs Boson
  and the Type-II Seesaw Model}},
  \href{https://doi.org/10.1007/JHEP03(2013)150,
  10.1007/JHEP05(2013)049}{\emph{JHEP} {\bfseries 03} (2013) 150},
  [\href{https://arxiv.org/abs/1301.3453}{{\ttfamily 1301.3453}}].

\bibitem{Kobakhidze:2013pya}
A.~Kobakhidze and A.~Spencer-Smith, \emph{{Neutrino Masses and Higgs Vacuum
  Stability}}, \href{https://doi.org/10.1007/JHEP08(2013)036}{\emph{JHEP}
  {\bfseries 08} (2013) 036},
  [\href{https://arxiv.org/abs/1305.7283}{{\ttfamily 1305.7283}}].

\bibitem{Datta:2013mta}
A.~Datta, A.~Elsayed, S.~Khalil and A.~Moursy, \emph{{Higgs vacuum stability in
  the $B-L$ extended standard model}},
  \href{https://doi.org/10.1103/PhysRevD.88.053011}{\emph{Phys. Rev.}
  {\bfseries D88} (2013) 053011},
  [\href{https://arxiv.org/abs/1308.0816}{{\ttfamily 1308.0816}}].

\bibitem{Chakrabortty:2013zja}
J.~Chakrabortty, P.~Konar and T.~Mondal, \emph{{Constraining a class of $B-L$
  extended models from vacuum stability and perturbativity}},
  \href{https://doi.org/10.1103/PhysRevD.89.056014}{\emph{Phys. Rev.}
  {\bfseries D89} (2014) 056014},
  [\href{https://arxiv.org/abs/1308.1291}{{\ttfamily 1308.1291}}].

\bibitem{Xiao:2014kba}
M.-L. Xiao and J.-H. Yu, \emph{{Stabilizing electroweak vacuum in a vectorlike
  fermion model}}, \href{https://doi.org/10.1103/PhysRevD.90.014007,
  10.1103/PhysRevD.90.019901}{\emph{Phys. Rev.} {\bfseries D90} (2014) 014007},
  [\href{https://arxiv.org/abs/1404.0681}{{\ttfamily 1404.0681}}].

\bibitem{Hamada:2014xka}
Y.~Hamada, H.~Kawai and K.-y. Oda, \emph{{Predictions on mass of Higgs portal
  scalar dark matter from Higgs inflation and flat potential}},
  \href{https://doi.org/10.1007/JHEP07(2014)026}{\emph{JHEP} {\bfseries 07}
  (2014) 026}, [\href{https://arxiv.org/abs/1404.6141}{{\ttfamily 1404.6141}}].

\bibitem{Khan:2014kba}
N.~Khan and S.~Rakshit, \emph{{Study of electroweak vacuum metastability with a
  singlet scalar dark matter}},
  \href{https://doi.org/10.1103/PhysRevD.90.113008}{\emph{Phys. Rev.}
  {\bfseries D90} (2014) 113008},
  [\href{https://arxiv.org/abs/1407.6015}{{\ttfamily 1407.6015}}].

\bibitem{Bambhaniya:2014hla}
G.~Bambhaniya, S.~Khan, P.~Konar and T.~Mondal, \emph{{Constraints on a seesaw
  model leading to quasidegenerate neutrinos and signatures at the LHC}},
  \href{https://doi.org/10.1103/PhysRevD.91.095007}{\emph{Phys. Rev.}
  {\bfseries D91} (2015) 095007},
  [\href{https://arxiv.org/abs/1411.6866}{{\ttfamily 1411.6866}}].

\bibitem{Khan:2015ipa}
N.~Khan and S.~Rakshit, \emph{{Constraints on inert dark matter from the
  metastability of the electroweak vacuum}},
  \href{https://doi.org/10.1103/PhysRevD.92.055006}{\emph{Phys. Rev.}
  {\bfseries D92} (2015) 055006},
  [\href{https://arxiv.org/abs/1503.03085}{{\ttfamily 1503.03085}}].

\bibitem{Salvio:2015cja}
A.~Salvio, \emph{{A Simple Motivated Completion of the Standard Model below the
  Planck Scale: Axions and Right-Handed Neutrinos}},
  \href{https://doi.org/10.1016/j.physletb.2015.03.015}{\emph{Phys. Lett.}
  {\bfseries B743} (2015) 428--434},
  [\href{https://arxiv.org/abs/1501.03781}{{\ttfamily 1501.03781}}].

\bibitem{Lindner:2015qva}
M.~Lindner, H.~H. Patel and B.~Radovcic, \emph{{Electroweak Absolute, Meta-,
  and Thermal Stability in Neutrino Mass Models}},
  \href{https://doi.org/10.1103/PhysRevD.93.073005}{\emph{Phys. Rev.}
  {\bfseries D93} (2016) 073005},
  [\href{https://arxiv.org/abs/1511.06215}{{\ttfamily 1511.06215}}].

\bibitem{Rose:2015fua}
L.~Delle~Rose, C.~Marzo and A.~Urbano, \emph{{On the stability of the
  electroweak vacuum in the presence of low-scale seesaw models}},
  \href{https://doi.org/10.1007/JHEP12(2015)050}{\emph{JHEP} {\bfseries 12}
  (2015) 050}, [\href{https://arxiv.org/abs/1506.03360}{{\ttfamily
  1506.03360}}].

\bibitem{Haba:2016zbu}
N.~Haba, H.~Ishida, N.~Okada and Y.~Yamaguchi, \emph{{Vacuum stability and
  naturalness in type-II seesaw}},
  \href{https://doi.org/10.1140/epjc/s10052-016-4180-z}{\emph{Eur. Phys. J.}
  {\bfseries C76} (2016) 333},
  [\href{https://arxiv.org/abs/1601.05217}{{\ttfamily 1601.05217}}].

\bibitem{Bambhaniya:2016rbb}
G.~Bambhaniya, P.~S. Bhupal~Dev, S.~Goswami, S.~Khan and W.~Rodejohann,
  \emph{{Naturalness, Vacuum Stability and Leptogenesis in the Minimal Seesaw
  Model}}, \href{https://doi.org/10.1103/PhysRevD.95.095016}{\emph{Phys. Rev.}
  {\bfseries D95} (2017) 095016},
  [\href{https://arxiv.org/abs/1611.03827}{{\ttfamily 1611.03827}}].

\bibitem{Khan:2016sxm}
N.~Khan, \emph{{Exploring Hyperchargeless Higgs Triplet Model up to the Planck
  Scale}},  \href{https://arxiv.org/abs/1610.03178}{{\ttfamily 1610.03178}}.

\bibitem{Garg:2017iva}
I.~Garg, S.~Goswami, V.~K. N. and N.~Khan, \emph{{Electroweak vacuum stability
  in presence of singlet scalar dark matter in TeV scale seesaw models}},
  \href{https://doi.org/10.1103/PhysRevD.96.055020}{\emph{Phys. Rev.}
  {\bfseries D96} (2017) 055020},
  [\href{https://arxiv.org/abs/1706.08851}{{\ttfamily 1706.08851}}].

\bibitem{Kusenko:1996bv}
A.~Kusenko, K.-M. Lee and E.~J. Weinberg, \emph{{Vacuum decay and internal
  symmetries}}, \href{https://doi.org/10.1103/PhysRevD.55.4903}{\emph{Phys.
  Rev.} {\bfseries D55} (1997) 4903--4909},
  [\href{https://arxiv.org/abs/hep-th/9609100}{{\ttfamily hep-th/9609100}}].

\bibitem{Coleman:1977th}
S.~R. Coleman, V.~Glaser and A.~Martin, \emph{{Action Minima Among Solutions to
  a Class of Euclidean Scalar Field Equations}},
  \href{https://doi.org/10.1007/BF01609421}{\emph{Commun. Math. Phys.}
  {\bfseries 58} (1978) 211}.

\bibitem{Blum:2016ipp}
K.~Blum, M.~Honda, R.~Sato, M.~Takimoto and K.~Tobioka, \emph{{O($N$)
  Invariance of the Multi-Field Bounce}},
  \href{https://doi.org/10.1007/JHEP05(2017)109,
  10.1007/JHEP06(2017)060}{\emph{JHEP} {\bfseries 05} (2017) 109},
  [\href{https://arxiv.org/abs/1611.04570}{{\ttfamily 1611.04570}}].

\bibitem{Fubini:1976jm}
S.~Fubini, \emph{{A New Approach to Conformal Invariant Field Theories}},
  \href{https://doi.org/10.1007/BF02785664}{\emph{Nuovo Cim.} {\bfseries A34}
  (1976) 521}.

\bibitem{Lipatov:1976ny}
L.~N. Lipatov, \emph{{Divergence of the Perturbation Theory Series and the
  Quasiclassical Theory}}, {\emph{Sov. Phys. JETP} {\bfseries 45} (1977)
  216--223}.

\bibitem{Gelfand:1959nq}
I.~M. Gelfand and A.~M. Yaglom, \emph{{Integration in functional spaces and it
  applications in quantum physics}},
  \href{https://doi.org/10.1063/1.1703636}{\emph{J. Math. Phys.} {\bfseries 1}
  (1960) 48}.

\bibitem{Dashen:1974ci}
R.~F. Dashen, B.~Hasslacher and A.~Neveu, \emph{{Nonperturbative Methods and
  Extended Hadron Models in Field Theory. 1. Semiclassical Functional
  Methods}}, \href{https://doi.org/10.1103/PhysRevD.10.4114}{\emph{Phys. Rev.}
  {\bfseries D10} (1974) 4114}.

\bibitem{Kirsten:2003py}
K.~Kirsten and A.~J. McKane, \emph{{Functional determinants by contour
  integration methods}},
  \href{https://doi.org/10.1016/S0003-4916(03)00149-0}{\emph{Annals Phys.}
  {\bfseries 308} (2003) 502--527},
  [\href{https://arxiv.org/abs/math-ph/0305010}{{\ttfamily math-ph/0305010}}].

\bibitem{Kirsten:2004qv}
K.~Kirsten and A.~J. McKane, \emph{{Functional determinants for general
  Sturm-Liouville problems}},
  \href{https://doi.org/10.1088/0305-4470/37/16/014}{\emph{J. Phys.} {\bfseries
  A37} (2004) 4649--4670},
  [\href{https://arxiv.org/abs/math-ph/0403050}{{\ttfamily math-ph/0403050}}].

\bibitem{Endo:2015ixx}
M.~Endo, T.~Moroi, M.~M. Nojiri and Y.~Shoji, \emph{{Renormalization-Scale
  Uncertainty in the Decay Rate of False Vacuum}},
  \href{https://doi.org/10.1007/JHEP01(2016)031}{\emph{JHEP} {\bfseries 01}
  (2016) 031}, [\href{https://arxiv.org/abs/1511.04860}{{\ttfamily
  1511.04860}}].

\bibitem{Burgess:2001tj}
C.~P. Burgess, V.~Di~Clemente and J.~R. Espinosa, \emph{{Effective operators
  and vacuum instability as heralds of new physics}},
  \href{https://doi.org/10.1088/1126-6708/2002/01/041}{\emph{JHEP} {\bfseries
  01} (2002) 041}, [\href{https://arxiv.org/abs/hep-ph/0201160}{{\ttfamily
  hep-ph/0201160}}].

\bibitem{Branchina:2013jra}
V.~Branchina and E.~Messina, \emph{{Stability, Higgs Boson Mass and New
  Physics}}, \href{https://doi.org/10.1103/PhysRevLett.111.241801}{\emph{Phys.
  Rev. Lett.} {\bfseries 111} (2013) 241801},
  [\href{https://arxiv.org/abs/1307.5193}{{\ttfamily 1307.5193}}].

\bibitem{Lalak:2014qua}
Z.~Lalak, M.~Lewicki and P.~Olszewski, \emph{{Higher-order scalar interactions
  and SM vacuum stability}},
  \href{https://doi.org/10.1007/JHEP05(2014)119}{\emph{JHEP} {\bfseries 05}
  (2014) 119}, [\href{https://arxiv.org/abs/1402.3826}{{\ttfamily 1402.3826}}].

\bibitem{Branchina:2014rva}
V.~Branchina, E.~Messina and M.~Sher, \emph{{Lifetime of the electroweak vacuum
  and sensitivity to Planck scale physics}},
  \href{https://doi.org/10.1103/PhysRevD.91.013003}{\emph{Phys. Rev.}
  {\bfseries D91} (2015) 013003},
  [\href{https://arxiv.org/abs/1408.5302}{{\ttfamily 1408.5302}}].

\bibitem{Branchina:2014usa}
V.~Branchina, E.~Messina and A.~Platania, \emph{{Top mass determination, Higgs
  inflation, and vacuum stability}},
  \href{https://doi.org/10.1007/JHEP09(2014)182}{\emph{JHEP} {\bfseries 09}
  (2014) 182}, [\href{https://arxiv.org/abs/1407.4112}{{\ttfamily 1407.4112}}].

\bibitem{Branchina:2015nda}
V.~Branchina and E.~Messina, \emph{{Stability and UV completion of the Standard
  Model}}, \href{https://doi.org/10.1209/0295-5075/117/61002}{\emph{EPL}
  {\bfseries 117} (2017) 61002},
  [\href{https://arxiv.org/abs/1507.08812}{{\ttfamily 1507.08812}}].

\bibitem{Branchina:2016bws}
V.~Branchina, E.~Messina and D.~Zappala, \emph{{Impact of Gravity on Vacuum
  Stability}}, \href{https://doi.org/10.1209/0295-5075/116/21001}{\emph{EPL}
  {\bfseries 116} (2016) 21001},
  [\href{https://arxiv.org/abs/1601.06963}{{\ttfamily 1601.06963}}].

\bibitem{Salvio:2016mvj}
A.~Salvio, A.~Strumia, N.~Tetradis and A.~Urbano, \emph{{On gravitational and
  thermal corrections to vacuum decay}},
  \href{https://doi.org/10.1007/JHEP09(2016)054}{\emph{JHEP} {\bfseries 09}
  (2016) 054}, [\href{https://arxiv.org/abs/1608.02555}{{\ttfamily
  1608.02555}}].

\bibitem{Bentivegna:2017qry}
E.~Bentivegna, V.~Branchina, F.~Contino and D.~Zappala, \emph{{Impact of New
  Physics on the EW vacuum stability in a curved spacetime background}},
  \href{https://doi.org/10.1007/JHEP12(2017)100}{\emph{JHEP} {\bfseries 12}
  (2017) 100}, [\href{https://arxiv.org/abs/1708.01138}{{\ttfamily
  1708.01138}}].

\bibitem{Patrignani:2016xqp}
{\scshape Particle Data Group} collaboration, C.~Patrignani et~al.,
  \emph{{Review of Particle Physics}},
  \href{https://doi.org/10.1088/1674-1137/40/10/100001}{\emph{Chin. Phys.}
  {\bfseries C40} (2016) 100001}.

\bibitem{Buttazzo:2013uya}
D.~Buttazzo, G.~Degrassi, P.~P. Giardino, G.~F. Giudice, F.~Sala, A.~Salvio
  et~al., \emph{{Investigating the near-criticality of the Higgs boson}},
  \href{https://doi.org/10.1007/JHEP12(2013)089}{\emph{JHEP} {\bfseries 12}
  (2013) 089}, [\href{https://arxiv.org/abs/1307.3536}{{\ttfamily 1307.3536}}].

\bibitem{Gorishnii:1990zu}
S.~G. Gorishnii, A.~L. Kataev, S.~A. Larin and L.~R. Surguladze,
  \emph{{Corrected Three Loop {QCD} Correction to the Correlator of the Quark
  Scalar Currents and gamma (Tot) (H0->Hadrons)}},
  \href{https://doi.org/10.1142/S0217732390003152}{\emph{Mod. Phys. Lett.}
  {\bfseries A5} (1990) 2703--2712}.

\bibitem{Tarasov:1980au}
O.~V. Tarasov, A.~A. Vladimirov and A.~{\relax Yu}. Zharkov, \emph{{The
  Gell-Mann-Low Function of QCD in the Three Loop Approximation}},
  \href{https://doi.org/10.1016/0370-2693(80)90358-5}{\emph{Phys. Lett.}
  {\bfseries 93B} (1980) 429--432}.

\bibitem{Gorishnii:1983zi}
S.~G. Gorishnii, A.~L. Kataev and S.~A. Larin, \emph{{Next Next-to-leading
  Perturbative {QCD} Corrections and Light Quark Masses}},
  \href{https://doi.org/10.1016/0370-2693(84)90315-0}{\emph{Phys. Lett.}
  {\bfseries 135B} (1984) 457--462}.

\bibitem{Kharchilava:1999yj}
A.~Kharchilava, \emph{{Top mass determination in leptonic final states with
  $J/\psi$}}, \href{https://doi.org/10.1016/S0370-2693(00)00120-9}{\emph{Phys.
  Lett.} {\bfseries B476} (2000) 73--78},
  [\href{https://arxiv.org/abs/hep-ph/9912320}{{\ttfamily hep-ph/9912320}}].

\bibitem{Hill:2005zy}
C.~S. Hill, J.~R. Incandela and J.~M. Lamb, \emph{{A Method for measurement of
  the top quark mass using the mean decay length of $b$ hadrons in $t \bar{t}$
  events}}, \href{https://doi.org/10.1103/PhysRevD.71.054029}{\emph{Phys. Rev.}
  {\bfseries D71} (2005) 054029},
  [\href{https://arxiv.org/abs/hep-ex/0501043}{{\ttfamily hep-ex/0501043}}].

\bibitem{Biswas:2010sa}
S.~Biswas, K.~Melnikov and M.~Schulze, \emph{{Next-to-leading order QCD effects
  and the top quark mass measurements at the LHC}},
  \href{https://doi.org/10.1007/JHEP08(2010)048}{\emph{JHEP} {\bfseries 08}
  (2010) 048}, [\href{https://arxiv.org/abs/1006.0910}{{\ttfamily 1006.0910}}].

\bibitem{Agashe:2012bn}
K.~Agashe, R.~Franceschini and D.~Kim, \emph{{Simple ''invariance'' of two-body
  decay kinematics}},
  \href{https://doi.org/10.1103/PhysRevD.88.057701}{\emph{Phys. Rev.}
  {\bfseries D88} (2013) 057701},
  [\href{https://arxiv.org/abs/1209.0772}{{\ttfamily 1209.0772}}].

\bibitem{Alioli:2013mxa}
S.~Alioli, P.~Fernandez, J.~Fuster, A.~Irles, S.-O. Moch, P.~Uwer et~al.,
  \emph{{A new observable to measure the top-quark mass at hadron colliders}},
  \href{https://doi.org/10.1140/epjc/s10052-013-2438-2}{\emph{Eur. Phys. J.}
  {\bfseries C73} (2013) 2438},
  [\href{https://arxiv.org/abs/1303.6415}{{\ttfamily 1303.6415}}].

\bibitem{Kawabataa:2014osa}
S.~Kawabata, Y.~Shimizu, Y.~Sumino and H.~Yokoya, \emph{{Weight function method
  for precise determination of top quark mass at Large Hadron Collider}},
  \href{https://doi.org/10.1016/j.physletb.2014.12.044}{\emph{Phys. Lett.}
  {\bfseries B741} (2015) 232--238},
  [\href{https://arxiv.org/abs/1405.2395}{{\ttfamily 1405.2395}}].

\bibitem{Ravasio:2018lzi}
S.~Ferrario~Ravasio, T.~Jezo, P.~Nason and C.~Oleari, \emph{{A Theoretical
  Study of Top-Mass Measurements at the LHC Using NLO+PS Generators of
  Increasing Accuracy}},  \href{https://arxiv.org/abs/1801.03944}{{\ttfamily
  1801.03944}}.

\bibitem{Horiguchi:2013wra}
T.~Horiguchi, A.~Ishikawa, T.~Suehara, K.~Fujii, Y.~Sumino, Y.~Kiyo et~al.,
  \emph{{Study of top quark pair production near threshold at the ILC}},
  \href{https://arxiv.org/abs/1310.0563}{{\ttfamily 1310.0563}}.

\bibitem{Espinosa:2007qp}
J.~R. Espinosa, G.~F. Giudice and A.~Riotto, \emph{{Cosmological implications
  of the Higgs mass measurement}},
  \href{https://doi.org/10.1088/1475-7516/2008/05/002}{\emph{JCAP} {\bfseries
  0805} (2008) 002}, [\href{https://arxiv.org/abs/0710.2484}{{\ttfamily
  0710.2484}}].

\bibitem{Lebedev:2012sy}
O.~Lebedev and A.~Westphal, \emph{{Metastable Electroweak Vacuum: Implications
  for Inflation}},
  \href{https://doi.org/10.1016/j.physletb.2012.12.069}{\emph{Phys. Lett.}
  {\bfseries B719} (2013) 415--418},
  [\href{https://arxiv.org/abs/1210.6987}{{\ttfamily 1210.6987}}].

\bibitem{Kobakhidze:2013tn}
A.~Kobakhidze and A.~Spencer-Smith, \emph{{Electroweak Vacuum (In)Stability in
  an Inflationary Universe}},
  \href{https://doi.org/10.1016/j.physletb.2013.04.013}{\emph{Phys. Lett.}
  {\bfseries B722} (2013) 130--134},
  [\href{https://arxiv.org/abs/1301.2846}{{\ttfamily 1301.2846}}].

\bibitem{Enqvist:2014bua}
K.~Enqvist, T.~Meriniemi and S.~Nurmi, \emph{{Higgs Dynamics during
  Inflation}}, \href{https://doi.org/10.1088/1475-7516/2014/07/025}{\emph{JCAP}
  {\bfseries 1407} (2014) 025},
  [\href{https://arxiv.org/abs/1404.3699}{{\ttfamily 1404.3699}}].

\bibitem{Herranen:2014cua}
M.~Herranen, T.~Markkanen, S.~Nurmi and A.~Rajantie, \emph{{Spacetime curvature
  and the Higgs stability during inflation}},
  \href{https://doi.org/10.1103/PhysRevLett.113.211102}{\emph{Phys. Rev. Lett.}
  {\bfseries 113} (2014) 211102},
  [\href{https://arxiv.org/abs/1407.3141}{{\ttfamily 1407.3141}}].

\bibitem{Kobakhidze:2014xda}
A.~Kobakhidze and A.~Spencer-Smith, \emph{{The Higgs vacuum is unstable}},
  \href{https://arxiv.org/abs/1404.4709}{{\ttfamily 1404.4709}}.

\bibitem{Kamada:2014ufa}
K.~Kamada, \emph{{Inflationary cosmology and the standard model Higgs with a
  small Hubble induced mass}},
  \href{https://doi.org/10.1016/j.physletb.2015.01.024}{\emph{Phys. Lett.}
  {\bfseries B742} (2015) 126--135},
  [\href{https://arxiv.org/abs/1409.5078}{{\ttfamily 1409.5078}}].

\bibitem{Herranen:2015ima}
M.~Herranen, T.~Markkanen, S.~Nurmi and A.~Rajantie, \emph{{Spacetime curvature
  and Higgs stability after inflation}},
  \href{https://doi.org/10.1103/PhysRevLett.115.241301}{\emph{Phys. Rev. Lett.}
  {\bfseries 115} (2015) 241301},
  [\href{https://arxiv.org/abs/1506.04065}{{\ttfamily 1506.04065}}].

\bibitem{Hook:2014uia}
A.~Hook, J.~Kearney, B.~Shakya and K.~M. Zurek, \emph{{Probable or Improbable
  Universe? Correlating Electroweak Vacuum Instability with the Scale of
  Inflation}}, \href{https://doi.org/10.1007/JHEP01(2015)061}{\emph{JHEP}
  {\bfseries 01} (2015) 061},
  [\href{https://arxiv.org/abs/1404.5953}{{\ttfamily 1404.5953}}].

\bibitem{Espinosa:2015qea}
J.~R. Espinosa, G.~F. Giudice, E.~Morgante, A.~Riotto, L.~Senatore, A.~Strumia
  et~al., \emph{{The cosmological Higgstory of the vacuum instability}},
  \href{https://doi.org/10.1007/JHEP09(2015)174}{\emph{JHEP} {\bfseries 09}
  (2015) 174}, [\href{https://arxiv.org/abs/1505.04825}{{\ttfamily
  1505.04825}}].

\bibitem{Ema:2016kpf}
Y.~Ema, K.~Mukaida and K.~Nakayama, \emph{{Fate of Electroweak Vacuum during
  Preheating}},
  \href{https://doi.org/10.1088/1475-7516/2016/10/043}{\emph{JCAP} {\bfseries
  1610} (2016) 043}, [\href{https://arxiv.org/abs/1602.00483}{{\ttfamily
  1602.00483}}].

\bibitem{Enqvist:2016mqj}
K.~Enqvist, M.~Karciauskas, O.~Lebedev, S.~Rusak and M.~Zatta,
  \emph{{Postinflationary vacuum instability and Higgs-inflaton couplings}},
  \href{https://doi.org/10.1088/1475-7516/2016/11/025}{\emph{JCAP} {\bfseries
  1611} (2016) 025}, [\href{https://arxiv.org/abs/1608.08848}{{\ttfamily
  1608.08848}}].

\bibitem{Kearney:2015vba}
J.~Kearney, H.~Yoo and K.~M. Zurek, \emph{{Is a Higgs Vacuum Instability Fatal
  for High-Scale Inflation?}},
  \href{https://doi.org/10.1103/PhysRevD.91.123537}{\emph{Phys. Rev.}
  {\bfseries D91} (2015) 123537},
  [\href{https://arxiv.org/abs/1503.05193}{{\ttfamily 1503.05193}}].

\bibitem{Kohri:2016wof}
K.~Kohri and H.~Matsui, \emph{{Higgs vacuum metastability in primordial
  inflation, preheating, and reheating}},
  \href{https://doi.org/10.1103/PhysRevD.94.103509}{\emph{Phys. Rev.}
  {\bfseries D94} (2016) 103509},
  [\href{https://arxiv.org/abs/1602.02100}{{\ttfamily 1602.02100}}].

\bibitem{East:2016anr}
W.~E. East, J.~Kearney, B.~Shakya, H.~Yoo and K.~M. Zurek, \emph{{Spacetime
  Dynamics of a Higgs Vacuum Instability During Inflation}},
  \href{https://doi.org/10.1103/PhysRevD.95.023526}{\emph{Phys. Rev.}
  {\bfseries D95} (2017) 023526},
  [\href{https://arxiv.org/abs/1607.00381}{{\ttfamily 1607.00381}}].

\bibitem{Ema:2017loe}
Y.~Ema, M.~Karciauskas, O.~Lebedev and M.~Zatta, \emph{{Early Universe Higgs
  dynamics in the presence of the Higgs-inflaton and non-minimal Higgs-gravity
  couplings}}, \href{https://doi.org/10.1088/1475-7516/2017/06/054}{\emph{JCAP}
  {\bfseries 1706} (2017) 054},
  [\href{https://arxiv.org/abs/1703.04681}{{\ttfamily 1703.04681}}].

\bibitem{Postma:2017hbk}
M.~Postma and J.~van~de Vis, \emph{{Electroweak stability and non-minimal
  coupling}}, \href{https://doi.org/10.1088/1475-7516/2017/05/004}{\emph{JCAP}
  {\bfseries 1705} (2017) 004},
  [\href{https://arxiv.org/abs/1702.07636}{{\ttfamily 1702.07636}}].

\bibitem{Joti:2017fwe}
A.~Joti, A.~Katsis, D.~Loupas, A.~Salvio, A.~Strumia, N.~Tetradis et~al.,
  \emph{{(Higgs) vacuum decay during inflation}},
  \href{https://doi.org/10.1007/JHEP07(2017)058}{\emph{JHEP} {\bfseries 07}
  (2017) 058}, [\href{https://arxiv.org/abs/1706.00792}{{\ttfamily
  1706.00792}}].

\bibitem{Fairbairn:2014zia}
M.~Fairbairn and R.~Hogan, \emph{{Electroweak Vacuum Stability in light of
  BICEP2}}, \href{https://doi.org/10.1103/PhysRevLett.112.201801}{\emph{Phys.
  Rev. Lett.} {\bfseries 112} (2014) 201801},
  [\href{https://arxiv.org/abs/1403.6786}{{\ttfamily 1403.6786}}].

\bibitem{Shkerin:2015exa}
A.~Shkerin and S.~Sibiryakov, \emph{{On stability of electroweak vacuum during
  inflation}},
  \href{https://doi.org/10.1016/j.physletb.2015.05.012}{\emph{Phys. Lett.}
  {\bfseries B746} (2015) 257--260},
  [\href{https://arxiv.org/abs/1503.02586}{{\ttfamily 1503.02586}}].

\bibitem{Ema:2017rkk}
Y.~Ema, K.~Mukaida and K.~Nakayama, \emph{{Electroweak Vacuum Metastability and
  Low-scale Inflation}},
  \href{https://doi.org/10.1088/1475-7516/2017/12/030}{\emph{JCAP} {\bfseries
  1712} (2017) 030}, [\href{https://arxiv.org/abs/1706.08920}{{\ttfamily
  1706.08920}}].

\bibitem{Ade:2015xua}
{\scshape Planck} collaboration, P.~A.~R. Ade et~al., \emph{{Planck 2015
  results. XIII. Cosmological parameters}},
  \href{https://doi.org/10.1051/0004-6361/201525830}{\emph{Astron. Astrophys.}
  {\bfseries 594} (2016) A13},
  [\href{https://arxiv.org/abs/1502.01589}{{\ttfamily 1502.01589}}].

\bibitem{Minkowski:1977sc}
P.~Minkowski, \emph{{$\mu \to e\gamma$ at a Rate of One Out of $10^{9}$ Muon
  Decays?}}, \href{https://doi.org/10.1016/0370-2693(77)90435-X}{\emph{Phys.
  Lett.} {\bfseries 67B} (1977) 421--428}.

\bibitem{Yanagida:1979as}
T.~Yanagida, \emph{{HORIZONTAL SYMMETRY AND MASSES OF NEUTRINOS}}, {\emph{Conf.
  Proc.} {\bfseries C7902131} (1979) 95--99}.

\bibitem{GellMann:1980vs}
M.~Gell-Mann, P.~Ramond and R.~Slansky, \emph{{Complex Spinors and Unified
  Theories}}, {\emph{Conf. Proc.} {\bfseries C790927} (1979) 315--321},
  [\href{https://arxiv.org/abs/1306.4669}{{\ttfamily 1306.4669}}].

\bibitem{Machacek:1983tz}
M.~E. Machacek and M.~T. Vaughn, \emph{{Two Loop Renormalization Group
  Equations in a General Quantum Field Theory. 1. Wave Function
  Renormalization}},
  \href{https://doi.org/10.1016/0550-3213(83)90610-7}{\emph{Nucl. Phys.}
  {\bfseries B222} (1983) 83--103}.

\bibitem{Machacek:1983fi}
M.~E. Machacek and M.~T. Vaughn, \emph{{Two Loop Renormalization Group
  Equations in a General Quantum Field Theory. 2. Yukawa Couplings}},
  \href{https://doi.org/10.1016/0550-3213(84)90533-9}{\emph{Nucl. Phys.}
  {\bfseries B236} (1984) 221--232}.

\bibitem{Machacek:1984zw}
M.~E. Machacek and M.~T. Vaughn, \emph{{Two Loop Renormalization Group
  Equations in a General Quantum Field Theory. 3. Scalar Quartic Couplings}},
  \href{https://doi.org/10.1016/0550-3213(85)90040-9}{\emph{Nucl. Phys.}
  {\bfseries B249} (1985) 70--92}.

\bibitem{Luo:2002ti}
M.-x. Luo, H.-w. Wang and Y.~Xiao, \emph{{Two loop renormalization group
  equations in general gauge field theories}},
  \href{https://doi.org/10.1103/PhysRevD.67.065019}{\emph{Phys. Rev.}
  {\bfseries D67} (2003) 065019},
  [\href{https://arxiv.org/abs/hep-ph/0211440}{{\ttfamily hep-ph/0211440}}].

\bibitem{Chankowski:1993tx}
P.~H. Chankowski and Z.~Pluciennik, \emph{{Renormalization group equations for
  seesaw neutrino masses}},
  \href{https://doi.org/10.1016/0370-2693(93)90330-K}{\emph{Phys. Lett.}
  {\bfseries B316} (1993) 312--317},
  [\href{https://arxiv.org/abs/hep-ph/9306333}{{\ttfamily hep-ph/9306333}}].

\bibitem{Avan:1985eg}
J.~Avan and H.~J. De~Vega, \emph{{INVERSE SCATTERING TRANSFORM AND INSTANTONS
  OF FOUR-DIMENSIONAL YUKAWA AND phi**4 THEORIES}},
  \href{https://doi.org/10.1016/0550-3213(86)90515-8}{\emph{Nucl. Phys.}
  {\bfseries B269} (1986) 621--664}.

\end{thebibliography}\endgroup

\end{document}